\documentclass[a4paper,11pt]{article}
\usepackage{jheppub}
\usepackage[T1]{fontenc}
\usepackage{hyperref}
\usepackage{tensor}
\usepackage{epstopdf}
\usepackage{epsfig}
\usepackage{subfigure}
\usepackage[dvipsnames]{xcolor}
\usepackage[title]{appendix}
\usepackage{amsmath,epsfig}
\usepackage{amssymb,amsfonts}
\usepackage{latexsym}
\usepackage{bbold}
\usepackage[latin1]{inputenc}
\usepackage{subeqnarray}
\usepackage{xcolor}
\usepackage{graphicx}
\usepackage{longtable}
\usepackage{color}
\usepackage{hyperref}
\usepackage{cleveref}
\usepackage{float}
\usepackage{pdfpages}
\usepackage{pdflscape}
\usepackage{caption}
\hypersetup{
       colorlinks=true,
      filecolor=black,
       anchorcolor=blue,
      linkcolor=blue,
      citecolor=blue,
      urlcolor=blue,
       linktocpage=true,
        plainpages=false,
        breaklinks=true,
        pdfstartview=FitH
           }

\relax
\DeclareMathOperator\arctanh{arctanh}

\renewcommand{\theequation}{\arabic{section}.\arabic{equation}}

\usepackage{changes}
	\definechangesauthor[name={Per cusse}, color=orange]{per}
\begin{document}
\title{\boldmath Interior Structure and Complexity Growth Rate of Holographic Superconductor from M-Theory}

\author{Yu-Sen An$^{1}$, Li Li$^{2,3,4,5}$, Fu-Guo Yang$^{2,3,4}$, Run-Qiu Yang$^{6}$}

\affiliation{$^1$ Department of Physics, Fudan University Shanghai 200433, China}

\affiliation{$^2$CAS Key Laboratory of Theoretical Physics, Institute of Theoretical Physics, Chinese Academy of Sciences, P.O. Box 2735, Beijing 100190, China.}

\affiliation{$^3$School of Physical Sciences, University of Chinese Academy of Sciences, No.19A Yuquan Road, Beijing 100049, China.}

\affiliation{$^4$School of Fundamental Physics and Mathematical Sciences, Hangzhou Institute for Advanced Study, UCAS, Hangzhou 310024, China.}

\affiliation{$^5$Peng Huanwu Collaborative Center for Research and Education, Beihang University, Beijing 100191, China.}

\affiliation{$^6$Center for Joint Quantum Studies and Department of Physics
School of Science, Tianjin University,Yaguan Road 135, Jinnan District, Tianjin 300350, China}

\emailAdd{anyusen@fudan.edu.cn}
\emailAdd{liliphy@itp.ac.cn}
\emailAdd{yangfuguo@ucas.ac.cn}
\emailAdd{aqiu@tju.edu.cn}

%\pacs{PACS}
%\keywords{keywords}
%\preprint{preprint}
%%%%%%%%%%%%%%%%%%%%%%%%%%%%%%%%%%%%%%
%\begin{abstract}
%%%%%%%%%%%%%%%%%%%%%%%%%%%%%%%%%%%%%%
\abstract{We study the interior dynamics of a top-down holographic superconductor from M-theory. The condense of the charged scalar hair necessarily removes the inner Cauchy horizon and the spacetime ends at a spacelike singularity.
Although there is a smooth superconducting phase transition at the critical temperature, the onset of superconductivity is accompanied by intricate interior dynamics, including the collapse of the Einstein-Rosen bridge, the Josephson oscillations of the condensate, and the final Kasner singularity. We obtain analytically the transformation rule for the alternation of different Kasner epochs. Thanks to the nonlinear couplings of the top-down theory, there is generically a never-ending chaotic alternation of Kasner epochs towards the singularity. We compute the holographic complexity using both the complexity-action and the complexity-volume dualities. In contrast to the latter, the complexity growth rate from the complexity-action duality has a discontinuity at the critical temperature, characterizing the sudden change of the internal structure before and after the superconducting phase transition.}

%\end{abstract}
%%%%%%%%%%%%%%%%%%%%%%%%%%%%%%%%%%%%%%
\maketitle
%\tableofcontents
\flushbottom

%%%%%%%%%%%%%%%%%%%%%%%%%%%%%%%%%%%%%%
\noindent
\newpage

\section{Introduction}\label{intro}

Although significant progress has been made toward the black hole interiors, it is still an open and interesting question to understand the geometry of spacetime inside generic black holes. It is obvious that the established picture for the exterior of a black hole is in dramatic contrast with its interior, for which two well-known features inside a black hole are the spacetime singularity and the inner Cauchy horizon-the boundary of the domain of dependence for Cauchy data prescribed in the black hole exterior. The asymptotic geometry near the singularity was studied in the seminal work by Belinski, Khalatnikov and Lifshitz (BKL)~\cite{Lifshitz:1963ps,Belinsky:1970ew,Belinski:1973zz}, and found to be characterized by chaotic oscillations of the Kasner epochs as approaching the singularity in many cases. The presence of the inner Cauchy horizon violates the predictability of the classical dynamics even far away from any singularity. This breakdown of predictability at the Cauchy horizon is prohibited by the strong cosmic censorship conjecture~\cite{Isenberg:2015rqa,Ringstrom:2015jza} which itself is still under investigation. Moreover, understanding the pre-asymptotic regime away from the singularity and the instability of the inner Cauchy horizon is a rather intricate and interesting question.

Motivated by the holographic duality, the interior structure of black holes with charged scalar hair has recently attracted much attention~\cite{Cai:2020wrp,Hartnoll:2020fhc,An:2021plu}. The case in asymptotic anti-de Sitter (AdS) spacetime is relevant to holographic superconductors~\cite{Hartnoll:2008kx} where the scalar hair is spontaneously generated below a critical temperature $T_c$ (for review see~\cite{Cai:2015cya}). It was shown that the presence of scalar hair generically destroys the inner Cauchy horizon, and thus the geometry inside the hairy black holes approaches a spacelike singularity. Remarkably, the onset of the scalar hair outside the event horizon results in significantly intricate dynamics inside~\cite{Hartnoll:2020fhc}. Slightly below $T_c$, the development of the scalar hair triggers an instability of the Cauchy horizon of the AdS Reissner-Nordstr\"{o}m (RN) black hole. Inside the superconducting black hole, the interior evolves through several distinct epochs, including a collapse of the Einstein-Rosen (ER) bridge, Josephson oscillations of the scalar condensate, and finally the Kasner singularity. Moreover, those behaviors depend on the nonlinear details of the model one considers. The presence of the Josephson oscillations might be removed by adding new couplings~\cite{Sword:2021pfm}. More recently, the generalization to an anisotropic black hole with vector hair (holographic $P$-wave superconductor~\cite{Cai:2013pda,Cai:2013aca}) was investigated in~\cite{Cai:2021obq} where some new features were identified.\footnote{See also ~\cite{Hartnoll:2020rwq,Grandi:2021ajl,Mansoori:2021wxf,VandeMoortel:2021gsp,Dias:2021afz,Devecioglu:2021xug,Wang:2020nkd,Henneaux:2022ijt} for other recent discussions of the interior of hairy black holes.}

In the present work, we are interested in the internal structure of a top-down holographic superconductor from string/M-theory. Our motivation is two-fold. On the one hand, all the above studies considered some effective models. In such a bottom-up approach, the effective theory is chosen to capture some desired features (\emph{e.g.} the spontaneously $U(1)$ symmetry breaking for the superconducting phase transition) and sometimes can give new conceptual insights into the physics we are interested in. Nevertheless, the Lagrangian is written down by hand and it is not clear whether such a bottom-up model is well-defined or can be embedded in a UV complete setting. In contrast, in the top-down approach, one starts with the UV-complete theory, allowing us to beyond the effective theory point of view. The precise dictionary between the field theory and its dual gravity theory is known. Moreover, although it is technically much more involved, the interactions and model parameters in a top-down model are completely fixed. On the other hand, our recent study~\cite{Cai:2020wrp,An:2021plu} suggested that if the scalar potential contains an exponential term\,\footnote{We consider the case in which the kinetic term of a scalar $\psi$ takes the canonical form, $-\frac{1}{2}(\partial_\mu\psi)^2$.}, there would be some non-trivial impact on the asymptotic geometry near the singularity. Such an exponential term typically appears in top-down models from the string/M-theory. The study of holographic superconductors in the top-down approach was initialed by~\cite{Gauntlett:2009dn, Gubser:2009qm} where the authors obtained the hairy solutions outside the event horizon numerically and discussed the spontaneous $U(1)$ symmetry breaking associated with the superconducting transition. The ground state solution at zero temperature was then constructed in~\cite{Gubser:2009gp,Gauntlett:2009bh}.

To be specific, we consider the four-dimensional top-down theory from a consistent truncation of M-theory~\cite{Gauntlett:2009dn,Gubser:2009gp} and study the interior dynamics of the hairy black hole that describes a superconducting phase from the viewpoint of the dual field theory. We first consider a generalized theory in St\"{u}ckelberg form and give a general proof of no inner horizon of these black holes, and thus the singularity inside is spacelike. Then we focus on the top-down holographic superconductor~\cite{Gauntlett:2009dn,Gubser:2009gp}. Near the critical temperature $T_c$, we find the collapse of the ER bridge and the Josephson oscillations of the scalar condensate. At the end of the Josephson oscillation, the scalar grows logarithmically and the system enters the era described by a Kasner universe~\eqref{kasner} with a single parameter $\alpha$ of~\eqref{kasnerep} controlling the Kasner exponents.

In contrast to previous studies where the system with scalar hairs ends up in a stable Kasner regime, in our case, the geometry enters into an endless number of Kasner regimes. There is an infinite number of alternations from one Kasner regime to another with new exponents. Moreover, taking advantage of both numerical and analytical approaches, we are able to obtain the transformation rule between two Kasner epochs, including the Kasner inversion and the Kanser transition. In the former, the values of $\alpha$ for two adjacent Kasner epochs are just the inverse of each other, while in the latter, the sum of two adjacent values of $\alpha$ is a constant. Interestingly, we find that our transformation law is different from the usual BKL type rules~\cite{Belinsky:1970ew} in four dimensions due to the particular exponential couplings in the top-down model. There is a chaotic feature of the Kasner exponents for which the underlying patterns are highly sensitive to initial conditions.

Holography also provides some useful probes to the interior of a black hole. A particularly interesting probe is the computational complexity. In quantum information, the complexity is defined by the minimal number of quantum gates of the unitary operator that evolves the initial states to the final states. There are two well-known holographic proposals of complexity. The first one is called ``complexity-volume (CV) duality'' which relates the size of a ``wormhole'' to the computational complexity of the dual quantum state~\cite{Stanford:2014jda}. The second is the ``complexity-action (CA) duality'' which relates the action of the bulk Wheeler-DeWitt (WdW) patch to the boundary complexity~\cite{Brown:2015bva, Brown:2015lvg, Lehner:2016vdi}. Both of them depend on the interior geometry of a black hole, thus it is difficult to compute the complexity of a black hole that does not allow an analytic solution. While there are many investigations of complexity for analytic solutions in the literature, see, \emph{e.g.}~\cite{An:2020tkn, An:2018xhv, Swingle:2017zcd,Ghodrati:2018hss,Cai:2016xho,Chapman:2016hwi,Jiang:2018pfk,Jiang:2019qea,Babaei-Aghbolagh:2021ast}, the complexity in holographic superconductor have been only considered by using the CV conjecture~\cite{Yang:2019gce} and sub-region CV conjecture~\cite{Lai:2022tjr}. The study of the holographic superconductor using the CA conjecture has not been done, as the WdW action relies on the whole internal geometry of a black hole. More recently, it was argued that the CV complexity generally fails to fully probe the interior geometry~\cite{Caceres:2022smh}. After knowing the interior structure clearly, we study the complexity of the top-down holographic superconductor using both the CA duality and the CV duality. Compared with the CV case, the complexity growth rate by CA conjecture is sensitive to the change of the inner structure of a black hole and could be a good probe to the black hole's interior dynamics.

The structure of this paper is as follows. In Section~\ref{sec:setup}, we introduce the model. In particular, we generalize the no inner Cauchy horizon theorem to the theory with St\"{u}ckelberg form. Section~\ref{sec:inner} is devoted to discussing the interior dynamics of the superconducting black hole. In Section~\ref{sec:CAconj}, we compute the holographic complexity using the CA conjecture. Finally, we conclude with some discussions in Section~\ref{diss-sum}. More examples of the configuration inside the hairy black holes are presented in Appendix~\ref{alpha-diagram}. Relation to the billiard approach is discussed in Appendix~\ref{billard}. The calculation of the complexity growth rate using the CV conjecture is presented in Appendix~\ref{CVconj}.

\section{The Setup}\label{sec:setup}
We begin with writing a general bulk theory describing the holographic superconductor in the generalized St\"{u}ckelberg form~\cite{Franco:2009yz,Kiritsis:2015hoa}:
\begin{equation}\label{modelst}
\mathcal{L}_{S}^{(d+1)}=-\frac{1}{2}(\partial_\mu\psi)^2-\mathcal{F}(\psi)(\partial_\mu\theta-q A_\mu)^2-V(\psi)-\frac{Z(\psi)}{4}F_{\mu\nu}F^{\mu\nu}\,,
\end{equation}
where $\psi$ and $\theta$ are two real scalars, and $A_\mu$ is the $U(1)$ gauge field with its strength $F_{\mu\nu}=\partial_\mu A_\nu-\partial_\nu A_\mu$. $\mathcal{F}$, $V$ and $Z$ are general functions of real scalar $\psi$ that takes in general any real value. One demands $\mathcal{F}$ and $Z$ to be positive to ensure positivity of the kinetic term for $\theta$ and $A_\mu$, respectively. The charged scalar action~\cite{Cai:2020wrp,Hartnoll:2020fhc,An:2021plu} can be written into the above St\"{u}ckelberg form by $\Psi=\psi e^{i\theta}$. Nevertheless, in this generalized class of theories, the two real degrees of freedom $\psi$ and $\theta$ are not necessarily associated with the magnitude and phase of a complex scalar, respectively.

\subsection{No smooth inner horizon for hairy black holes}
Before going to a specific model, we now show that a hairy black hole of the following form cannot have a smooth inner horizon.
\begin{equation}\label{bkansatzst}
\mathrm{d}s^2=\frac{1}{z^2}\left[-f(z)\mathrm{e}^{-\chi(z)}\mathrm{d}t^2+\frac{\mathrm{d}z^2}{f(z)}+\mathrm{d}\Sigma^2_{d-1,k}\right],\quad \psi=\psi(z),\quad A=A_t(z) \mathrm{d}t\,.
\end{equation}
where $\mathrm{d}\Sigma^2_{d-1,k}$ denotes the standard metric of unit sphere $(k=1)$, planar $(k=0)$ or unit hyperbolic plane $(k=-1)$.
For black hole solutions with a regular event horizon at $z_H$, we have $f(z_H)=0$ and all the functions are continuous near the horizon. The Hawking temperature is given by
\begin{equation}\label{HawkingT}
  T=-\frac{\mathrm{e}^{-\chi(z_H)/2}f'(z_H)}{4\pi}\,.
\end{equation}
The equations of motion are given as follows.
\begin{equation}\label{eomst}
\begin{aligned}
\theta'&=0\,,\\
\frac{z^{d+1}}{\mathrm{e}^{-\chi/2}}\left(\frac{\mathrm{e}^{-\chi/2}}{z^{d-1}}f\psi'\right)'&=\frac{\mathrm{d}\mathcal{F}}{\mathrm{d}\psi}q^2z^2\mathrm{e}^{\chi}fA_t^2-\frac{\mathrm{d}V}{\mathrm{d}\psi}+\frac{\mathrm{d}Z}{\mathrm{d}\psi}\frac{z^4\mathrm{e}^{\chi}}{2}A_t'^2\,,\\
\frac{z^{d+1}}{\mathrm{e}^{-\chi/2}}\left(\frac{\mathrm{e}^{\chi/2}}{z^{d-3}}ZA_t'\right)'&=\frac{q^2z^2\mathrm{e}^{\chi}}{f}A_t^2\,,\\
(d-1)\chi'&=z\psi'^2+\frac{2q^2z\mathrm{e}^{\chi}}{f^2}\mathcal{F}A_t^2\,,\\
\frac{z^{d+1}}{\mathrm{e}^{-\chi/2}}\left(\frac{\mathrm{e}^{-\chi/2}}{z^{d}}f\right)'&=\frac{1}{d-1}\left(-k(d-2)z^2-V+\frac{z^4\mathrm{e}^{\chi}}{2}ZA_t'^2\right)\,,\\
\end{aligned}
\end{equation}
where the prime denotes the derivative with respect to the radial coordinate $z$ and $Z$ are general functions of real scalar $\psi$ in (\ref{modelst}). Without loss of generality, we shall choose $\theta=0$.

Following~\cite{Cai:2020wrp}, we can obtain from the equations of motion~\eqref{eomst} that there is a radially conserved quantity given by
\begin{equation}\label{myQ}
\mathcal{Q}(z)=z^{3-d}\mathrm{e}^{\chi/2}\left[z^{-2}(f\mathrm{e}^{-\chi})'-ZA_t A_t'\right]+2k (d-2)\int^z y^{-d+1}\mathrm{e}^{-\chi(y)/2}\mathrm{d}y\,,
\end{equation}
\emph{i.e.} $\mathcal{Q}'(z)=0$. For the planar case ($k=0$), this conserved quantity is the Noether charge associated with a particular scaling symmetry. In contrast, there is no symmetry associated with $k=\pm 1$. Nevertheless, we have found an interesting way to obtain the above radially conserved quantity using the geometrical construction~\cite{Yang:2021civ}\,\footnote{The geometrical construction for a radially conserved quantity is valid for general static spacetimes, see~\cite{Yang:2021civ} for more details.}.

Besides the event horizon at $z_H$ with $f(z_H)=0$ and $f'(z_H)<0$,  we now assume an inner horizon located at $z=z_I>z_H$ for which $f(z_I)=0$ and $f'(z_I)>0$. Moreover, the smoothness of a horizon demands $A_t$ to be vanishing at both horizons for the black hole with non-vanishing scalar hair. By evaluating $\mathcal{Q}$ at both horizons, we then obtain
\begin{equation}\label{mQ}
\mathcal{Q}(z_H)-\mathcal{Q}(z_I)=\frac{\mathrm{e}^{-\chi(z_H)/2}f'(z_H)}{z_H^{d-1}}-\frac{\mathrm{e}^{-\chi(z_I)/2}f'(z_I)}{z_I^{d-1}}-2k (d-2)\int^{z_I}_{z_H} y^{-d+1}\mathrm{e}^{-\chi(y)/2}\mathrm{d}y\,.
\end{equation}
We have the following two cases.
\begin{itemize}
  \item For $k=0$ and $k=1$, the right-hand side of the above equation is negative, while the left-hand side is vanishing since $\mathcal{Q}$ is conserved. This is a contradiction.
  \item For the hyperbolic case ($k=-1$), both sides of~\eqref{mQ} have the same sign, so we are not able to rule out the inner horizon. Nevertheless, as discussed in our previous work~\cite{An:2021plu}, the inner horizon of the hyperbolic case can be removed if considering the null energy condition.
\end{itemize}
Therefore, most hairy black holes~\eqref{bkansatzst} do not have Cauchy horizons in the generalized St\"{u}ckelberg theory~\eqref{modelst}.

\subsection{Holographic superconductor from M-theory}
Many top-down models from consistent truncation of supergravity and superstring theories take the above St\"{u}ckelberg form. As concrete examples, we show two holographic superconductor models obtained from the top-down approach as follows.

The five-dimensional model is given by~\cite{Gubser:2009qm}\,\footnote{To have a standard normalization for the kinetic term of $U(1)$ sector, we have rescaled $A_\mu\rightarrow\frac{\sqrt{3}}{2L} A_\mu$ of~\cite{Gubser:2009qm}. We also redefined $\eta=\psi$.}
\begin{equation}\label{modelst5D}
\mathcal{L}_{S}^{(5)}=-\frac{1}{2}(\partial_\mu\psi)^2-\frac{\sinh^2\psi}{2}\left(\partial_\mu\theta-\frac{\sqrt{3}}{L}A_\mu\right)^2+\frac{3}{L^2}\cosh^2\frac{\psi}{2}(5-\cosh\psi)-\frac{1}{4}F_{\mu\nu}F^{\mu\nu}\,,
\end{equation}
which can be lifted to a class of solutions of type IIB supergravity, based on D3-branes at the tip of a Calabi-Yau cone. The four-dimensional one reads~\cite{Gauntlett:2009dn}\,\footnote{The original paper~\cite{Gauntlett:2009dn} used a different form of Lagrangian from~\eqref{modelst4D}. The notation of~\cite{Gauntlett:2009dn} is related to ours by $\hat{A}_1=A$ and $\hat{\chi}=\sqrt{2}e^{i\theta}\tanh\frac{\psi}{2}$.}
\begin{equation}\label{modelst4D}
\mathcal{L}_{S}^{(4)}=-\frac{1}{2}(\partial_\mu\psi)^2-\frac{\sinh^2\psi}{2}\left(\partial_\mu\theta-\frac{1}{L}A_\mu\right)^2+\frac{1}{L^2}\cosh^2\frac{\psi}{2}(7-\cosh\psi)-\frac{1}{4}F_{\mu\nu}F^{\mu\nu}\,,
\end{equation}
derived as a consistent truncation of M-theory with $F\wedge F=0$. In addition to ensuring a consistent underlying quantum theory and its dual, one particular advantage of a top-down model is that all parameters are fixed. The planar hairy black hole solutions that are holographically dual to superconductors for the above two top-down theories were constructed in~\cite{Gubser:2009qm,Gauntlett:2009dn}. There are also zero-temperature domain wall solutions of these supergravity theories, interpolating between two copies of AdS space, one of which preserves an $U(1)$ gauge symmetry while the other breaks it~\cite{Gubser:2009gp,Gauntlett:2009bh}.

In the present work, we focus on the second model~\eqref{modelst4D} and study the interior dynamics of the superconducting black holes. Therefore, we consider the planar black hole, $k=0$, and set $L=1$ without loss of generality. The equations of motion are given as follows.
\begin{equation}\label{eom-psi}
  \psi''=-\left(\frac{1}{z}+\frac{h'}{h}\right)\psi'-\frac{1}{2}\sinh 2\psi \frac{A_t^2}{z^6h^2L^2}+\frac{1}{2}\frac{\mathrm{e}^{-\chi/2}}{z^5hL^2}\left(\sinh 2\psi-6\sinh\psi\right),
\end{equation}

\begin{equation}\label{eom-at}
  (\mathrm{e}^{\chi/2}A_t')'=\frac{A_t}{z^5hL^2}\sinh^2 \psi,
\end{equation}

\begin{equation}\label{eom-chi}
   2\chi'=z\psi'^2+\frac{A_t^2}{z^5h^2L^2}\sinh^2\psi,
\end{equation}

\begin{equation}\label{eom-h}
  h'=\frac{\mathrm{e}^{-\chi/2}}{2}\left[-\frac{1}{2z^4L^2}(1+\cosh \psi)(7-\cosh\psi)+\frac{1}{2}\mathrm{e}^{\chi}A_t'^2\right].
\end{equation}
Here we have introduced $h=z^{-3}\mathrm{e}^{-\chi/2}f$ for the later convenience.

To solve these coupled equations of motion, we need to impose appropriate boundary conditions both at the horizon, $z=z_H$, and the UV boundary, $z=0$. To break the $U(1)$ symmetry spontaneously, we require no deformation of the boundary system by the operator $\mathcal{O}$ dual to $\psi$, \emph{i.e.} near the AdS boundary\,\footnote{We have considered the so-called standard quantization for which the leading source term of a bulk field is identified to be the source of the dual operator in the boundary theory. The UV expansion~\eqref{UVpsi} means there is no explicit source for the scalar operator $\mathcal{O}$ and thus the breaking of the $U(1)$ symmetry is spontaneously.}
\begin{equation}\label{UVpsi}
\psi(z)=z^2 \left<\mathcal{O}\right>+\dots\,,
\end{equation}
where the expectation value $ \left<\mathcal{O}\right>$ is the order parameter for the superconducting phase. The gauge potential $A_t$ for small $z$ is given by
\begin{equation}
A_t(z)=\mu-\rho z+\dots,
\end{equation}
with $\mu$ and $\rho$ the chemical potential and the charge density, respectively. Moreover, in order to fix the normalization of the boundary time coordinate, we demand $\chi(z=0)=0$, for which the temperature at the UV boundary is equal to the standard Hawking temperature~\eqref{HawkingT}. At the black hole horizon $z=z_H$, we require the smoothness of the geometry, in particular, $A_t(z=z_H)=0$. We shall work in the grand canonical ensemble with the chemical potential $\mu$ fixed.
\begin{figure}[H]
\centering
\includegraphics[width=0.48\textwidth]{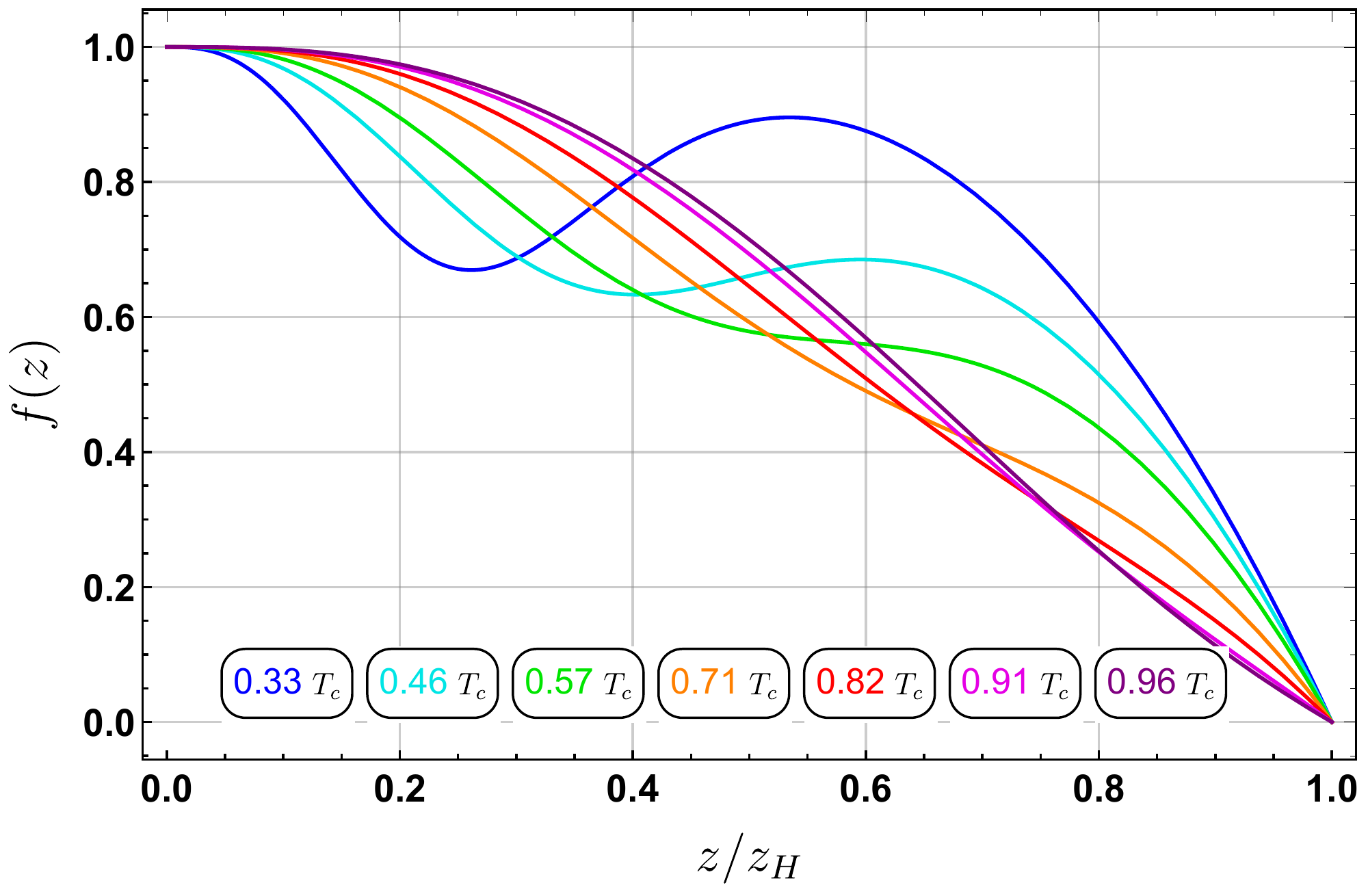}\quad
\includegraphics[width=0.48\textwidth]{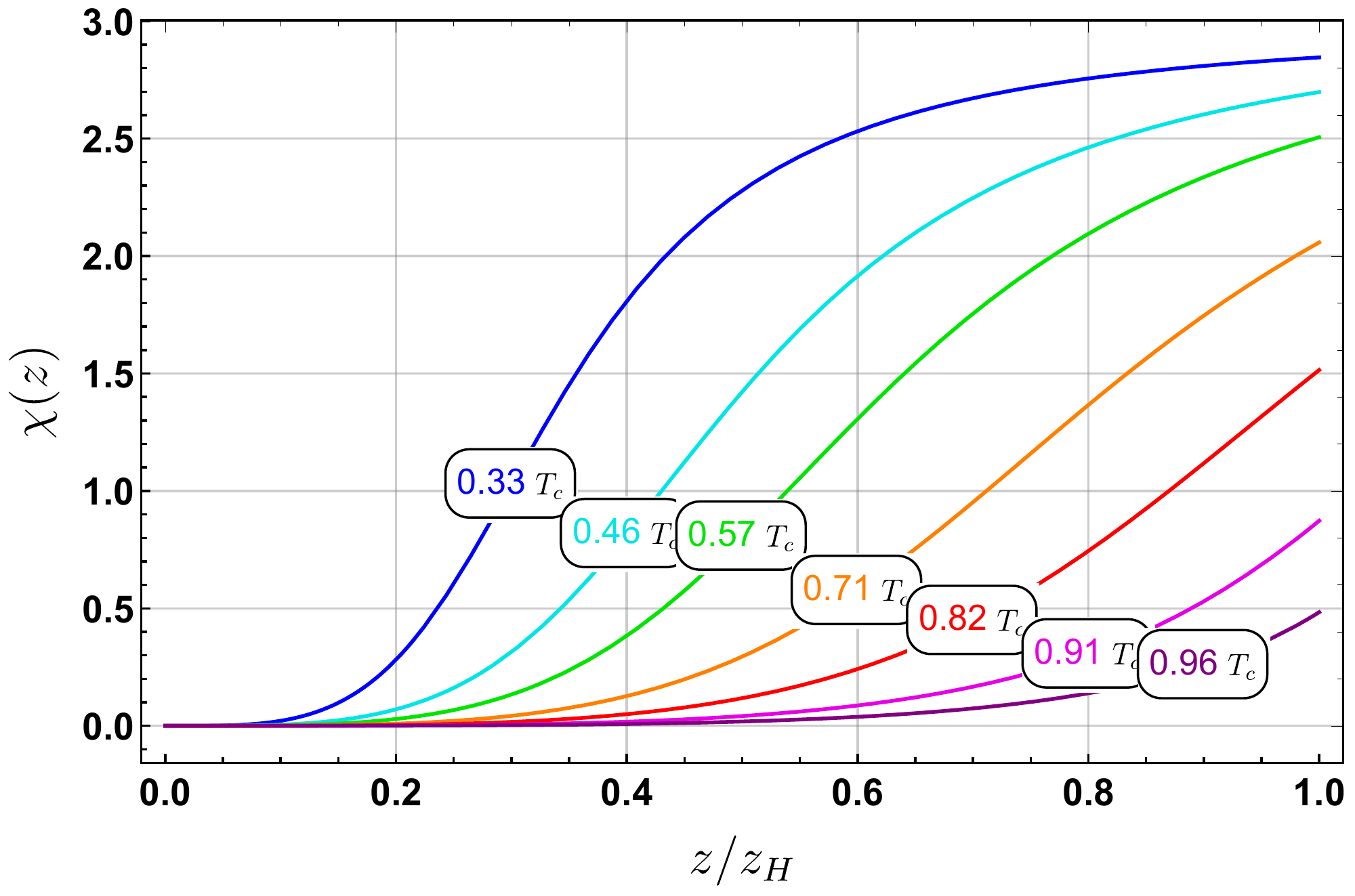}\\
\includegraphics[width=0.48\textwidth]{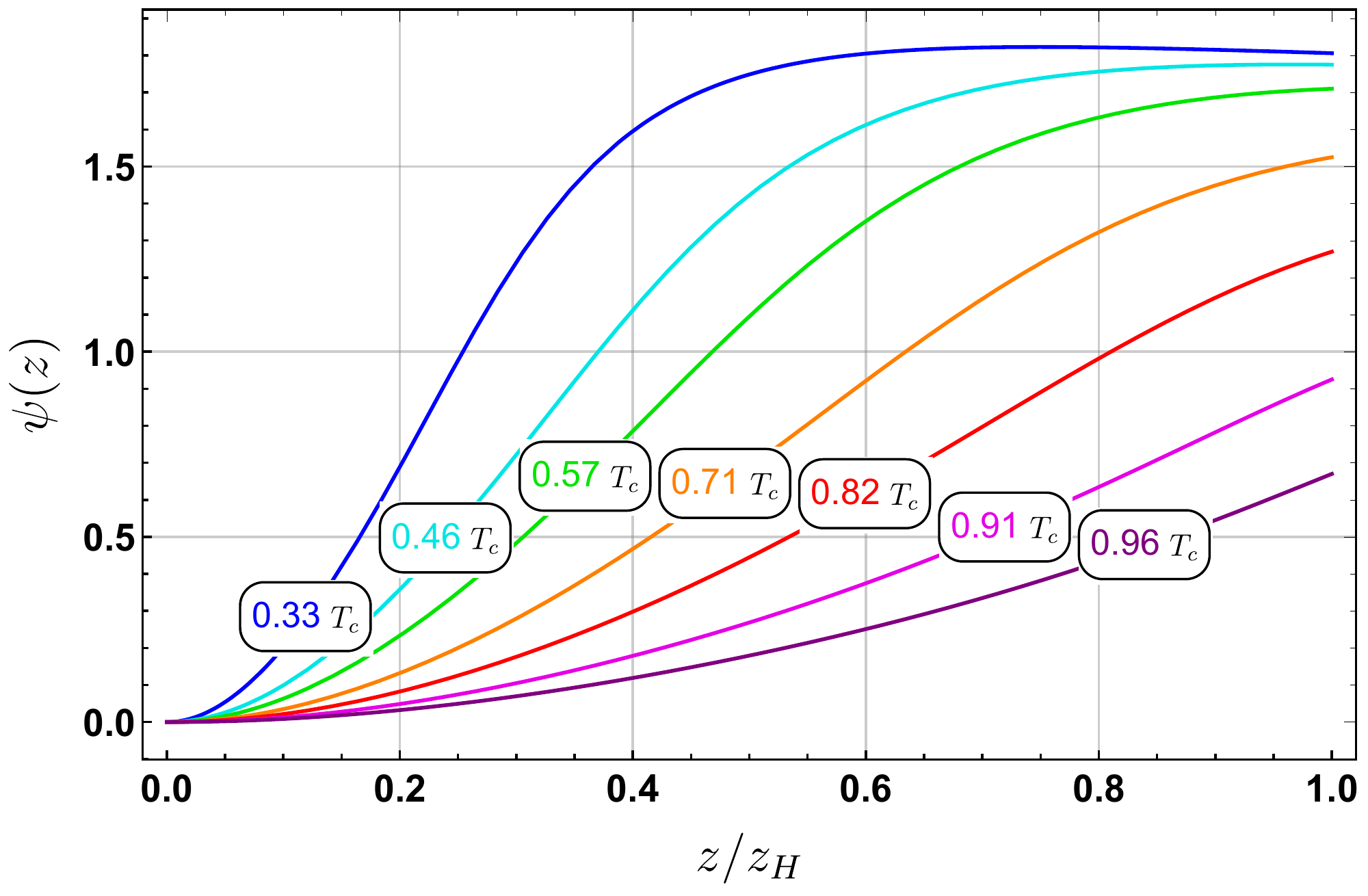}\quad
\includegraphics[width=0.48\textwidth]{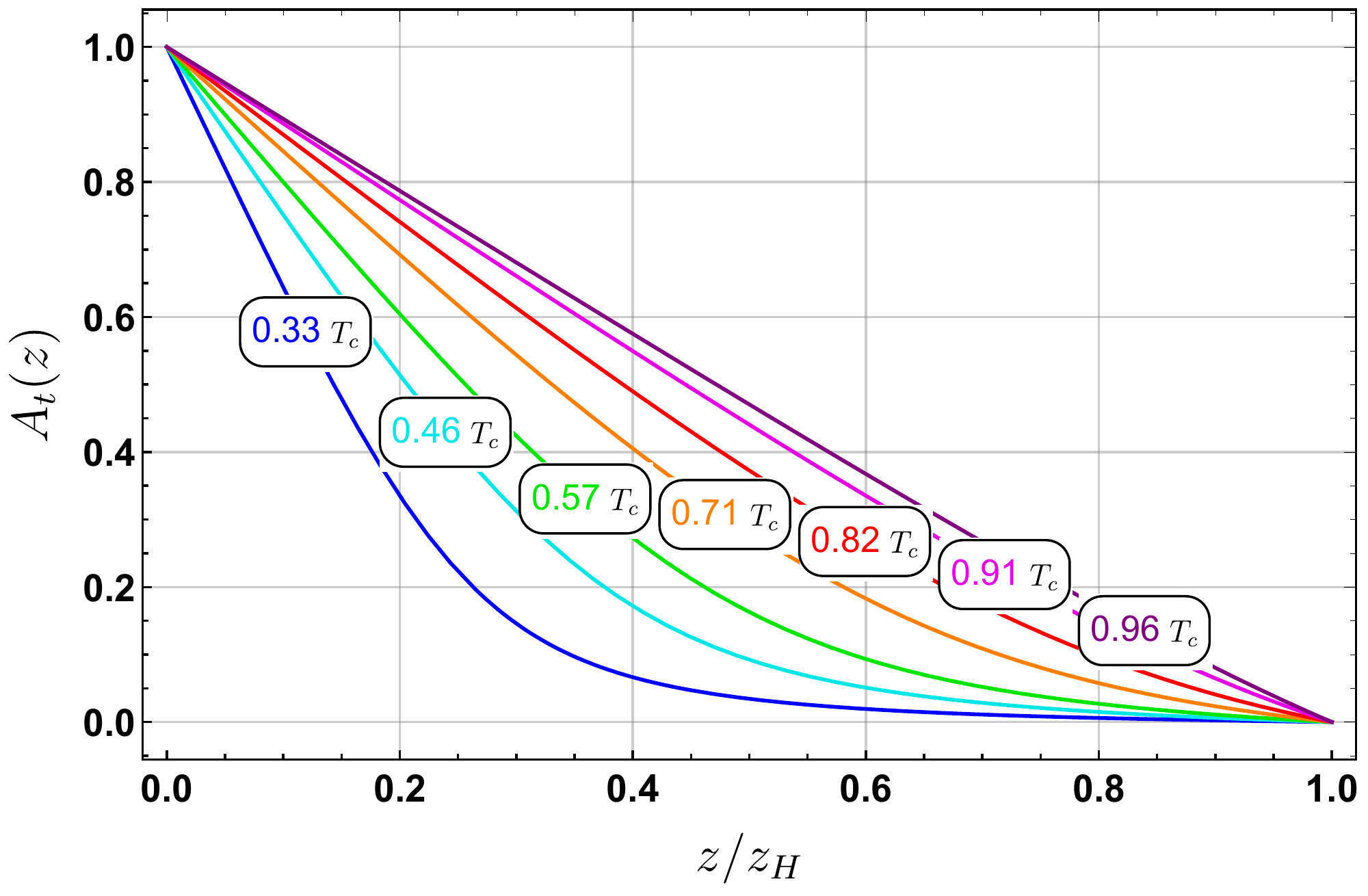}\\
\caption{The configuration of the charge black holes with scalar hair from the event horizon, $z=z_H$, to the AdS boundary, $z=0$, at different temperatures. We have fixed the chemical potential $\mu=1$.}\label{Fig:ex-solution}
\end{figure}

\begin{figure}[H]
\centering
\includegraphics[width=0.60\textwidth]{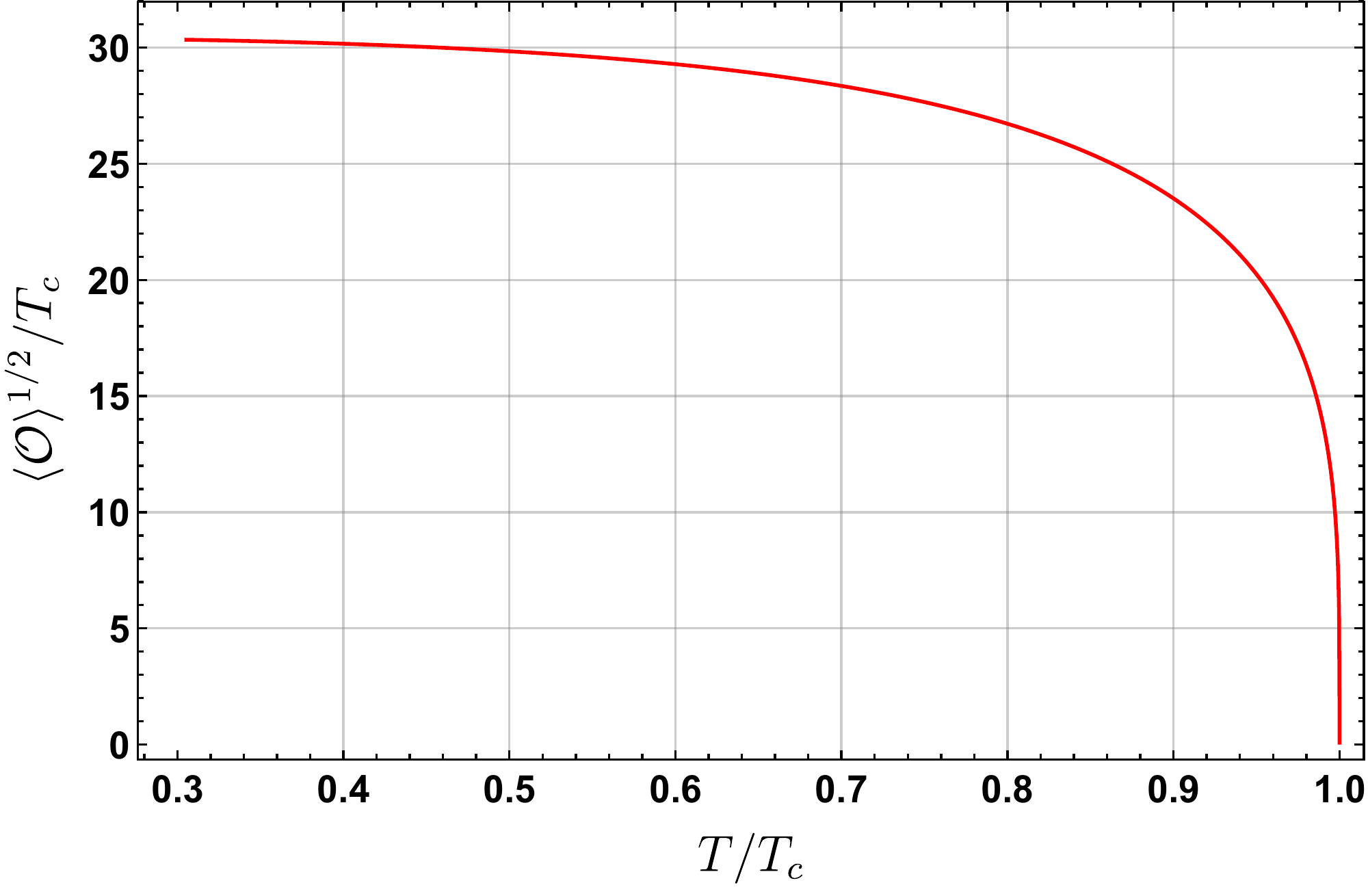}
  \caption{The superconducting condensate $\left<\mathcal{O}\right>$ as a function of temperature. Below $T_{c}\approx0.0208\mu$, $\left<\mathcal{O}\right>$ appears spontaneously from a second order phase transition. We work in the grand canonical ensemble.}\label{Fig:phase-diagram}
\end{figure}

At high temperatures, the charged black hole of~\eqref{eom-psi}-\eqref{eom-h} does not have scalar hair and is given by the AdS RN solution:
\begin{equation}
h=\frac{1}{z^3}-\frac{1}{z_H^3}+\frac{\mu^2}{4z_H^2}(z-z_H),\quad  A_t=\mu\left(1-\frac{z}{z_H}\right) \,,
\end{equation}
with $\psi(z)=\chi(z)=0$. Besides the event horizon, there is an inner Cauchy horizon at $z=z_I>z_H$. At a temperature below $T_c$, there are charged black hole solutions with non-trivial scalar hair. Some hairy black hole solutions at different temperatures are presented in Fig.~\ref{Fig:ex-solution}, using the numerical techniques described in~\cite{Hartnoll:2008kx}. We show the condensate $\left<\mathcal{O}\right>$ as a function of the temperature in Fig.~\ref{Fig:phase-diagram}. One can find that $\left<\mathcal{O}\right>$ develops smoothly and increases monotonically as the temperature is decreased. This corresponds to a second-order phase transition from the normal phase (RN black hole) to the superconducting phase (hairy black hole), known as the holographic superconductor.

%
%%%%%%%%%%%%%%%%%%%%%%%%%%%%%%
%%%%%%%%%%%%%%%%%%%

\section{Interior Dynamical Epochs}\label{sec:inner}
After knowing the exterior configuration, we proceed to study the interior dynamics of the top-down holographic superconductor~\eqref{modelst4D}. The interior geometry can be obtained straightforwardly by numerically solving the equations of motion~\eqref{eom-psi}-\eqref{eom-h} inside the event horizon towards the spacelike singularity.\footnote{One may worry about that the coordinate patch of~\eqref{bkansatzst} cannot cover both the exterior and interior of the hairy black holes. This issue can be fixed by simply switching to the ingoing coordinates by using $v=t-\int ^z e^{\chi(s)/2}/f(s) ds$ instead of $t$ (see~\eqref{nullcord} below), for which the equations of motion~\eqref{eom-psi}-\eqref{eom-h} do not change. }

\begin{figure}[H]
\centering
\includegraphics[width=0.49\textwidth]{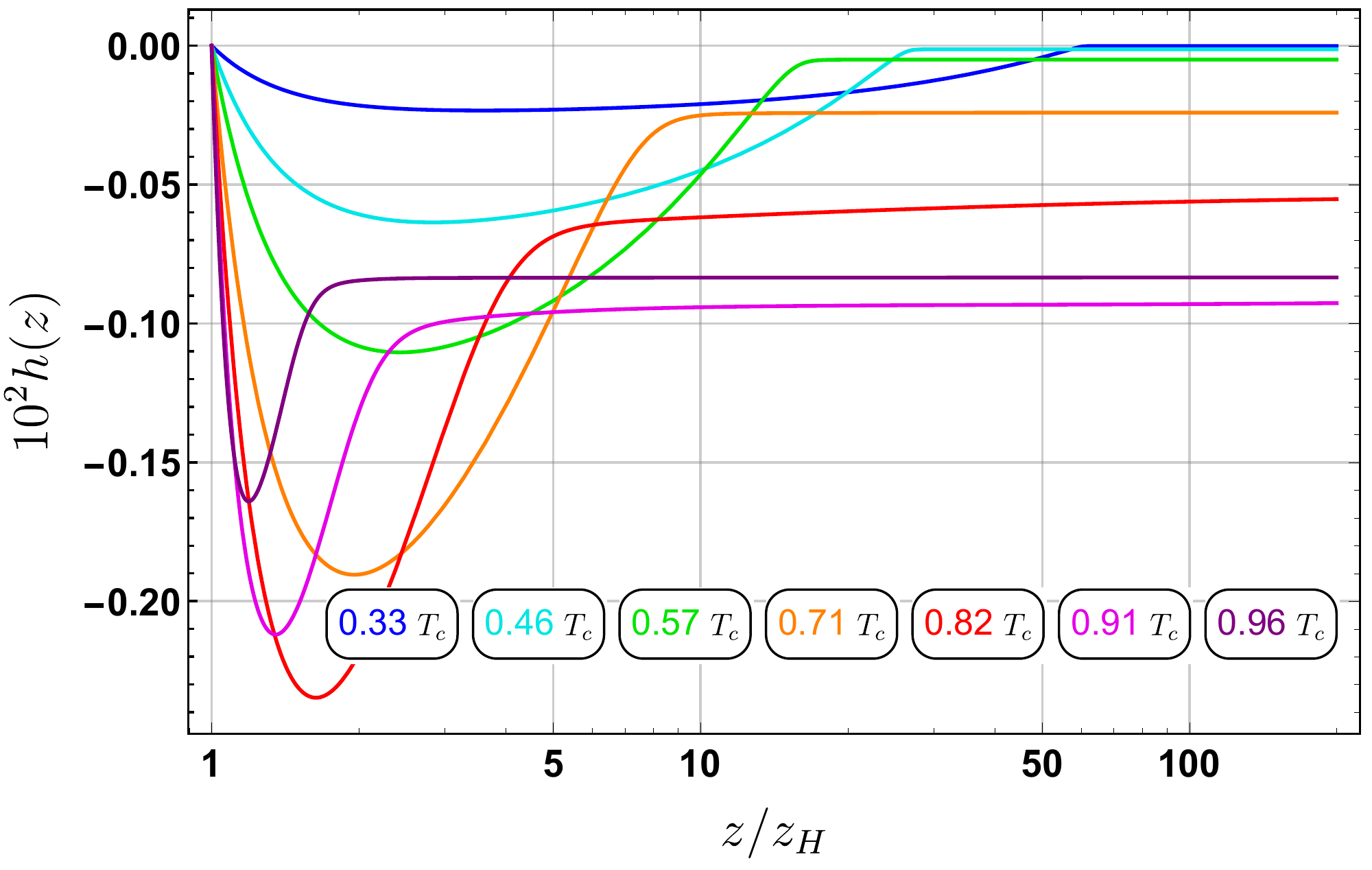}
\includegraphics[width=0.50\textwidth]{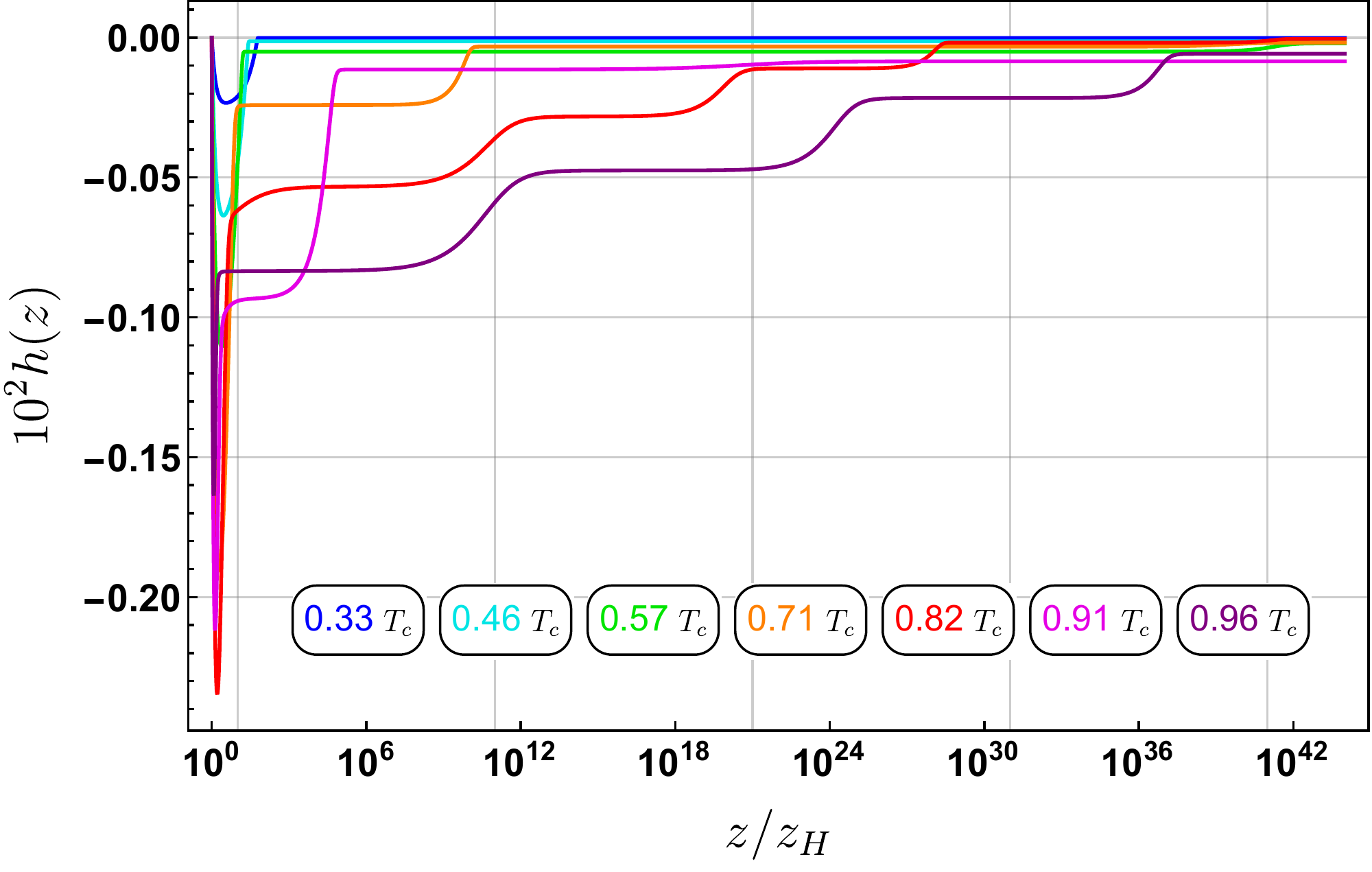}
\caption{The behaviors of $h$ in function of $z/z_H$ inside the event horizon for the hairy black holes. There are strong non-linear effects at the near-horizon regimes below $T_c$ (left) and a sequence of alternations of different plateaus in the deep interior (right).}\label{Fig:in-h}
\end{figure}

\begin{figure}[H]
\centering
\includegraphics[width=0.48\textwidth]{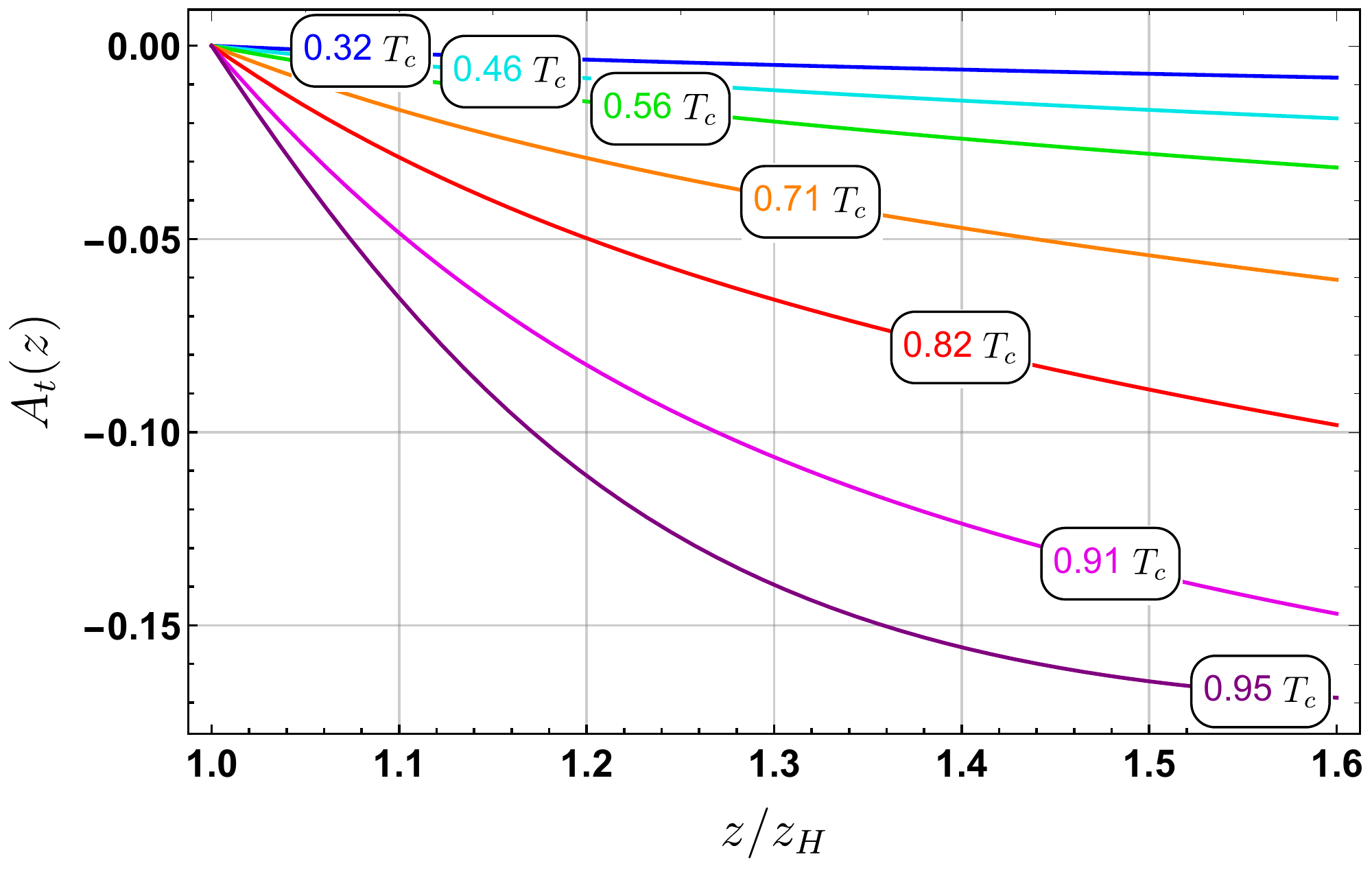}
\includegraphics[width=0.48\textwidth]{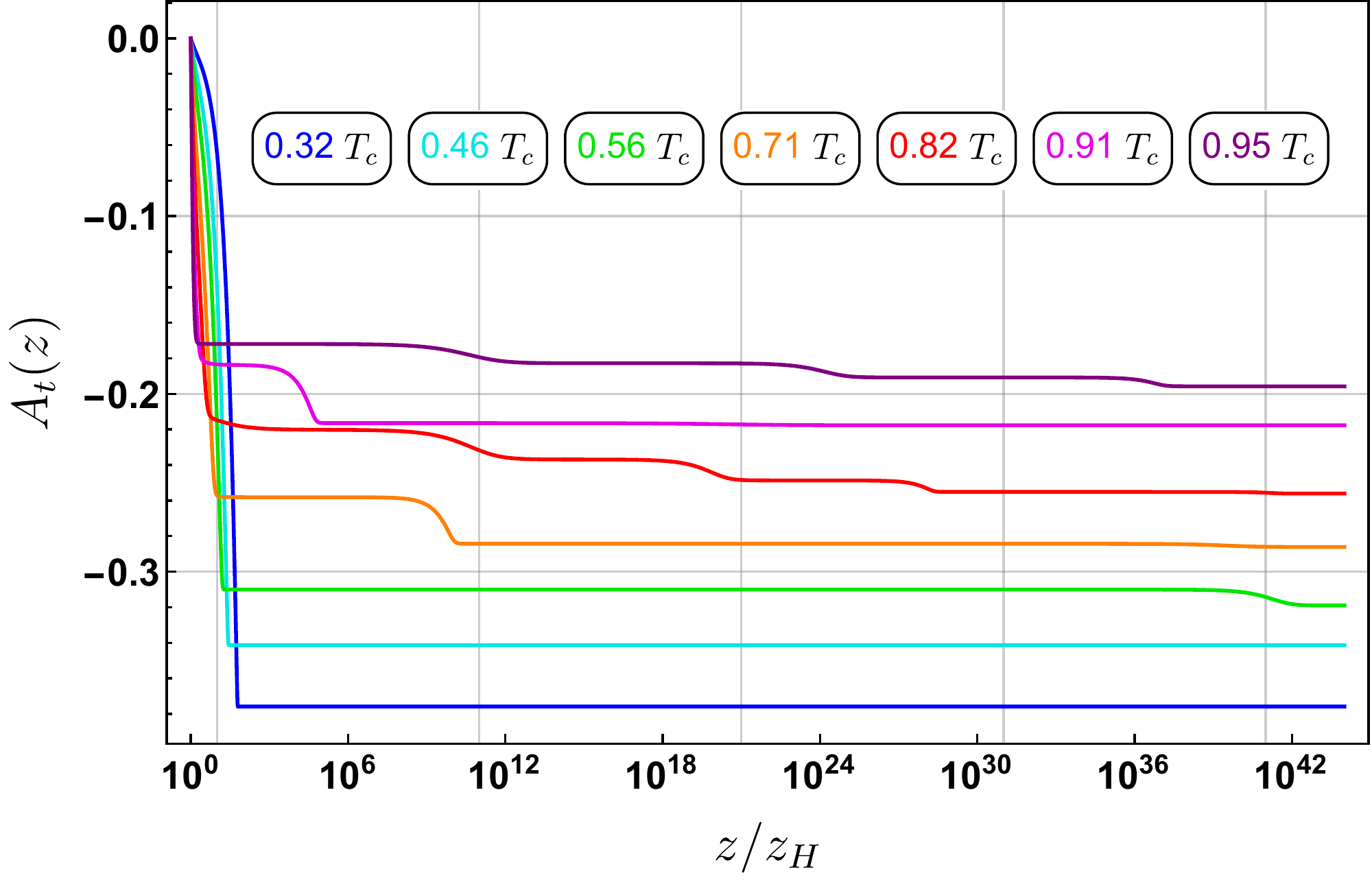}
\caption{The dynamics of $A_t$ with respect to $z/z_H$ inside the hairy black holes at different temperatures. Towards the singularity, $A_t$ decreases monotonically with many alternations of plateaus in the deep interior.}
\label{Fig:in-At}
\end{figure}

In Figures.~\ref{Fig:in-h} and~\ref{Fig:in-At}, we present the behaviors of $h$ and $A_t$ at different temperatures behind the event horizon, which can be regarded as a holographic flow triggered spontaneously by the charged scalar hair. The flow interpolates from a UV radial scaling to a timelike scaling towards a late-time singularity inside the black hole. One can find some interesting behaviors in the presence of scalar hair. In particular, no matter how small the scalar hair is, it has a strong non-linear effect closed to the would-be inner Cauchy horizon of the AdS RN black hole. Moreover, in the deep interior towards the singularity, there are many alternations of plateaus that, as we will show, correspond to different Kasner epochs. See Appendix~\ref{alpha-diagram} for more examples.

\subsection{ER collapse and Josephson oscillation}
We first check if the dynamical epochs associated with the instability of the inner Cauchy horizon persist in this top-down model, including the ER collapse and the Josephson oscillations of the scalar condensate. Recently, it has been shown that the latter might be sensitive to the couplings~\cite{Sword:2021pfm}.

\begin{figure}[H]
\centering
\includegraphics[width=0.48\textwidth]{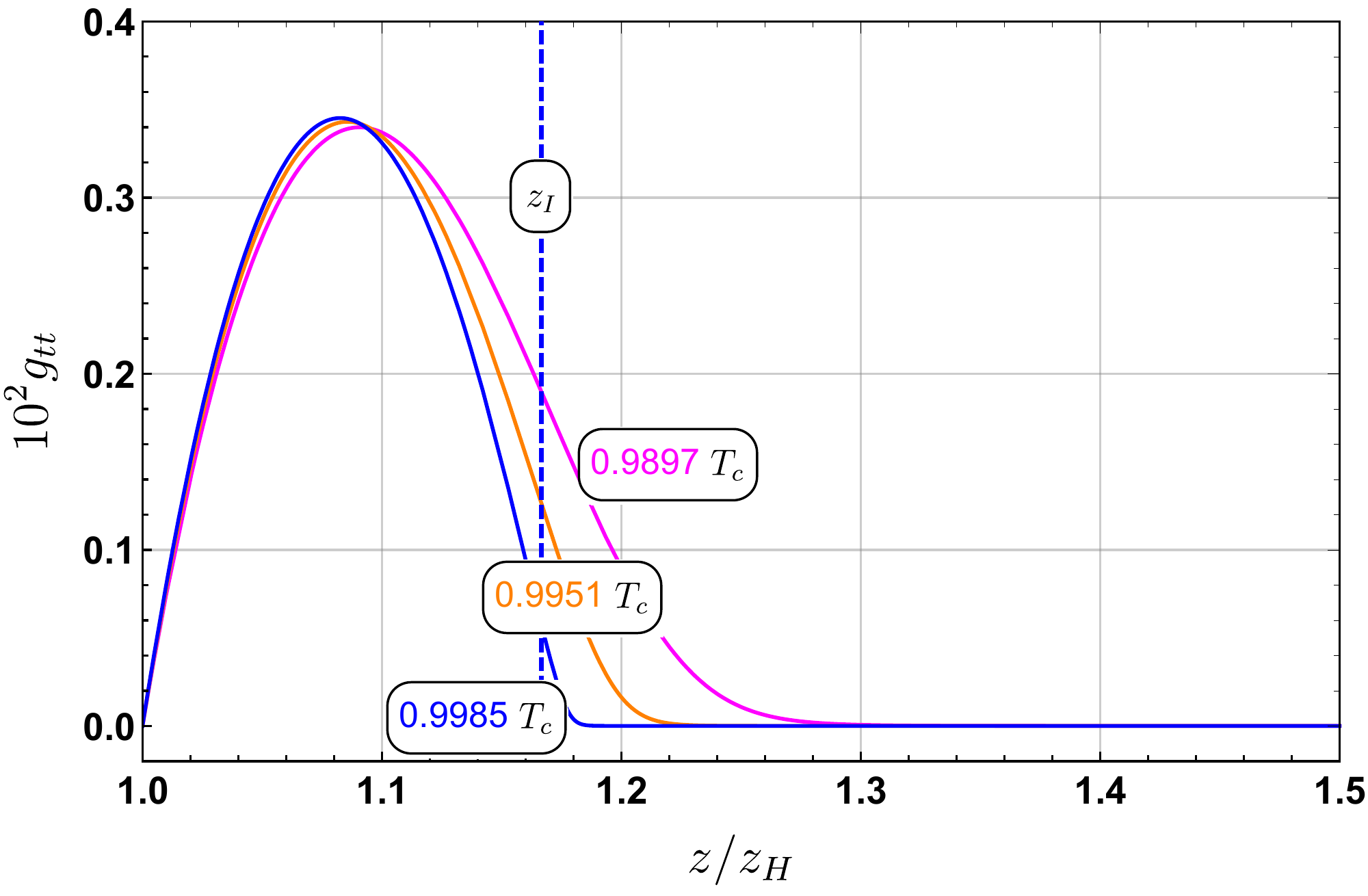}
\includegraphics[width=0.49\textwidth]{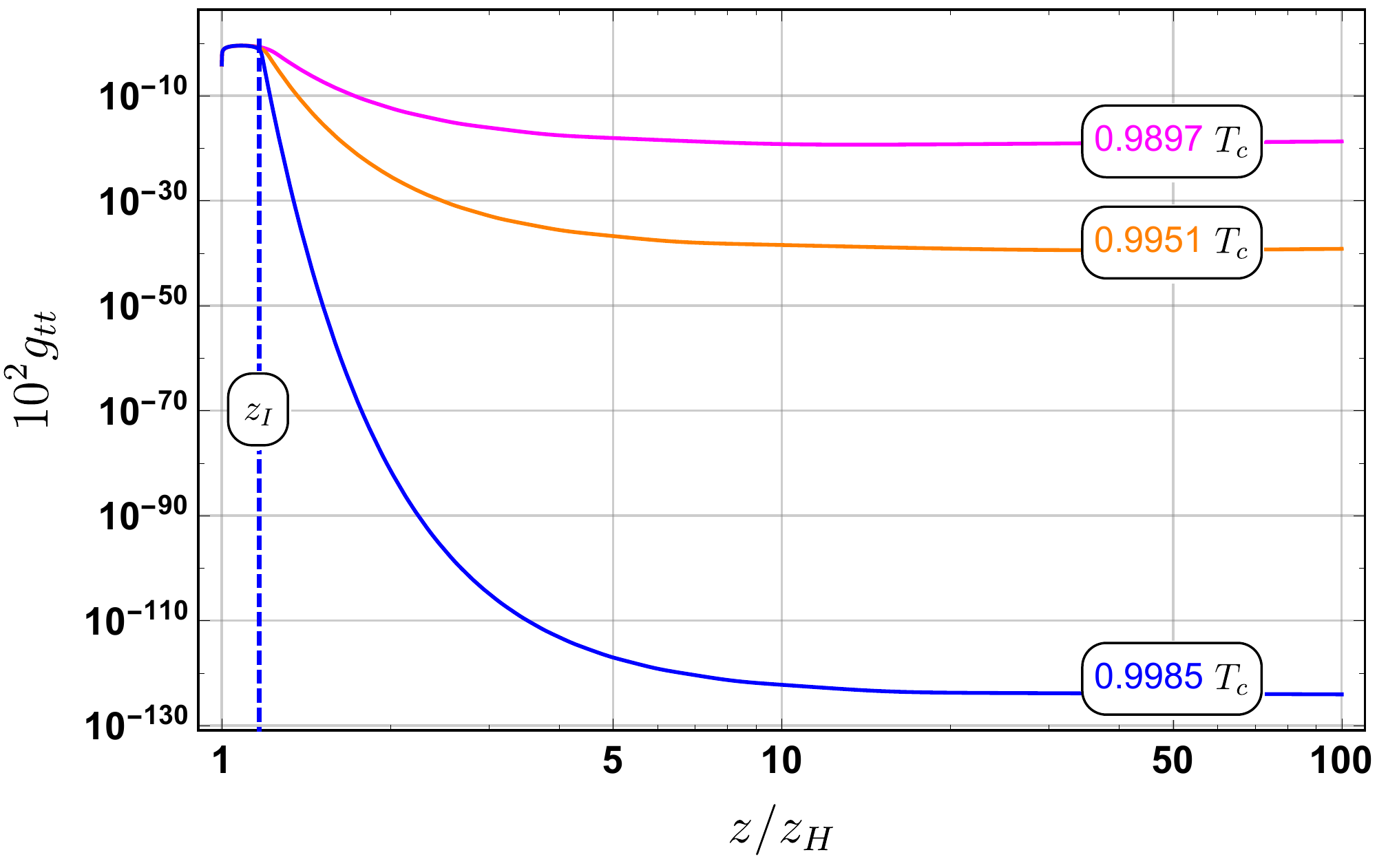}
\caption{Interior epoch for the ER collapse at different temperatures. $g_{tt}$ experiences a sudden drop to an exponentially small value in the vicinity of the would-be innner horizon $z_I/z_H\approx1.1667$ at $T_c$ (the vertical dashed line). We set the chemical potential $\mu=1$.}\label{Fig:er-coll}
\end{figure}

The ER collapse is characterized by the quick decrease of  $g_{tt}$ which, behind the event horizon, is the measure for the spatial $t$ coordinate that runs along the wormhole connecting the two exteriors of the black hole. As shown in Fig.~\ref{Fig:er-coll}, slightly below $T_c$, there is a rapid collapse of $g_{tt}$ to an exponentially small value in the vicinity of $z_I\approx1.1667z_H$ (the location of the inner horizon of the AdS RN black hole at $T_c$). The collapse becomes stronger as the temperature approaches $T_c$ from below (for which the scalar itself is vanishing small), revealing the highly nonlinear nature of this rapid transition.

\begin{figure}[H]
\centering
\includegraphics[width=0.49\textwidth]{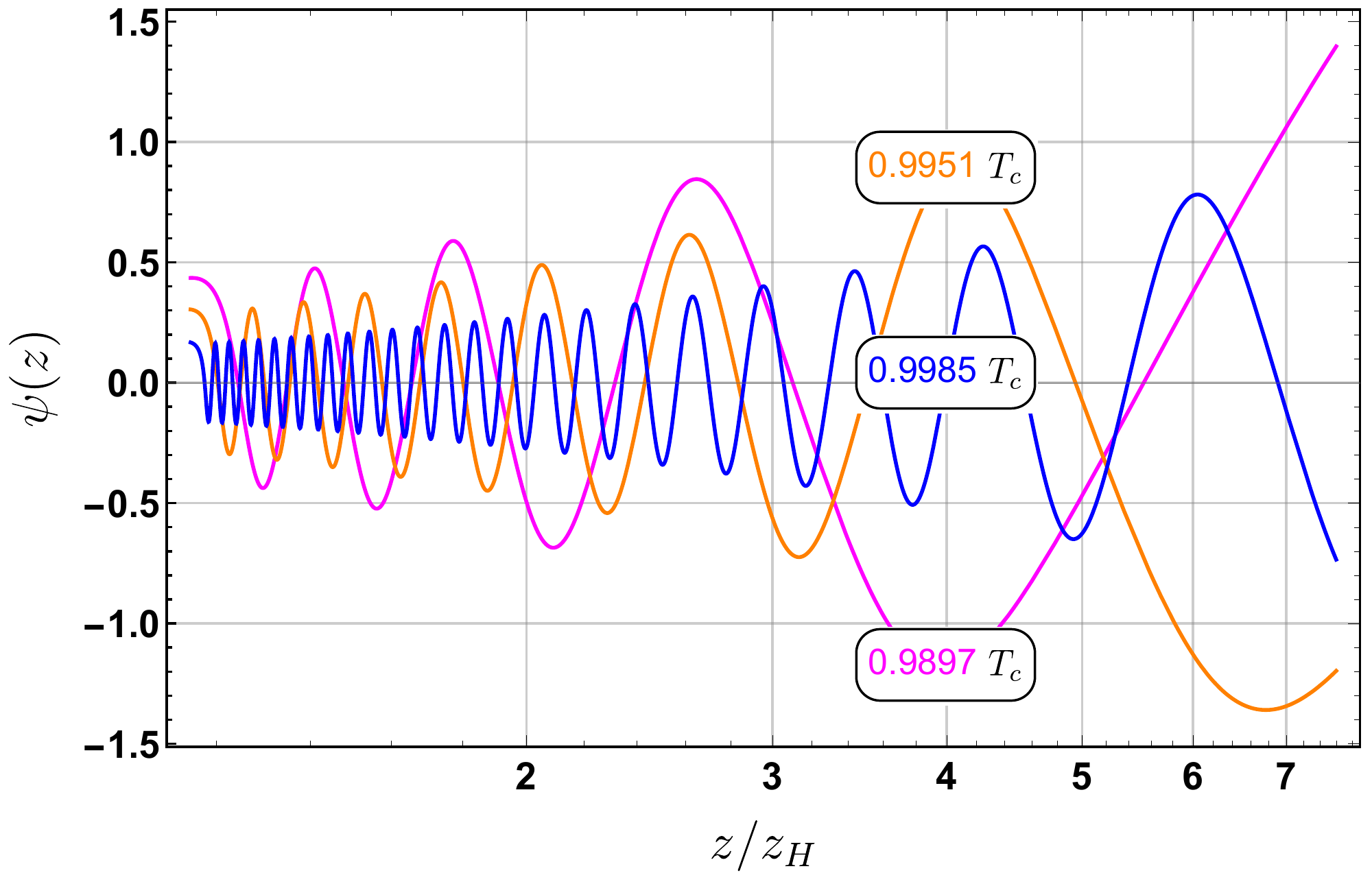}
\includegraphics[width=0.49\textwidth]{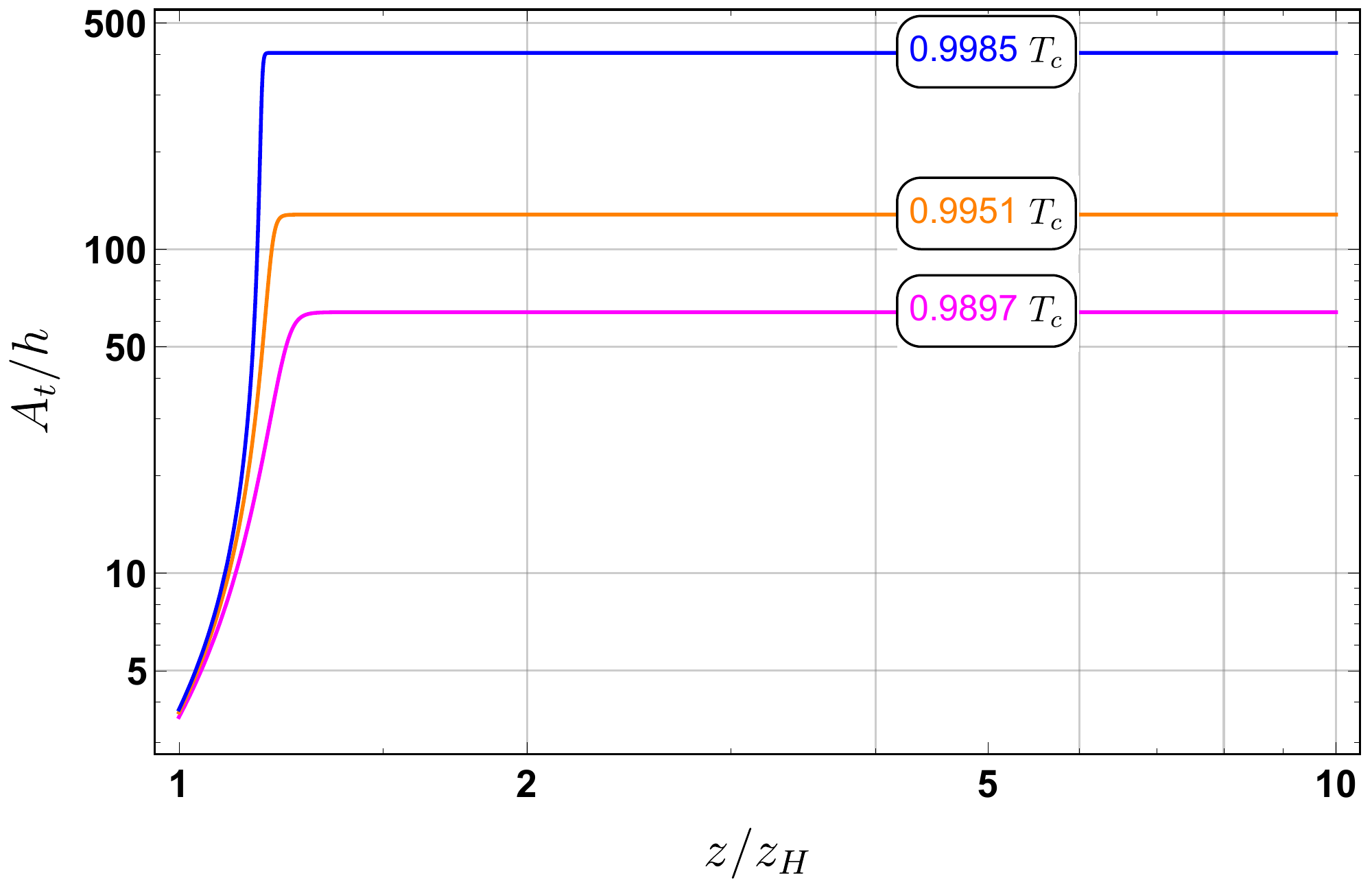}
\caption{Interior dynamics in the Josephson oscillation epoch at different temperatures. \textbf{Left: } The scalar oscillation as a function of $z/z_H$. \textbf{Right: } The behavior of $A_{t}/h$ in function of $z/z_H$. After a short period of growth, $A_{t}/h$ quickly saturates to a constant.}\label{Fig:psi}
\end{figure}

In the aftermath of the ER collapse, we observe the oscillation of the scalar condensate which starts in the collapse of the ER bridge epoch and propagate continuously into the interior, see the left panel of Fig.~\ref{Fig:psi}. The oscillation becomes faster as close to $T_c$, meanwhile its amplitude decreases. This Josephson oscillation epoch can be understood as follows. We find numerically that $\mathrm{e}^{-\chi(z)}$ becomes quite small due to the ER collapse. Therefore, we can drop the last term of~\eqref{eom-psi}.
\begin{equation}\label{app-psi1}
\psi''=-\left(\frac{1}{z}+\frac{h'}{h}\right)\psi'-\frac{1}{2} \sinh 2\psi \frac{A_{t}^{2}}{z^{6}h^{2}}.
\end{equation}
Moreover, the charge term on the right-hand side of the Maxwell equation~\eqref{eom-at} can be neglected, from which we obtain $A_{t}' \sim \mathrm{e}^{-\chi/2}$. By combining with the equation of motion~\eqref{eom-h}, one can easily find that, after the ER collapse, $A_{t}/h$ is approximately a constant (see the right panel of Fig.~\ref{Fig:psi}). 

\begin{figure}[H]
\centering
\includegraphics[width=0.85\textwidth]{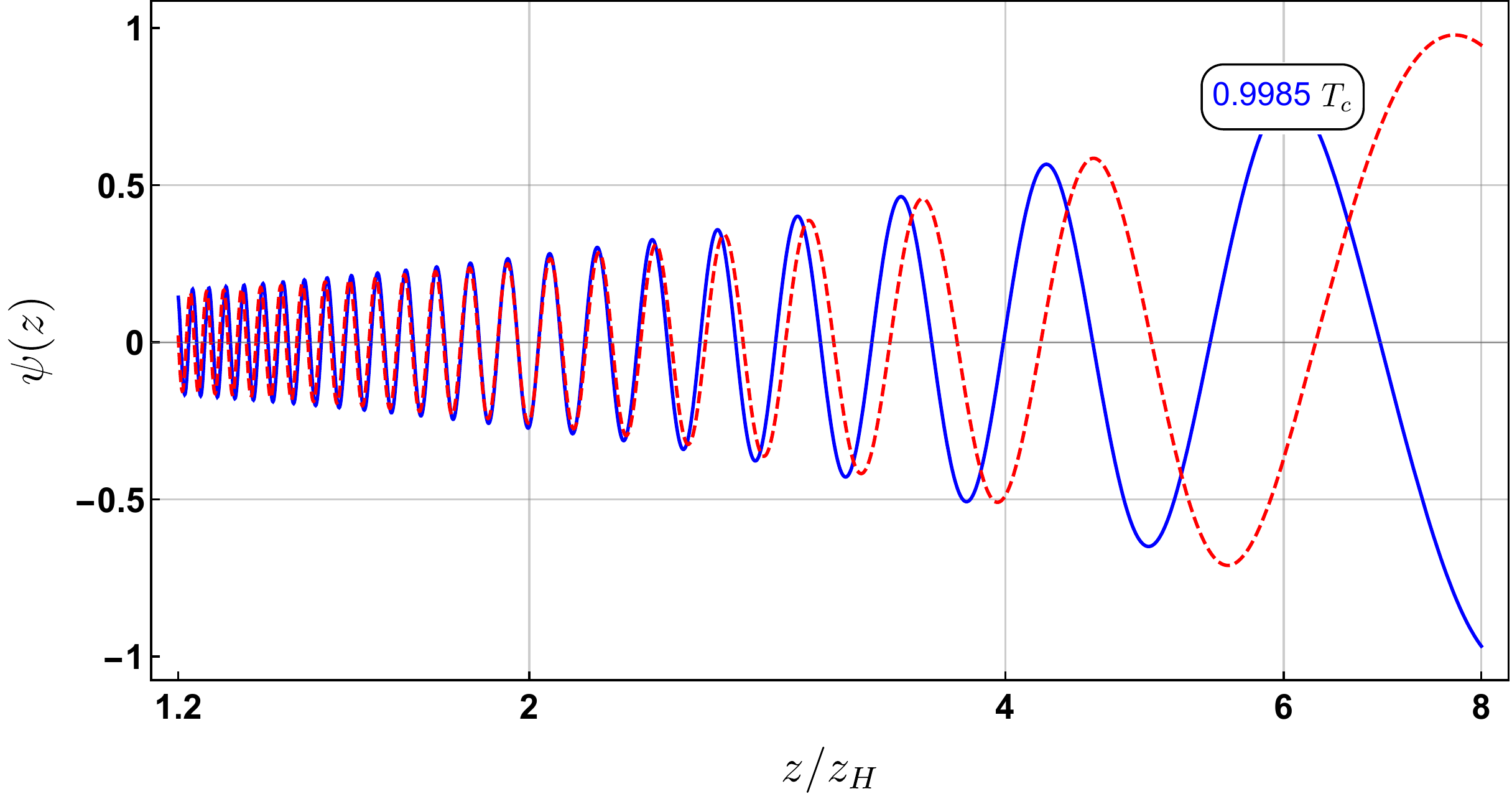}
\caption{A comparison of the numerical solutions (solid blue curve) and the analytical description~\eqref{psiBJ} (red dashed curve) for Josephson oscillations. The approximation is in excellent agreement to the numerical solution when $\psi$ is small ($z/z_H\lesssim 3$), and loses its effectiveness at large $z$ for which $\psi$ is not small. We choose $T=0.9985T_c$ with $\mu=1$.}\label{Fig:psi-fit}
\end{figure}

Thus, the equation~\eqref{app-psi1} at small $\psi$ becomes
\begin{equation}\label{app-psi2}
\frac{1}{z}(z\psi')'=-\frac{A_{t}^{2}}{z^{6}h^{2}}\psi\,,
\end{equation}
which can be solved explicitly in terms of Bessel functions.
\begin{equation}\label{psiBJ}
  \psi=c_J \mathrm{J}_0\left(\frac{A_t}{2hz^{2}}\right)+c_Y\mathrm{Y}_0\left(\frac{A_t}{2hz^{2}}\right),
\end{equation}
with $c_{J}$ and $c_{Y}$ integration constants. The analytical approximation~\eqref{psiBJ} is compared with the numerical data for the scalar field in Fig.~\ref{Fig:psi-fit}. One can find that the analytical description (red dashed line) for the Josephson oscillation epoch is in good agreement ($z/z_H\lesssim 3$) with the numerical solution (as $T\rightarrow T_c$ for which $\psi$ is small). Due to the existence of $\sinh(2\psi)$ term, the approximation in~\eqref{app-psi1} will fail when $\psi$ becomes large. That's why the scalar oscillation behavior is only observed when the temperature is sufficiently closed to $T_c$. Moreover, note that the amplitude of $\psi$ increases as $z$ is increased, thus the analytical approximation~\eqref{psiBJ} will fail when $z$ is large enough (see Fig.~\ref{Fig:psi-fit} at large $z$).

Both the collapse of the ER bridge and the Josephson oscillations are similar to the free charge scalar model of~\cite{Hartnoll:2020fhc}. We will show that this will no longer be the case for the Kanser region we discuss below.

\subsection{Kasner behavior}
At the end of the Josephson oscillations, our numerics suggest a logarithmic growth of $\psi$, which indicates the onset of a Kasner regime. As is evidence from Figures.~\ref{Fig:alpha-low} and~\ref{Fig:alpha-high} in Appendix~\ref{alpha-diagram}, the large $z$ behaviors of $z\psi'$, $\sqrt{z\chi'}$ and $\sqrt{1-zg'_{tt}/g_{tt}}$ yield a number of plateaus. In each plateau, many terms in the equations of motion~\eqref{eom-psi}-\eqref{eom-h} are negligible, which is first confirmed by numerical solutions and will be checked by comparing the resulting analytic solutions to the full numerical one. 

After dropping those terms, the equations of motion~\eqref{eom-psi}-\eqref{eom-h} can be simplified to be
\begin{equation}\label{kas-equ}
  \psi''\simeq -\frac{1}{z}\psi',\quad \mathrm{e}^{\chi/2}A_t'\simeq C_0,\quad 2\chi'\simeq z\psi'^2,\quad h'\simeq \frac{C_0}{4}\mathrm{e}^{-\chi/2},
\end{equation}
where $C_{0}$ is an integration constant.  The above equations can be solved explicitly yielding
\begin{equation}\label{kasner}
\begin{split}
&\mathrm{d}s^{2}=\frac{1}{z^{2}}\left[h_s z^{3-\frac{\alpha^{2}}{4}}\mathrm{d}t^{2}-\frac{1}{h_s z^{3+\frac{\alpha^{2}}{4}}}\mathrm{d}z^{2}+\mathrm{d}\Sigma^{2}_{2,0}\right]\,,\\
&\psi(z)\sim\alpha\ln(z),\quad A_{t}'\sim z^{-\frac{\alpha^{2}}{4}}\,.
\end{split}
\end{equation}
at large $z$. Here $\alpha$ is a free constant and $h_{s}$ is the value of $|h|$ in this plateau.

One finds that all metric components are power laws of $z$ and the scalar field is logarithmic. By the coordinate transformation to the proper time $\tau \sim z^{-(3+\frac{\alpha^{2}}{4})/2}$, we obtain
\begin{equation}\label{Kasner}
\begin{split}
&\mathrm{d}s^{2}=-\mathrm{d}\tau^{2}+c_{t}\tau^{2p_{t}} \mathrm{d}t^{2}+c_{s} \tau^{2p_{s}} \mathrm{d}\Sigma^{2}_{2,0}\,,\\
&\psi(\tau)\sim-\sqrt{2} p_\psi\ln(\tau)\,,
\end{split}
\end{equation}
with
\begin{equation}\label{kasnerep}
p_{t}=\frac{\frac{\alpha^{2}}{4}-1}{\frac{\alpha^{2}}{4}+3},\quad p_{s}=\frac{2}{\frac{\alpha^{2}}{4}+3},\quad p_{\psi}=\frac{\sqrt{2}\alpha}{\frac{\alpha^{2}}{4}+3}\,.
\end{equation}
Here $c_t$ and $c_s$ are two positive constants. One can see that the above exponents satisfy the following relation.
\begin{equation}
p_{t}+2p_{s}=1,\quad p_{t}^{2}+2 p_{s}^{2}+p_{\psi}^{2}=1\,.
\end{equation}
So, the geometry takes a Kasner form. The Kasner exponents are determined by the parameter $\alpha$ that can be only obtained by solving the full equations of motion. Note that inside the event horizon $t$ becomes a spatial coordinate and runs along the ER bridge. As time evolves (with $\tau=0$ corresponding to $z\rightarrow\infty$), the ER bridge grows when $|\alpha|<2$ and contracts when $|\alpha|>2$.

For a certain class of black holes with charge scalar hair, it was recently argued that~\cite{Henneaux:2022ijt} the interior dynamics falls within the scope of the ``cosmological billiard'' description and the corresponding hyperbolic billiard region has infinite volume so that the system ultimately settles down to a final Kasner regime. Interestingly, as we will show below, in the top-down model we consider, the final dynamic behavior can not settle down to a Kasner regime to the singularity at $z\rightarrow\infty$.

\subsection{Alternation of Kasner epochs}
From the numerical results, one has found that there exist many alternations between different Kasner epochs (see Figures.~\ref{Fig:in-h} and~\ref{Fig:in-At}). It is obvious that the approximation obtaining~\eqref{kas-equ} fails in the alternation region between adjacent Kasner epochs. 
To understand those alternations, we need to consider the sub-leading terms (which are neglected during a single Kasner region) in the equations of motion~\eqref{eom-psi}-\eqref{eom-h}. We find the following two cases. 
\begin{itemize}
  \item Case 1: $|\alpha|>2$. In this case, we need to consider the term $\frac{A_t^2}{h^2}\frac{\sinh2\psi}{z^6}$ in~\eqref{eom-psi}, because its order could be larger than the order of the term $\psi'/z$. Meanwhile, the term $h'/h$ is still not important compared to $1/z$ since $\mathrm{e}^{-\chi/2}$ is much smaller than $1/z$ and $h$ is bounded. We shall call this case ``Kasner transition''.

  \item Case 2: $|\alpha|<2$. In this case, we can neglect the term $\frac{A_t^2}{h^2}\frac{\sinh2\psi}{z^6}$ as its order is smaller than the order of $\psi'/z$. But the term $h'/h$ becomes important compared to $1/z$ because $h$ will have a notable change. For this case, we will call it ``Kasner inversion''.
\end{itemize}

\paragraph{Kasner Transition}
According to the analysis above, for $|\alpha|>2$, we can obtain the approximated equation of motion for $\psi$:
\begin{equation}\label{kas-tran-psi}
  \frac{1}{z}(z\psi')'=-\frac{A_t^2}{2h^2L^2}\frac{\sinh2\psi}{z^6}.
\end{equation}
To solve this differential equation, it is convenient to write the solution as
\begin{equation}\label{alpha-psi}
  \psi=\int^z\frac{\tilde{\alpha}(s)}{s} ds\,.
\end{equation}
Taking derivative of~\eqref{kas-tran-psi} and using~\eqref{alpha-psi}, one obtains that
\begin{equation}\label{equ-alpha}
\begin{aligned}
\alpha>2:\quad z \tilde{\alpha}''+5\tilde{\alpha}'-2\tilde{\alpha}'\tilde{\alpha}=&0\,,\\
\alpha<-2:\quad z \tilde{\alpha}''+5\tilde{\alpha}'+2\tilde{\alpha}'\tilde{\alpha}=&0\,.
\end{aligned}
\end{equation}
We have also used the fact that $A_{t}/h$ is observed to be nearly constant over the transition region. This feature is visible in Figures.~\ref{Fig:in-h} and~\ref{Fig:in-At} at low temperautres, $T=0.32 T_c$ (blue), $T=0.46T_c$ (cyan) and $T=0.56T_c$ (green), for which there are at least two Kanser transitions for each temperature. Note also that $\psi(z)\sim\alpha\ln(z)$ at large $z$, for which one has $\sinh 2\psi\sim \mathrm{e}^{2\psi}$ when $\alpha>2$ and $\sinh 2\psi \sim \mathrm{e}^{-2\psi}$ when $\alpha <-2$. Moreover, our numerics shows that $\tilde{\alpha}\tilde{\alpha}'<0$ in the transition region (see Figures.~\ref{Fig:alpha-low} and~\ref{Fig:alpha-high} in Appendix~\ref{alpha-diagram}).

Solving~(\ref{equ-alpha}), one can obtain analytically that
\begin{equation}\label{sol-alpha}
\begin{aligned}
\alpha>2:\quad \tilde{\alpha}&=2-\sqrt{b}\tanh(\sqrt{b}\ln z/z_{tr}),\\
\alpha<-2:\quad \tilde{\alpha}&=-2+\sqrt{d}\tanh(\sqrt{d}\ln z/z_{tr}),
\end{aligned}
\end{equation}
where $b$ and $d$ are integration constants, and $z_{tr}$ denotes the position in the transition region at which $|\tilde{\alpha}|=2$\,\footnote{The location of $z_{tr}$ is not important in our discussion. One can choose a different $z_{tr}$ in the transition region.}. Considering the limit for which $z/z_{tr}\ll 1$, one has
\begin{equation}
\begin{aligned}
\alpha>2:\quad \lim_{z/z_{tr}\rightarrow 0}\tilde{\alpha}&=2+\sqrt{b}\,,\\
\alpha<-2:\quad \lim_{z/z_{tr}\rightarrow 0}\tilde{\alpha}&=-2-\sqrt{d}\,,
\end{aligned}
\end{equation}
which corresponds to the value of $\alpha$ in the Kasner epoch before the Kasner transition.
Taking the opposite limit of large $z/z_{tr}$ yields
\begin{equation}
\begin{aligned}
\alpha>2:\quad \lim_{z/z_{tr}\rightarrow \infty}\tilde{\alpha}&=2-\sqrt{b}\,,\\
\alpha<-2:\quad \lim_{z/z_{tr}\rightarrow \infty}\tilde{\alpha}&=-2+\sqrt{d}\,,
\end{aligned}
\end{equation}
which is exactly the value of $\alpha$ in the Kasner epoch after the Kasner transition. Therefore, we obtain the transformation rule for the \textbf{Kasner transition} between two adjacent Kasner epochs.
\begin{equation}\label{trasi-algebra}
\begin{aligned}
\alpha>2:\quad \alpha+\alpha_T=&4,\\
\alpha<-2:\quad \alpha+\alpha_T=&-4,
\end{aligned}
\end{equation}
where $\alpha$ and $\alpha_{T}$ are, respectively, the value of $z\psi'$~\eqref{kasner} before and after the Kasner Transition process. 

\begin{figure}[H]
\centering
\includegraphics[width=0.49\textwidth]{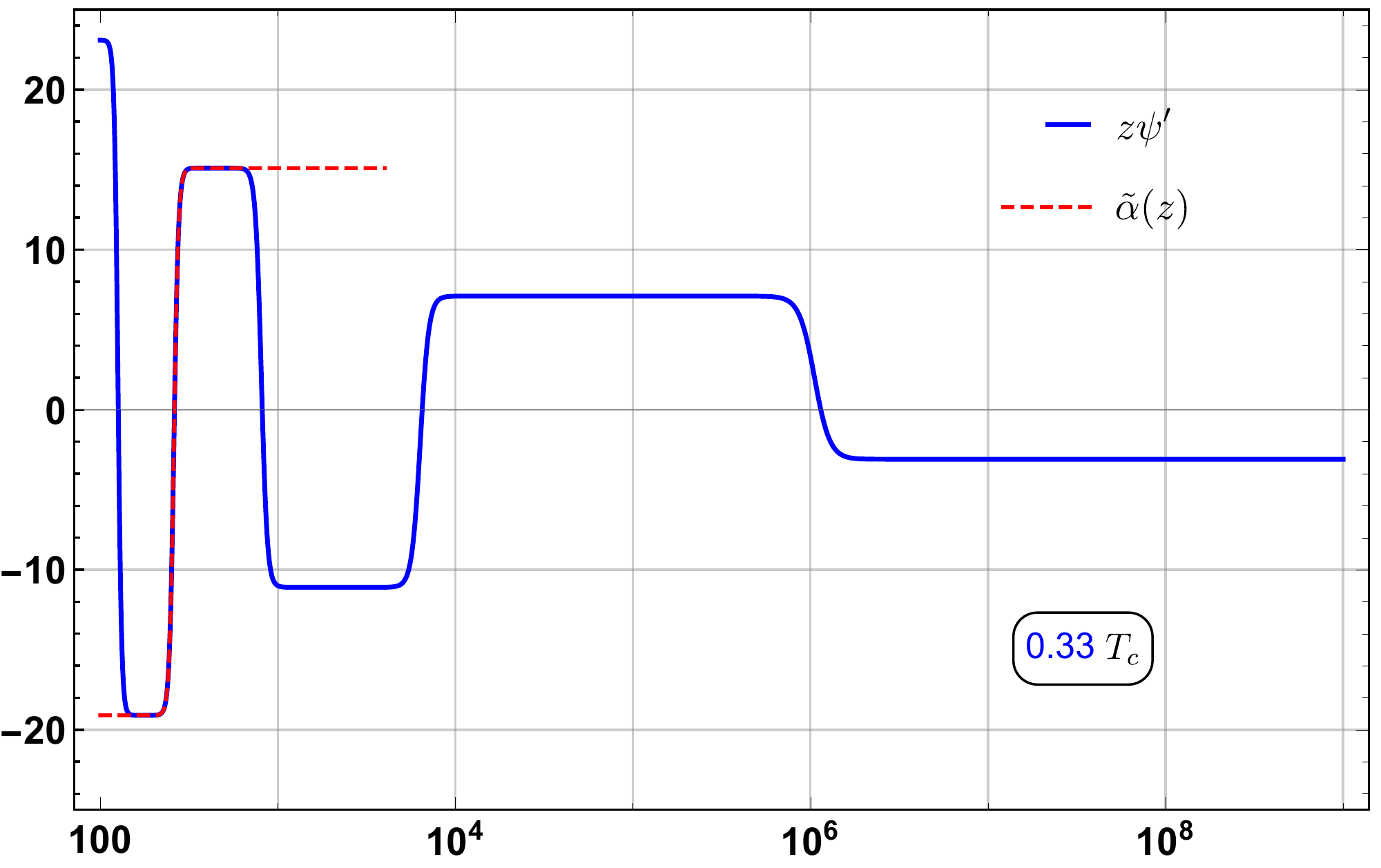}
\includegraphics[width=0.49\textwidth]{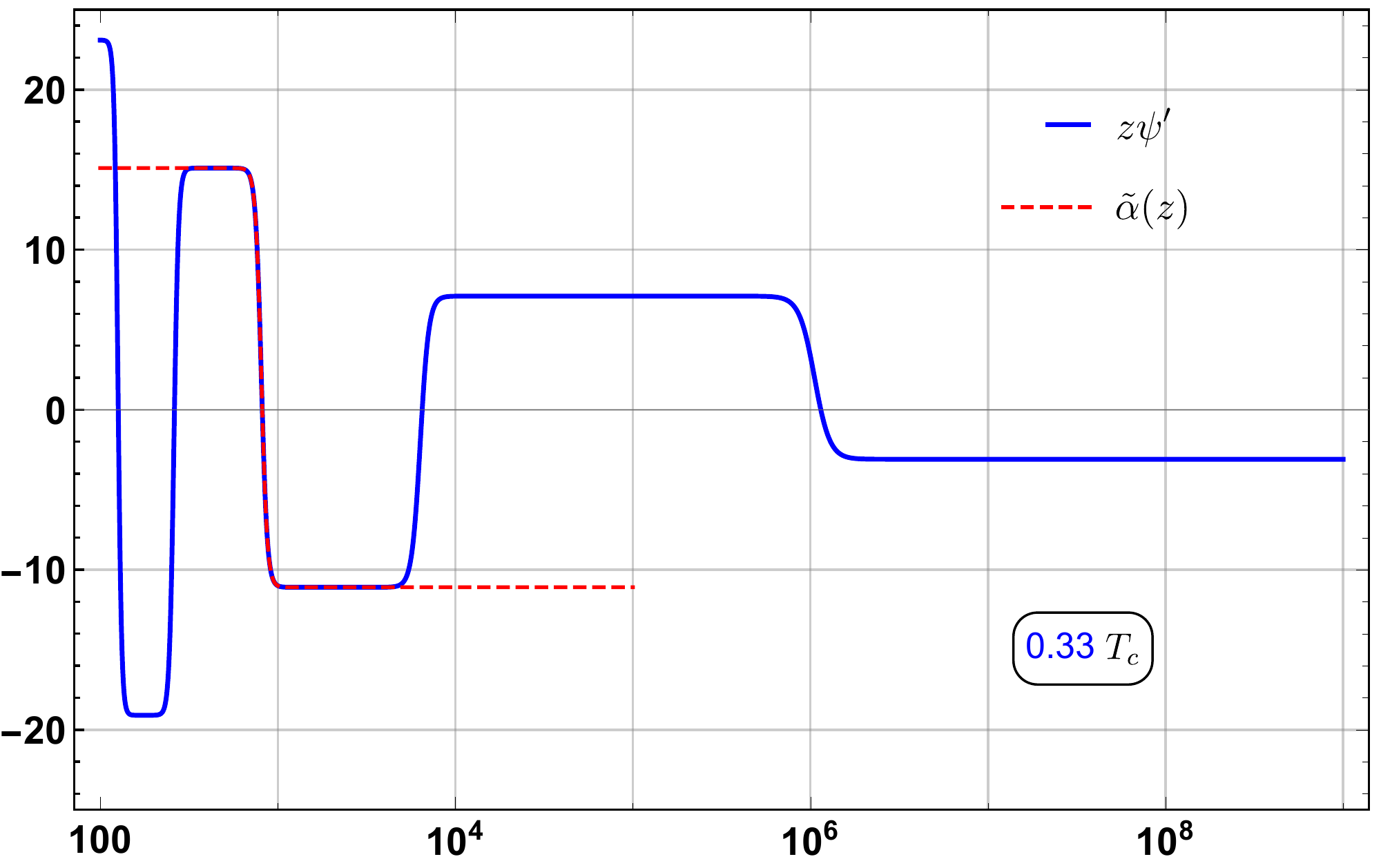}\\
\includegraphics[width=0.49\textwidth]{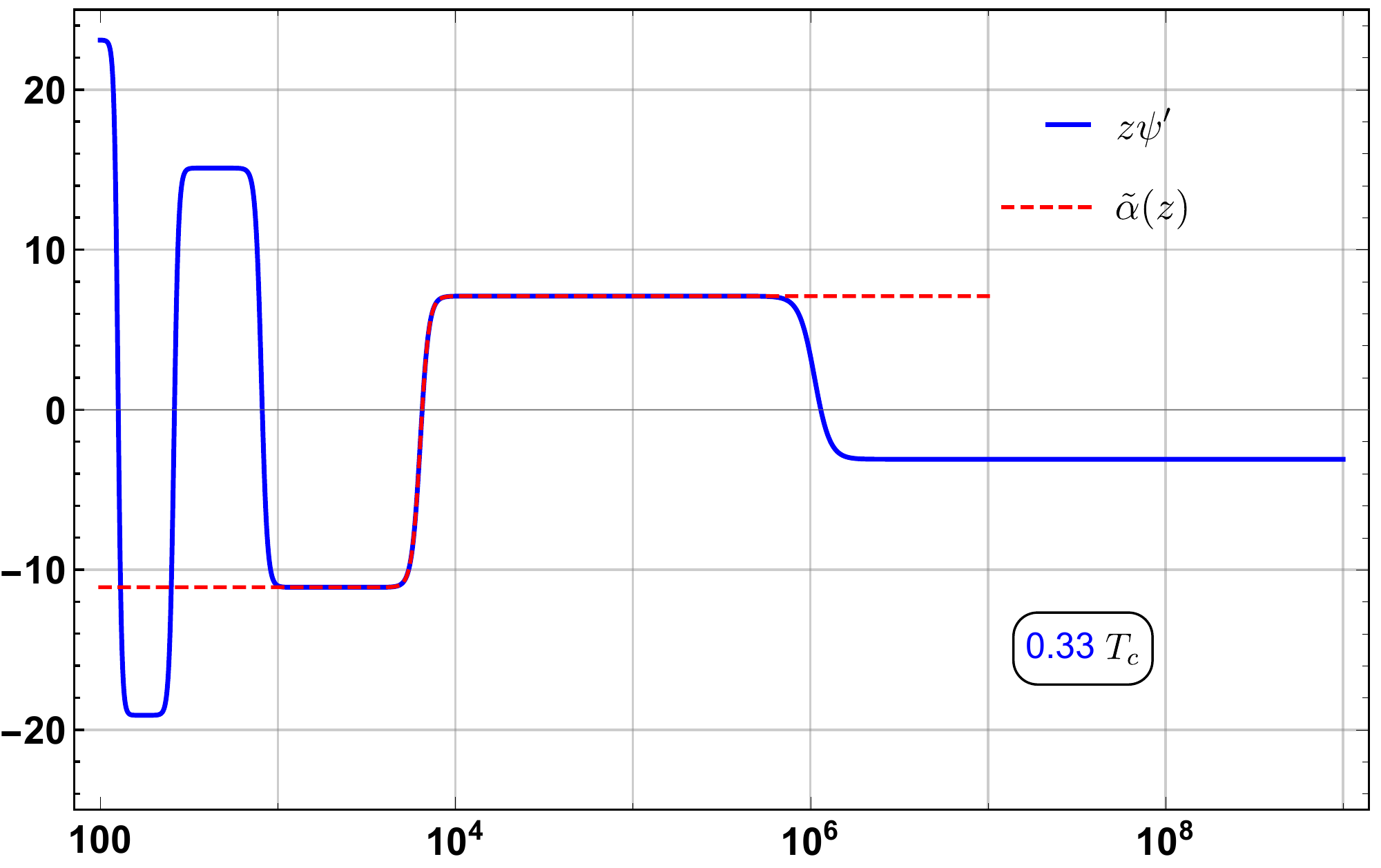}
\includegraphics[width=0.49\textwidth]{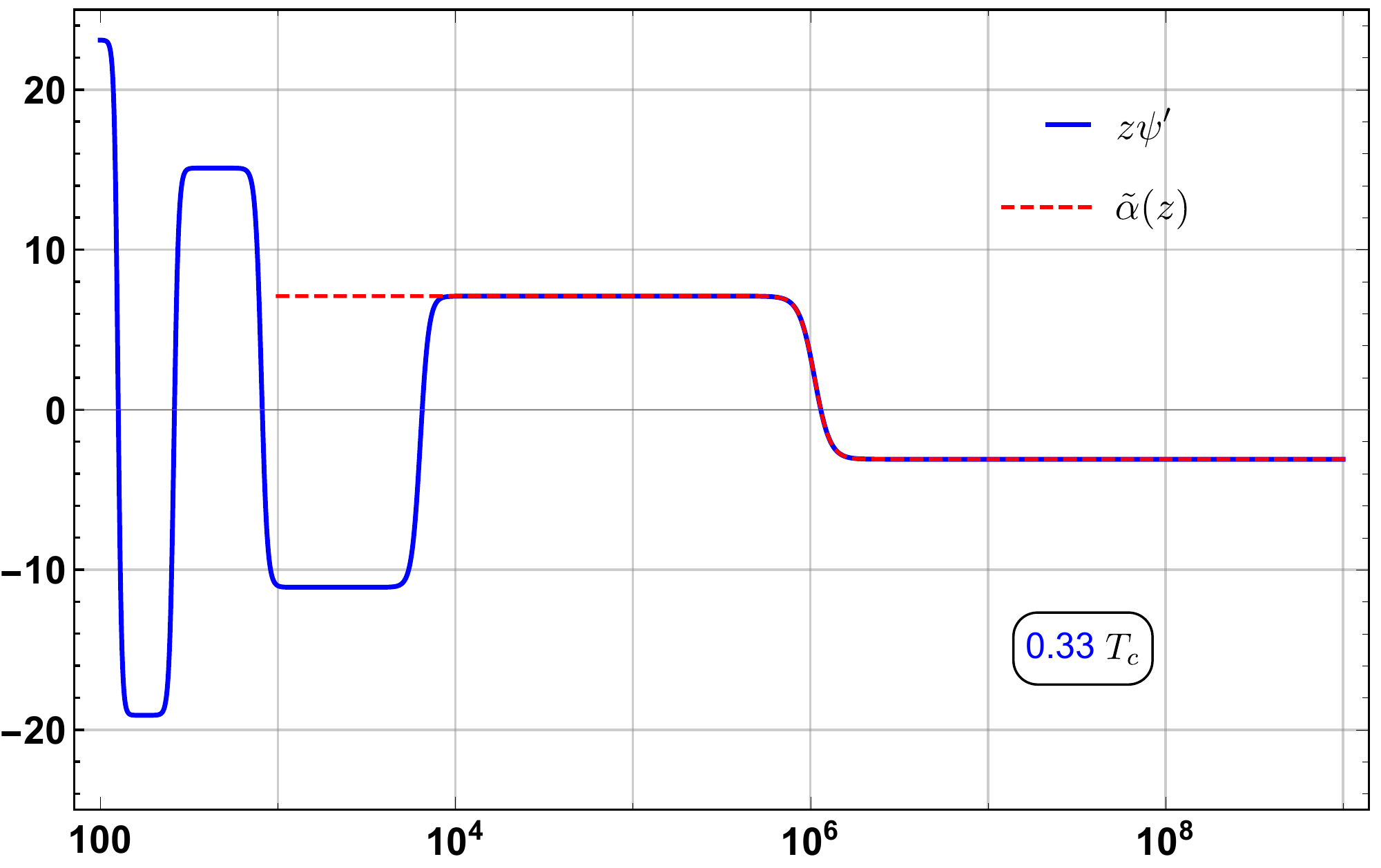}\\
\caption{A comparison of the analytical description~\eqref{sol-alpha} (red dashed curve) and the numerical one (solid blue curve) for Kasner transitions. We choose $T=0.33 T_c$ for which there are at least four Kanser transitions. The approximation~\eqref{sol-alpha} is in excellent agreement to the numerical solution of the full equations of motion~\eqref{eom-psi}-\eqref{eom-h}.}\label{Fig:tran-fit}
\end{figure}

We now check if the analytical solution~\eqref{sol-alpha} can capture all the important effects describing the Kasner transition. To fix the free parameters in~\eqref{sol-alpha}, we choose a value $z=z_i$ in the middle of the first Kasner epoch and compute $\tilde{\alpha}$ and $\tilde{\alpha}'$ at $z=z_i$. In Fig.~\ref{Fig:tran-fit}, we compare the profile of $\tilde{\alpha}=z\psi'$ from the analytical solution~\eqref{sol-alpha} with the numerical one of the full equations of motion. When $\tilde{\alpha}$ is constant away from a transition region, it describes a Kasner epoch with $\tilde{\alpha}=\alpha$ on the left side and $\tilde{\alpha}=\alpha_T$ on the right side. It is clear from Fig.~\ref{Fig:tran-fit} that our analytic solution~\eqref{sol-alpha} describes the Kasner transition very well. 

\paragraph{Kasner Inversion: }
The Kasner inversion behavior was first discussed in~\cite{Hartnoll:2020fhc} for a free charged scalar and also found in~\cite{Sword:2021pfm, Cai:2021obq}. For $|\alpha|<2$, we can obtain the approximate equation of motion for $\psi$ and $h$:
\begin{equation}\label{kas-inver-psi}
\frac{1}{z}(z\psi')'=-\frac{h'}{h}\psi', \quad h'=\frac{(\mathrm{e}^{\chi/2}A_t')^2}{4}\mathrm{e}^{-\chi/2}\,.
\end{equation}
We still use the constant variant method~\eqref{alpha-psi} to solve the differential equations~\eqref{kas-inver-psi}. Substituting~(\ref{alpha-psi}) into~(\ref{kas-inver-psi}), one obtains the following equation for $\tilde{\alpha}(z)$:
\begin{equation}\label{alpha-inver}
4z\tilde{\alpha}''\tilde{\alpha}-8z\tilde{\alpha}'^2+\tilde{\alpha}'\tilde{\alpha}^3=0\,,
\end{equation}
where we have used the fact that $\mathrm{e}^{\chi/2}A_t'$ is approximately a constant during the Kasner inversion. The analytic solutions will be compared to the full numerical ones later.

Let us consider the case with $0<\alpha<2$ in the Kasner epoch before the alternation. Solving the equation~(\ref{alpha-inver}) yields
\begin{equation}\label{inversion}
\begin{split}
\ln(z/z_{in})+\frac{c_1}{\sqrt{c_1^2-1}}\arctanh\left[\frac{2c_1-\tilde{\alpha}}{2\sqrt{c_1^2-1}}\right]+\ln (\tilde{\alpha}/2)-\frac{1}{2}\ln\left[\frac{4-4c_1\tilde{\alpha}+\tilde{\alpha}^2}{8-8c_1}\right]=\\
\frac{c_1}{\sqrt{c_1^2-1}}\arctanh\left[\frac{c_1-1}{\sqrt{c_1^2-1}}\right]\,,
\end{split}
\end{equation}
where $c_1>1$ is an integration constant and $z_{in}$ denotes the position in the inversion region at which $\tilde{\alpha}=2$. Considering the limit for which $z/z_{in}\ll 1$, $\tilde{\alpha}$ approaches a constant that is nothing but the value of $\alpha$ in the Kasner epoch before the Kasner inversion. Under the limit $z/z_{in}\rightarrow 0$, one obtains from~\eqref{inversion} that 
\begin{equation}
\lim_{z/z_{in}\rightarrow 0} \tilde{\alpha}=2\left(c_1-\sqrt{c_1^2-1}\right)\,.
\end{equation}
Taking the opposite limit $z/z_{in}\rightarrow \infty$, $\tilde{\alpha}$ should go to a constant that corresponds to the Kasner epoch after the Kasner inversion. Therefore, one has
\begin{equation}
\lim_{z/z_{in}\rightarrow \infty} \tilde{\alpha}=2\left(c_1+\sqrt{c_1^2-1}\right)\,.
\end{equation}
We then obtain the transformation rule for the \textbf{Kasner inversion} between two adjacent Kasner epochs.
\begin{equation}\label{in-algebra}
\alpha\,\alpha_I=4,\quad 0<\alpha<2\,,
\end{equation}
where $\alpha$ and $\alpha_{I}$ are, respectively, the value of $z\psi'$~\eqref{kasner} before and after the Kasner inversion process. Using a similar discussion, we find that the transformation rule for the case $-2<\alpha<0$ is the same as~\eqref{in-algebra}.  

\begin{figure}[H]
\centering
\includegraphics[width=0.65\textwidth]{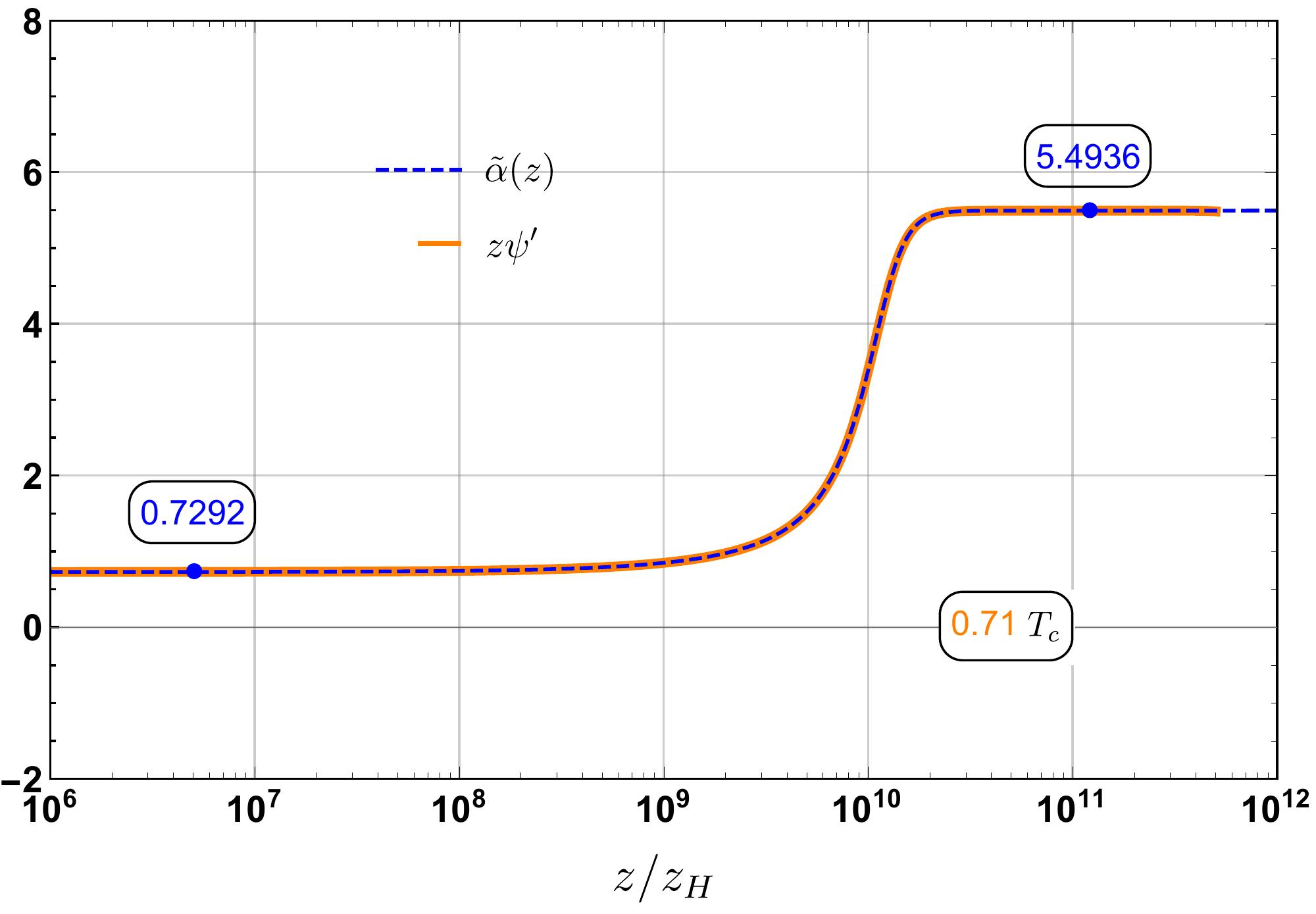}
\caption{A comparison of the analytical description~\eqref{inversion} (dashed blue curve) and the full numerical solutions (solid red curve) for Kasner inversion. We choose $T=0.71 T_c$ and show the value of $\alpha=z\psi'$ for each Kanser epoch. The approximation~\eqref{sol-alpha} is in good agreement to the numerical solution of the full equations of motion.}\label{Fig:inver-fit}
\end{figure}

In Fig.~\ref{Fig:inver-fit}, we compare the analytical solution~\eqref{inversion} with the numerical solution of the full equations of motion for $T=0.71T_c$. Away from the inversion region, the configuration corresponds to a Kasner epoch with $\tilde{\alpha}=\alpha\approx0.7292$ on the left side and $\tilde{\alpha}=\alpha_I\approx5.4936$ on the right side. The analytical solution fits the numerical one quite well. Moreover, $\alpha\,\alpha_T\approx4.0059$ agrees with~\eqref{in-algebra} excellently. 

\begin{figure}[H]
\centering
\includegraphics[width=0.65\textwidth]{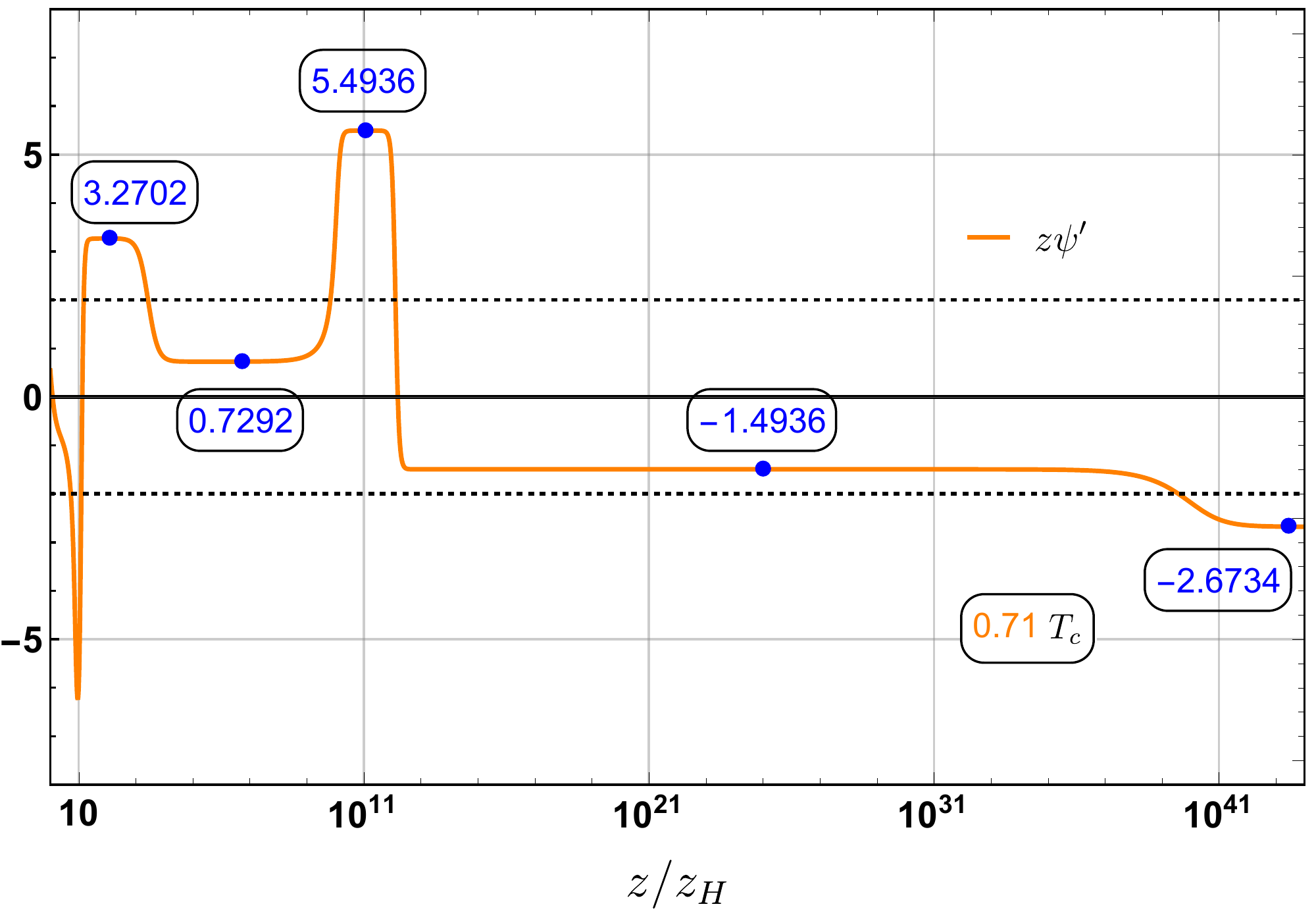}
\caption{Check the validity of the transformation rule for alternation of Kasner epochs. We choose $T=0.71 T_c$ for which two Kanser transitions and two Kasner inversions are shown. The value of $\alpha=z\psi'$ for each Kanser epoch is labeled explicitly. Two horizontal dashed lines correspond to $|\alpha|=2$. The transformation rule~\eqref{kas-law} is confirmed quantitatively.}\label{Fig:kas-emp}
\end{figure}

To summarize, we obtain the following transformation rule for the alternation of different Kasner epochs:
\begin{equation}\label{kas-law}
\begin{cases}
\mathrm{Kasner\ Transition}: \alpha+\alpha_T=4,\quad \alpha>2, \\
\mathrm{Kasner\ Inversion}: \quad \alpha\,\alpha_{I}=4,\quad 0<|\alpha|<2,\\
\mathrm{Kasner\ Transition}: \alpha+\alpha_T=-4,\quad \alpha<-2.\\
\end{cases}
\end{equation}
Let us first check the validity of the transformation rule~\eqref{kas-law}. The case for $T=0.71T_c$ is presented in Fig.~\ref{Fig:kas-emp} where we label the value of $\alpha$ in each Kasner epoch by fitting the numerical solution of the full equations of motion. There are two Kasner inversions and two Kasner transitions in Fig.~\ref{Fig:kas-emp}. For the first Kasner epoch, $\alpha_1\approx 3.2702>2$ for which there should be a Kanser transition according to~\eqref{kas-law}. One finds the value  $\alpha$ of the second epoch is $\alpha_2\approx0.7292$. So one has $\alpha_1+\alpha_2\approx 3.9994$ which agrees with~\eqref{kas-law} very well. Since $\alpha_2\approx0.7292<2$, the rule~\eqref{kas-law} suggests a Kasner inversion with $\alpha_2\, \alpha_3=4$. We obtain from our numerics that $\alpha_3\approx5.4936$ from which we have $\alpha_2\, \alpha_3\approx4.0059$. Then, there is a Kasner transition since $\alpha_3>2$, and the value of $\alpha$ for the fourth Kasner epoch is found to be $\alpha_4\approx-1.4936$ from which $\alpha_3+\alpha_4\approx4.0000$. After that, the fifth Kasner epoch develops from a Kasner inversion as $|\alpha_4|<2$. The predicted value of $\alpha_5=4/\alpha_4\approx-2.6781$ agrees perfectly with the numerical one $\alpha_5\approx-2.6734$. According to~\eqref{kas-law}, one then has the sixth Kasner epoch with $\alpha_6=-4-\alpha_5\approx-6.6734$ due to a Kasner inversion. In order to see more alternations of Kasner epochs, one needs to solve the equations of motion~\eqref{eom-psi}-\eqref{eom-h} to the far interior. Due to the limitation of computing power, we are not able to obtain the numerical solutions for sufficiently large $z$. Nevertheless, we have checked various numerical examples with a sequence of alternation of Kasner epochs, all of which agree with our transformation rule~\eqref{kas-law}.
\begin{figure}[H]
\centering
\includegraphics[width=1.0\textwidth]{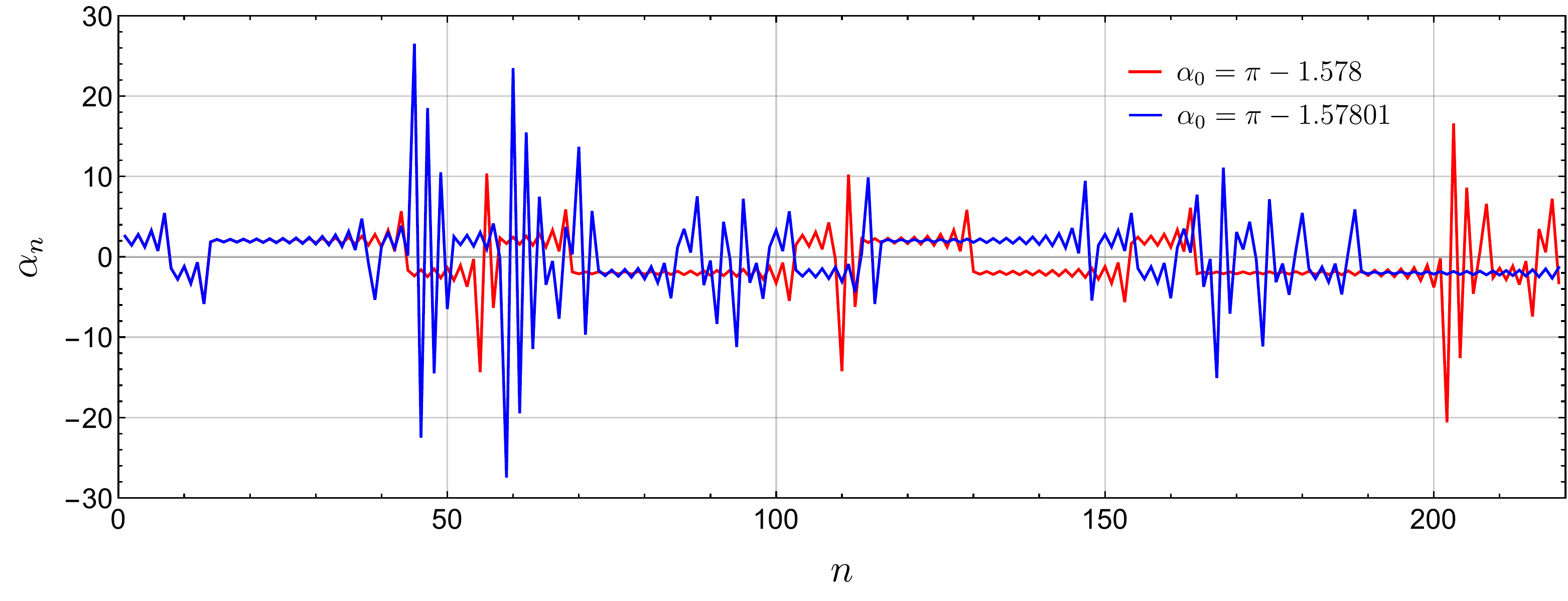}
\caption{The patten of $\alpha$ generated by the transformation rule~\eqref{kas-law} due to the Kanser transition and the Kanser inversion. The initial value of $\alpha$ is chosen to be $\alpha_0=\pi-1.578$ (red) and $\alpha_0=\pi-1.57801$ (blue). It is clear that a slight difference in initial values causes significant changes after about $n=43$ times alternation of Kasner epoches.}\label{Fig:chaos}
\end{figure}
We emphasize that the presence of an exponential form of couplings ($\sinh \psi, \cosh\psi$) plays important role in triggering the alternation of Kasner epochs with the transformation rule~\eqref{kas-law}. 
Instead of a monotonic Kasner behavior all the way to the singularity $z\rightarrow\infty$, our transformation rule~\eqref{kas-law} suggests that there would be generically a never-ending alternation of Kasner epochs towards the singularity. Once giving the value of $\alpha$ in the first Kanser epoch, one can easily obtain all the Kasner exponents for the next series of Kasner epochs. Indeed, the transformation rule~\eqref{kas-law} yields a chaotic behavior of the component $\alpha$ for which the underlying patterns are highly sensitive to initial conditions. The pattern of $\alpha$ with $220$ times of the alternation of Kasner epochs is presented in Fig.~\ref{Fig:chaos} where the initial conditions for $\alpha$ are changed by $0.00001$. As can be seen from Fig.~\ref{Fig:chaos}, even the slightest difference in initial values causes significant changes, exhibiting sensitive dependence on initial conditions.

Our result is different from the result in~\cite{Henneaux:2022ijt} where the author discussed the interior dynamical behavior of the metric using the ``cosmological billiard'' description~\cite{Damour:2002et,Henneaux:2007ej}. It was found that the corresponding hyperbolic billiard region has infinite volume so that the system will settle down to a final Kasner regime. As our Lagrangian is different, the resulting billiard table is different. The study of~\cite{Henneaux:2022ijt} only considered a massive charged scalar with a non-minimal term $\psi^2 F_{\mu\nu}F^{\mu\nu}$. As shown in Appendix~\ref{billard}, the billiard table structure of~\cite{Henneaux:2022ijt} may not apply to our case due to the presence of the exponential form of couplings $\sinh \psi$ and $\cosh\psi$ in the top-down theory~\eqref{modelst4D}. It is an interesting question to see if our result can be obtained from the billiard approach. Moreover, the analysis of the original BKL paper~\cite{Belinsky:1970ew} suggests that the power substitution rule of the adjacent Kasner epochs is given by
\begin{equation}\label{BKL}
p_t\rightarrow \frac{|p_{t}|}{1-2|p_{t}|}\,, \quad p_{s}\to \frac{p_{s}-2|p_{t}|}{1-2|p_{t}|}.
\end{equation}
One can check explicitly that while the Kasner inversion of~\eqref{kas-law} satisfies this rule, the Kasner trantiton~\eqref{trasi-algebra} does violate it~\eqref{BKL}.

\section{Complexity in CA Conjecture}\label{sec:CAconj}
In the previous section, we have proven that there exists no inner horizon for the black holes with non-trivial charged scalar hair, and the hairy black holes approach a spacelike singularity. After knowing the interior structure of the hairy black holes, we will proceed to compute the complexity growth rate of the holographic superconductor. From the field theory point of view, we consider the thermo-field double (TFD) state
\begin{equation}
|\mathrm{TFD}\rangle=\sum_{i}\mathrm{e}^{-\beta E_{i}/2}|E_{i}\rangle_{L} |E_{i}\rangle_{R}\,,
\end{equation}
which is dual to an eternal AdS black hole~\cite{Maldacena:2001kr} with $\beta$ the inverse of temperature. We want to know if this quantum information measure can probe the dynamics inside black holes.

Two well-known holographic proposals for the computation of the complexity are the CV duality~\cite{Stanford:2014jda} and the CA duality~\cite{Brown:2015lvg}. More recently, it was argued that the CV complexity generally fails to fully probe the interior geometry~\cite{Caceres:2022smh}. Moreover, an infinite class of gravitational descriptions of the complexity from volume are proposed~\cite{Belin:2021bga}, while the complexity from the action of the WdW patch is uniquely defined. In this section, we focus on the CA conjecture to compute the complexity growth rate.  So far, there is no study on the behavior of complexity of holographic superconductor using the CA conjecture as this computation requires detailed knowledge of the interior structure of a hairy black hole that has to be constructed numerically. The complexity of CV is presented in Appendix~\ref{CVconj} and is found to be similar to the free charged scalar case studied in~\cite{Yang:2019gce}.

\begin{figure}[H]
\begin{minipage}{6in}
  \centering
  $\vcenter{\hbox{\includegraphics[width=0.44\textwidth]{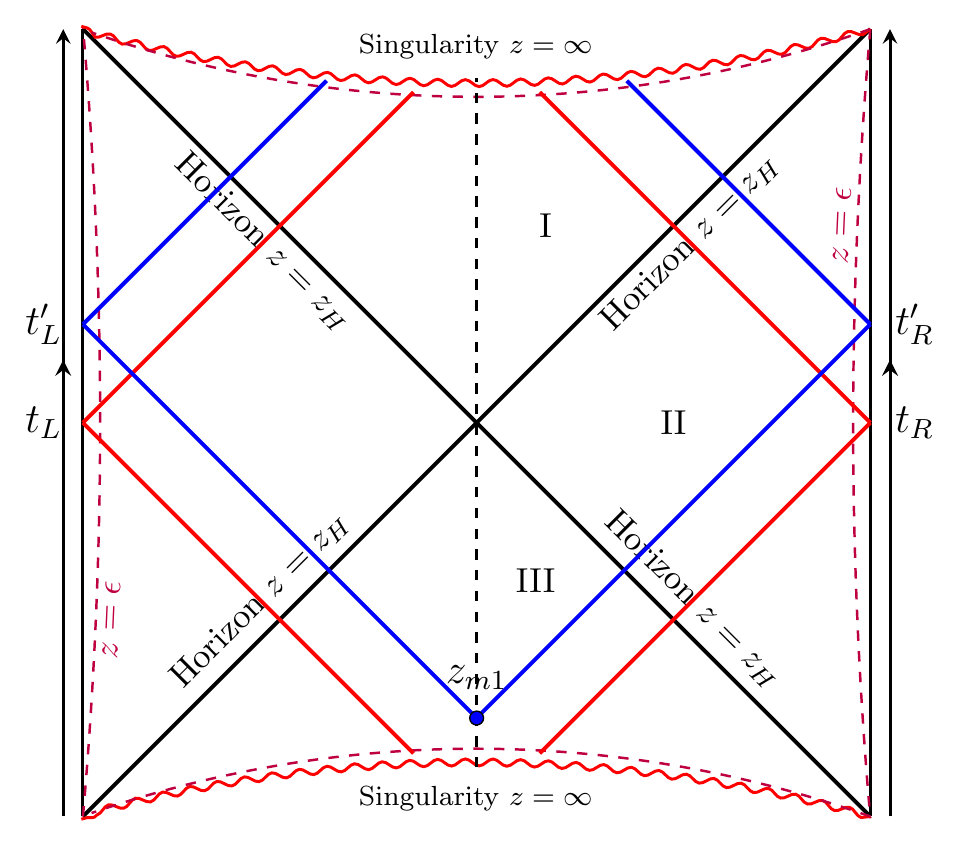}}}$
  \hspace*{.2in}
  $\vcenter{\hbox{\includegraphics[width=0.44\textwidth]{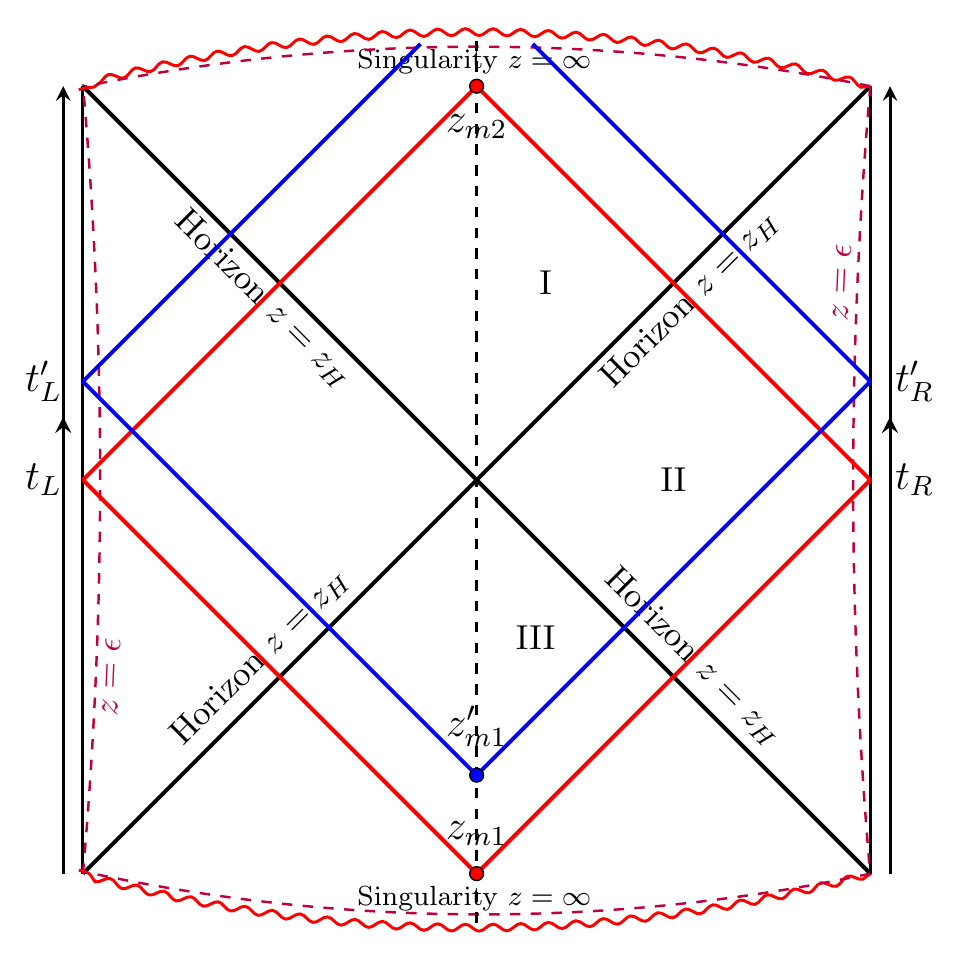}}}$
\end{minipage}
\caption{The WdW patch at the boundary time $t=0$ (red line) and $t>0$ (blue line). The critical time $t_{c}$ is the time where the null past joint of WdW patch locates exactly at the past singularity. \textbf{Left: }the case with $t_c>0$. \textbf{Right: }the case with $t_c<0$. }
\label{penrose-ca}
\end{figure}

In the CA conjecture, one needs to compute the on-shell action of the WdW patch obtained by shooting null rays from a constant-$t$ boundary slice into the bulk, see Fig.~\ref{penrose-ca}. Note that the dual state depends on two times $t_L$ and $t_R$ with subscripts $L$ and $R$ representing, respectively, the left and right boundary times. In the present work, we are interested in the symmetric configuration with $t=t_L=t_R$. The complexity from CA is then given by
\begin{equation}
\mathcal{C}_{A}=\frac{S_{\mathrm{WdW}}}{\pi \hbar}\,.
\end{equation}
It is manifest from Fig.~\ref{penrose-ca} that the WdW patch can probe the region near the singularity of a black hole. Therefore, one needs to know the whole structure of the spacetime, in particular, the geometry near the singularity. Such a difficulty does not appear in the CV conjecture because the extremal hypersurface will not touch the singularity. Our numerical results suggest that the spacetime near the singularity takes the Kasner form, which we will see later is important when computing the Gibbons-Hawking-York (GHY) boundary term in the action. In the following context, we will use the ingoing and outgoing coordinates
\begin{equation}\label{in-out}
    \ v=t-F(z),\quad u=t+F(z),\quad  F(z)=\int^z \frac{\mathrm{e}^{\chi(s)/2}}{f(s)}\mathrm{d}s.
\end{equation}
By doing the coordinate transformation to ingoing coordinate $v$ in~(\ref{in-out}), one can obtain the metric
\begin{equation}\label{nullcord}
\mathrm{d}s^{2}=\frac{1}{z^{2}}\left(-f(z)\mathrm{e}^{-\chi(z)} \mathrm{d}v^{2}-2 \mathrm{e}^{-\chi/2} \mathrm{d}v \mathrm{d}z+\mathrm{d}\Sigma_{2,0}^{2}\right)\,,
\end{equation}
where $\Sigma_{2,0}$ denotes the two-dimensional transverse space.

\subsection{Evaluating the action}
Based on the analysis of~\cite{Carmi:2017jqz}, one can directly neglect the Hayward joint terms and the null boundary terms when the computation only focuses on the growth rate of complexity since these terms are time-independent. We write the action of the WdW patch in the following form.
 \begin{equation}\label{WdW}
\begin{split}
S_{\text{WdW}} = & \frac{1}{16 \pi G_{N}} \int_\mathcal{M} \mathrm{d}^{4} x \sqrt{-g} \left(\mathcal R+\mathcal{L}_{S}^{(4)} \right) \\
&+ \frac{1}{8\pi G_{N}} \left(\int_{\mathcal{B}} \mathrm{d}^3 x \sqrt{|h|} K + \int_{\Sigma'} \mathrm{d}^{2} x \sqrt{\sigma} \textbf{a} + \int_{\mathcal{B'}} d\lambda \mathrm{d}^{2}x \sqrt{\gamma} \Theta \ln (\ell_{c t}\Theta)\right)\,.
\end{split}
\end{equation}
Here $h$ and $\sigma$ are the induced metrics for the GHY surface and the transverse space of the null boundary, respectively. $K$ is the trace of the extrinsic curvature of the GHY surface and $\textbf{a}$ relates the normals on the intersection of the null boundary segments, $\Theta=\partial_{\lambda} \ln \sqrt{\gamma}$ is the expansion scalar of the null boundary generators, and $\ell_{c t}$ is an arbitrary length scale that will be set to be the AdS radius $L$. The two terms in the first line are the bulk Einstein-Hilbert term. The three terms in the second line are the GHY term, the null joint term, and its corresponding counter term sequentially. The counter term is added to avoid the ambiguity of arbitrary normalization of the normal vectors. We now show how to compute the growth rate $\mathrm{d}S_{\text{WdW}} /\mathrm{d}t$.
\paragraph{Bulk contribution} For the top-down theory~\eqref{modelst4D}, the bulk action reads
\begin{equation}
    S_{\mathrm{bulk}}=\frac{1}{16\pi G_N}\int \mathrm{d}^{4}x\sqrt{-g}(R+\mathcal{L}_{S}^{(4)})\,.
\end{equation}
The energy momentum tensor is
\begin{equation}
  T_{\mu\nu}=-\frac{\delta\mathcal{L}_{S}^{(4)}}{\delta g^{\mu\nu}}+\frac{1}{2}g_{\mu\nu}\mathcal{L}_{S}^{(4)}\,.
\end{equation}
Considering the Einstein's equation, one can derive
\begin{equation}
R_{\mu \nu}-\frac{1}{2}g_{\mu\nu}R=T_{\mu\nu}\ \Rightarrow \ R=g^{\mu\nu}\frac{\delta\mathcal{L}_{S}^{(4)}}{\delta g^{\mu\nu}}-2\mathcal{L}_{S}^{(4)}\,.
\end{equation}
Therefore, the on-shell action of the bulk term is given by
\begin{equation}
S_{\mathrm{bulk}}=\frac{1}{16\pi G_N}\int \mathrm{d}^{4}x\sqrt{-g}I(z),\quad \sqrt{-g}=\frac{1}{z^{4}}\mathrm{e}^{-\chi(z)/2}\,,
\end{equation}
with
\begin{equation}
I(z)=\frac{1}{2}z^{4}\mathrm{e}^{\chi(z)}A_{t}'(z)^{2}-\frac{1}{2}(1+\cosh\psi)(7-\cosh \psi)\,.
\end{equation}
\paragraph{GHY surface contribution} Next, we consider the GHY boundary term. The normal vector of the future singularity is
\begin{equation}
n^a=\frac{1}{z\sqrt{-f}}\left(\frac{\partial}{\partial z}\right)^a\,.
\end{equation}
Then, one can obtain the extrinsic curvature
\begin{equation}
    K=\nabla_a n^a=\frac{1}{\sqrt{-g}} \partial_{z}\left(\sqrt{-g} n^{z}\right)=-z^{4}\mathrm{e}^{\chi/2} \partial_{z}\left(\frac{\mathrm{e}^{-\chi/2}}{z^{3}}\sqrt{-f}\right)\,.
\end{equation}
Therefore, the boundary term action is given by
\begin{equation}
    S_{bdy}=\frac{1}{8\pi G_N} \int K\mathrm{d}\Sigma=-\frac{\Omega_{2}}{8\pi G_{N}} z\sqrt{-f}\partial_{z}\left(\frac{\mathrm{e}^{-\chi/2}}{z^{3}}\sqrt{-f}\right)\left(\frac{t}{2}+F(\infty)-F(0)\right)\,,
\end{equation}
and its time derivative is
\begin{equation}
    \frac{\mathrm{d}S_{bdy}}{\mathrm{d}t}=-\frac{\Omega_{2}}{8\pi G_{N}} z \sqrt{-f}\partial_{z}\left(\frac{1}{z^{3}}\sqrt{-f}\mathrm{e}^{-\chi/2}\right)\bigg|_{z=\infty}\,.
\end{equation}
Here $\Omega_2$ is the spatial volume for the transverse space of~\eqref{nullcord}.

We note here that there is a close relationship between the Kasner behavior near the singularity $z\rightarrow\infty$ and the GHY boundary term of the action. It's direct to check that the action growth rate of the boundary term is a finite constant for the Kasner singularity:
\begin{equation}
\frac{\mathrm{d}S_{bdy}}{\mathrm{d}t}=-\frac{\Omega_{2}}{16\pi G_{N}}\left(3+\frac{\alpha^2}{4}\right)h(\infty)\,.
\end{equation}
\paragraph{Null joint and counter term contribution} The null vectors are given by
\begin{equation}
k_{R}=c \left(\mathrm{d} t+\frac{\mathrm{e}^{\chi/2}}{f(z)}\mathrm{d}z\right),\quad k_{L}=c \left(-\mathrm{d} t+\frac{\mathrm{e}^{\chi/2}}{f(z)}\mathrm{d}z\right)\,,
\end{equation}
where we have used $\sqrt{\sigma}=1/z^{2}$ for the transverse space and $c$ is a constant. Then, the joint term of action for both past and future joints is
\begin{equation}\label{i-jnt}
S_{jnt} =\frac{1}{8 \pi G_N} \int_{\Sigma} \mathrm{d}^{2} x \sqrt{\sigma} \ln \left(\frac{1}{2} k_{R} \cdot k_{L}\right) =-\frac{\Omega_{2}}{8\pi G_{N}z_{m}^{2}}\ln \frac{g_{tt}(z_{m})}{c^2}\,.
\end{equation}
To deal with the dependence of the normalization factor, we need to add a counter term for the joint term. The counter term is given by
\begin{equation}
S_{cnt}=\frac{1}{8\pi G_{N}} \int \mathrm{d}\lambda \mathrm{d}^{2}x \sqrt{\gamma} \Theta \ln\Theta\,.
\end{equation}
For our study, we take the parameter to be $\lambda=1/cz$, although the result does not depend on the particular parametrization. Here, we have $\Theta=\partial_{\lambda}\ln \sqrt{\sigma}=2 c z $. Hence, we can obtain
\begin{equation}
S_{cnt}=-\frac{\Omega_{2}}{4\pi G} \int_{\mathcal{B'}}\frac{\mathrm{d}z}{z^{3}}\ln 2cz=-\frac{\Omega_{2}}{4\pi G z_{m}^{2}}\left(\frac{1}{2}+\ln (2c z_{m})\right)\,,
\end{equation}
where $\mathcal{B'}$ means the null hypersurface accompanying the joint. One can see that the dependence on the normalization factor $c$ precisely cancels. 

We denote the past null boundary joint as $z_{m1}$ and the future null boundary joint as $z_{m2}$ (see Fig.~\ref{penrose-ca}). Both are related to the boundary time by the following equations:
\begin{equation}\label{bdytime}
  \frac{t}{2}+F(0)-F\left(z_{m1}\right)=0, \quad \frac{t}{2}-F(0)+F\left(z_{m 2}\right)=0\,.
\end{equation}
Therefore, one can obtain
\begin{equation}
\frac{\mathrm{d} z_{m 1}}{\mathrm{d} t}=-\frac{1}{2} g_{t t} z_{m 1}^{2} \mathrm{e}^{\chi / 2}, \quad \frac{\mathrm{d} z_{m 2}}{\mathrm{d} t}=\frac{1}{2} g_{t t} z_{m 2}^{2} \mathrm{e}^{\chi / 2}\,.
\end{equation}
Therefore, the time derivative of the action~\eqref{WdW} from the null joint term and its corresponding counter term for both past and future joints are, respectively,
\begin{equation}
\begin{aligned}
\frac{\mathrm{d}S_{jnt_{1}}}{\mathrm{d}t}+\frac{\mathrm{d}S_{cnt_{1}}}{\mathrm{d}t}&=\frac{\Omega_{2}}{8\pi G_{N}}\left[-\frac{\mathrm{d}}{\mathrm{d}z_{m1}}\left(\frac{\ln g_{tt}}{z_{m1}^{2}}\right) \frac{\mathrm{d}z_{m1}}{\mathrm{d}t}+2h(z_{m1})\ln(2z_{m1})\right]\,,\\
\frac{\mathrm{d}S_{jnt_{2}}}{\mathrm{d}t}+\frac{\mathrm{d}S_{cnt_{2}}}{\mathrm{d}t}&=\frac{\Omega_{2}}{8\pi G_{N}}\left[-\frac{\mathrm{d}}{\mathrm{d}z_{m2}}\left(\frac{\ln g_{tt}}{z_{m2}^{2}}\right)\frac{\mathrm{d}z_{m2}}{\mathrm{d}t}- 2h(z_{m2})\ln(2z_{m2})\right]\,,
\end{aligned}
\end{equation}
where we have used $g_{tt}$ and $h$ in those equations for a concise form.
\subsection{Complexity growth rate}
\subsubsection{The shape of WdW patch}
Due to the time-reversal symmetry, we only need to calculate the complexity growth rate for $t\geq0$. First, we should determine the structure of the WdW patch at $t=0$ and check whether it will intersect the singularity or not. To do that, we compute the critical time $t_{c}$ which is the time where the null past joint of the WdW patch locates exactly at the past singularity. More precisely, it is given by the equation of constant $u$ of~\eqref{in-out} along the past null sheet.
\begin{equation}
\frac{t_{c}}{2}+F(0)=F(\infty)\,,
\end{equation}
which determines the boundary time when the null joint intersects the past singularity.

\begin{figure}[H]
\centering
\includegraphics[width=0.65\textwidth]{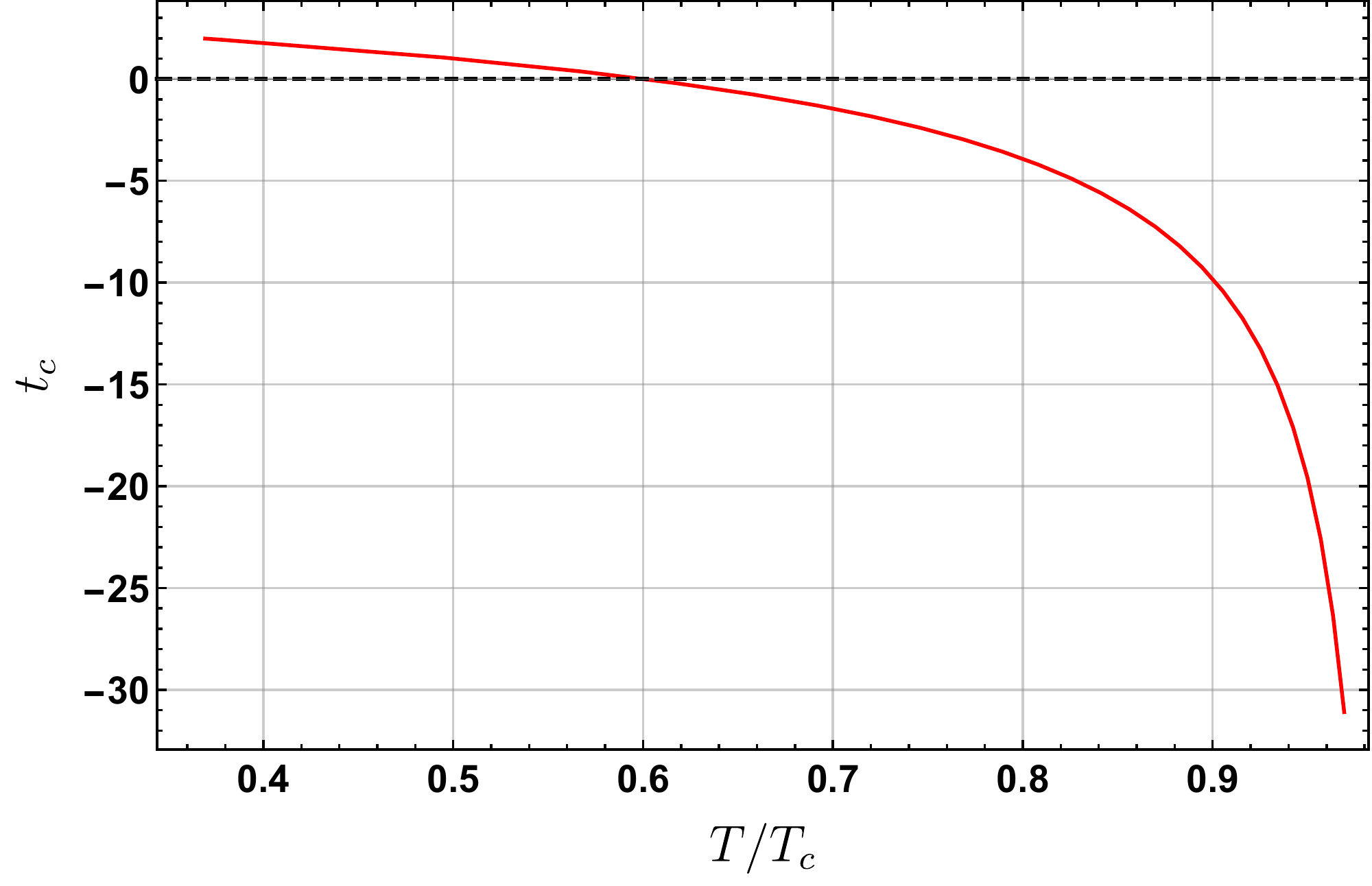}
\caption{Relation between the critical time $t_c$ and the temperature of the superconducting black hole of~\eqref{modelst4D}. $t_c$ changes sign at $T\approx0.6T_c$. We have fixed the chemical potential $\mu=1$.}
\label{ctime-temp}
\end{figure}

For some analytical solutions, such as BTZ black hole, the critical time can be evaluated analytically. While for our numerical background, to obtain $t_c$, we need to handle the divergence at the event horizon where $h(z_{H}) =0$. Fortunately, the divergence will cancel each other between exterior and interior by noting that $h(z)=h'(z_{H})(z-z_{H})$ near $z_H$. To see this point more clearly, let us take the small cutoff $\delta$ near the event horizon. In the small $\delta$ limit, one has
\begin{equation}\label{cut-crit}
\int_{z_{H}(1-\delta)}^{z_{H}} \frac{1}{h'(z_{H})(z-z_{H})z_{H}^{3}}\mathrm{d}z \simeq -\int_{z_{H}}^{z_{H}(1+\delta)} \frac{1}{h'(z_{H})(z-z_{H})z_{H}^{3}}\mathrm{d}z\,.
\end{equation}
Therefore, one can safely take a small cutoff to calculate the critical time in practice. The value of $t_c$ in the function of the temperature of the hairy black holes is presented in Fig.~\ref{ctime-temp}. One can find that the value of $t_c$ will change from negative to positive as the temperature is decreased. Therefore, we have to consider two kinds of WdW patch corresponding to the positive critical time (relatively low temperature, left panel of Fig.~\ref{penrose-ca}) and the negative critical time (relatively high temperature, right panel of Fig.~\ref{penrose-ca}), respectively.

\subsubsection{The complexity evolution}
\paragraph{The case of positive critical time}
For a given temperature, when the critical time $t_c>0$, the WdW patch intersects the singularity at the boundary time $t=0$, see the left panel of Fig.~\ref{penrose-ca}. In this case, at first, the GHY term does not depend on time as the WdW patch intersects both the future and the past singularities. It can be easily shown that the total action growth rate (the complexity growth rate) vanishes until the boundary time past the critical time, \emph{i.e.} $t>t_c$.

After the critical time $t_{c}$, the terms that contribute to the complexity growth rate by the CA conjecture are the bulk term, the joint term at $z_{m1}$, and the corresponding counter term on the null surface. First, we decompose the bulk region into three portions:
\begin{equation}
\begin{aligned}
S_{\mathrm{bulk}}^{\mathrm{I}}&=\frac{2\Omega_{2}}{16\pi G_{N}} \int_{z_H}^{\infty} \frac{1}{z^{4}} \mathrm{e}^{-\chi(z)/2} I(z)\left(\frac{t}{2}+F(z)-F(0)\right)\mathrm{d}z\,,\\
S_{\mathrm{bulk}}^{\mathrm{II}}&=\frac{4\Omega_{2}}{16\pi G_{N}} \int_{0}^{z_H} \frac{1}{z^{4}} \mathrm{e}^{-\chi(z)/2} I(z)\left(F(0)-F(z)\right) \mathrm{d}z\,,\\
S_{\mathrm{bulk}}^{\mathrm{III}}&=\frac{2\Omega_{2}}{16\pi G_{N}} \int_{z_{m1}}^{z_H} \frac{1}{z^{4}} \mathrm{e}^{-\chi(z)/2} I(z)\left(\frac{t}{2}-F(z)+F(0)\right)\mathrm{d}z\,.
\end{aligned}
\end{equation}
Then, we can obtain
\begin{equation}
    \frac{\mathrm{d}S_{\mathrm{bulk}}}{\mathrm{d}t}=\frac{\Omega_{2}}{16\pi G_{N}} \int_{z_{m1}}^{\infty} \frac{1}{z^{4}}I(z)\mathrm{e}^{-\chi(z)/2}\mathrm{d}z\,.
\end{equation}
Considering other contributions from the boundary term, the past joint term and its corresponding counter term, we can get the total complexity growth rate

\begin{equation}
\begin{aligned}
\frac{\mathrm{d}\mathcal{C}_{A}}{\mathrm{d}t}=\frac{\Omega_{2}}{16\pi^2 G_{N}} \bigg[\int_{z_{m1}}^{\infty} & \frac{1}{z^{4}}I(z)\mathrm{e}^{-\chi(z)/2}\mathrm{d}z-\left(3+\frac{\alpha^2}{4}\right)h(\infty)\\
&+\frac{\mathrm{d}}{\mathrm{d}z_{m1}}\left(-\frac{\ln g_{tt}^2}{z_{m1}^{2}}\right)\frac{\mathrm{d}z_{m1}}{\mathrm{d}t}+4h(z_{m1}) \ln (2z_{m1})\bigg]\,,
\end{aligned}
\end{equation}
where we choose $\hbar=1$ from now on.

\begin{figure}[H]
\centering
\includegraphics[width=0.49\textwidth]{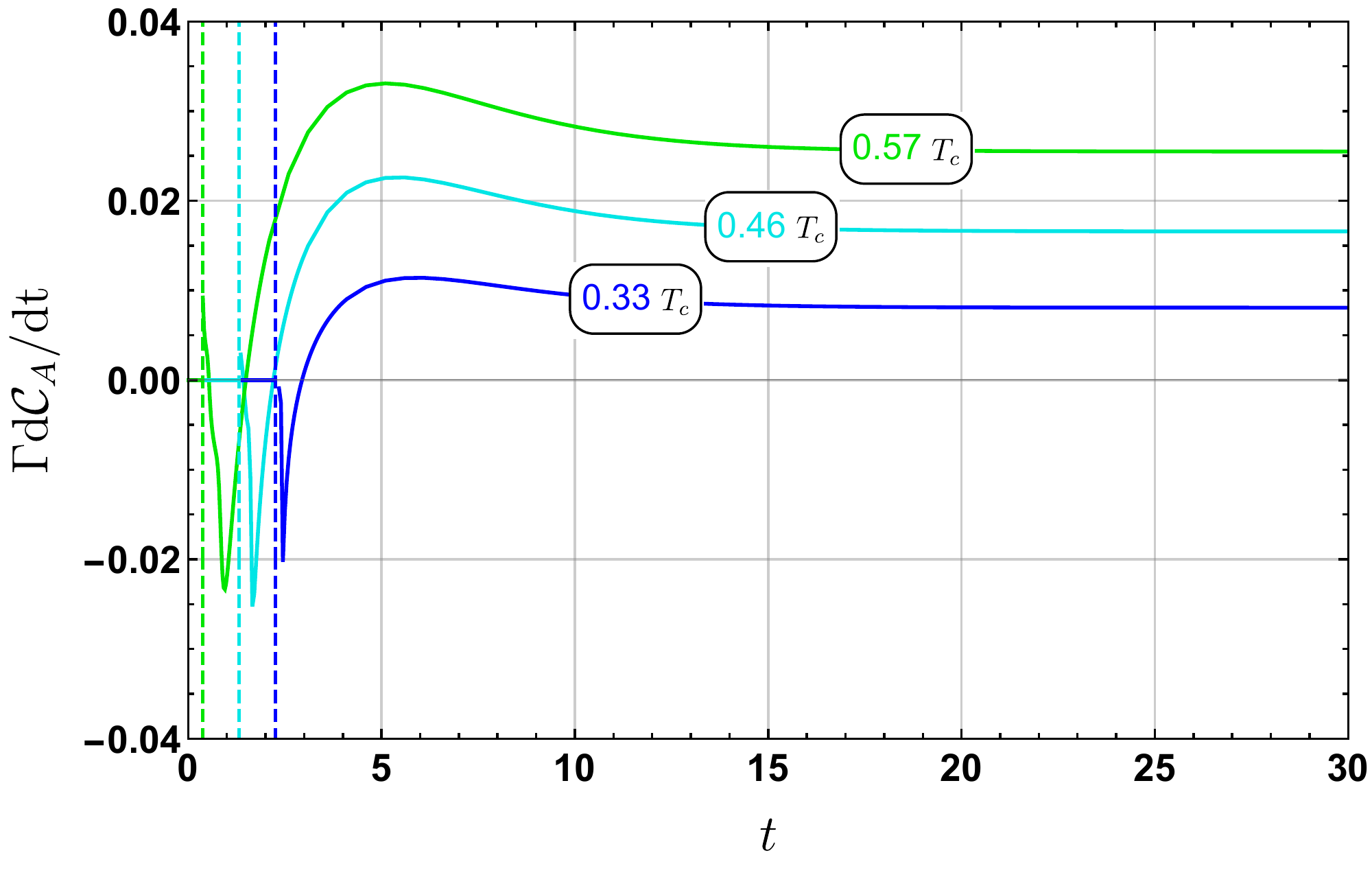}
\includegraphics[width=0.48\textwidth]{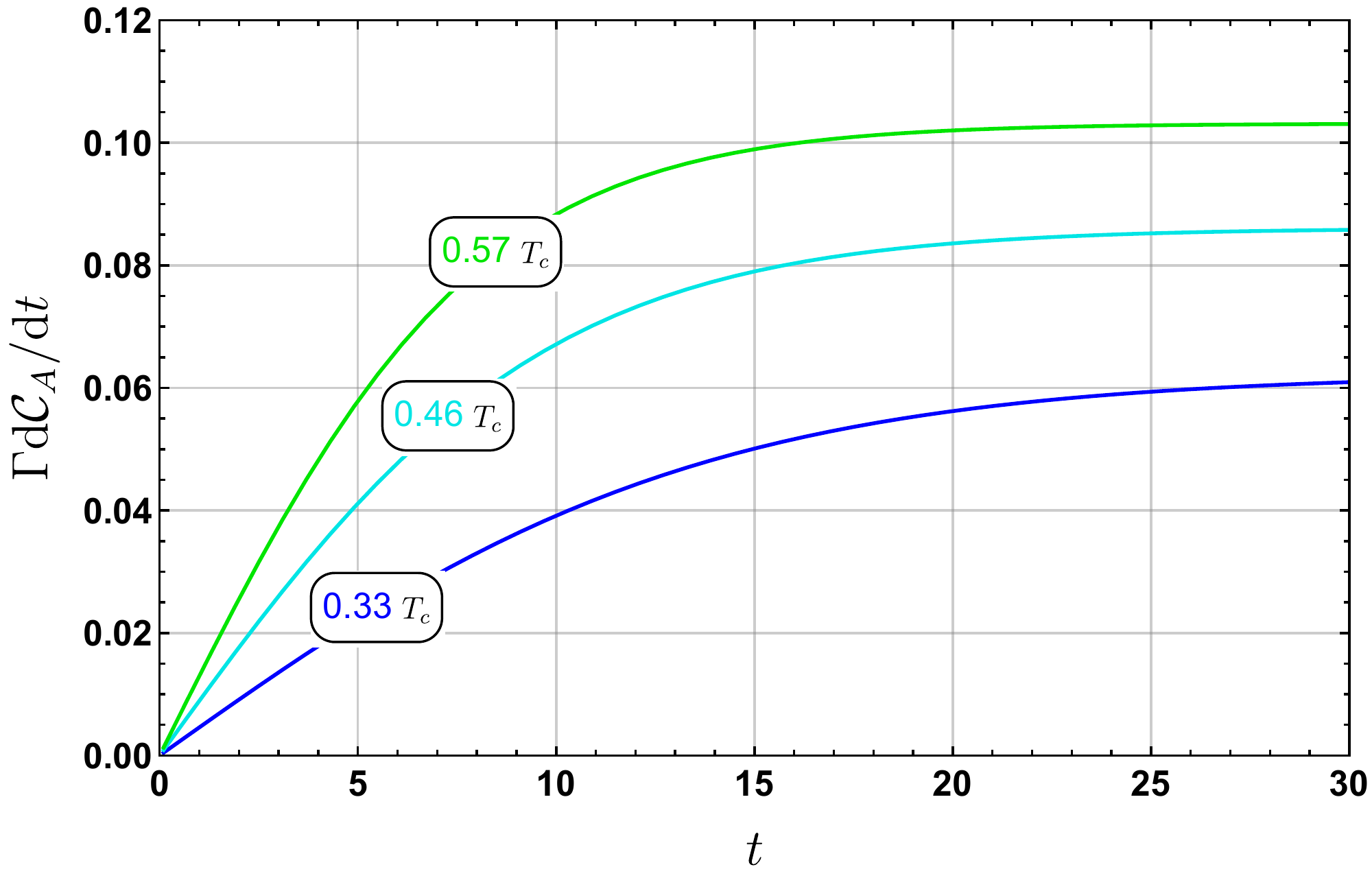}
\caption{The complexity growth rate of the superconducting phase for the positive critical time $t_c>0$ (left panel). As a comparison, the case for the normal phase at the same temperature is presented in the right panel. Here $\Gamma= 16\pi^2 G_{N}/\Omega_{2}$ and $\mu=1$.}\label{Fig:grca-1}
\end{figure}

We first discuss the late time limit, where the null joint at $z_{m1}$ is extremely close to the horizon. For $z_{m1} \to z_H$, we can derive from the equation of motion of $h(z)$ that
\begin{equation}
\lim_{z_{m}\to z_H}\int_{z_{m}}^{\infty} \frac{1}{z^{4}}I(z)\mathrm{e}^{-\chi(z)/2}\mathrm{d}z=\lim_{z_{m}\to z_H}\int_{z_{m}}^{\infty} 2h'(z) \mathrm{d}z=2h(\infty)\,.
\end{equation}
We also have
\begin{equation}
\lim_{z_{m1}\to z_H}\frac{\mathrm{d}}{\mathrm{d}z_{m1}}\left(-\frac{\ln g_{tt}^2}{z_{m1}^{2}}\right)\frac{\mathrm{d}z_{m1}}{\mathrm{d}t}= \lim_{z_{m1}\to z_H} \mathrm{e}^{\chi/2}g_{tt}'=\frac{4\pi T}{z_H^2}\,.
\end{equation}
Moreover, considering the conserved charge~\eqref{myQ} with $\mathcal{Q}(\infty)=\mathcal{Q}(z_H)$, one can obtain
\begin{equation}\label{Th}
 \frac{4\pi T}{z_H^2}= -h(\infty)\left(3-\frac{\alpha^2}{4}\right)+ (\rho A_t)|_{\infty}\,,
\end{equation}
where $\rho=\mathrm{e}^{\chi/2}A_t'$ characterizes the charge degrees of freedom behind the surface generating a nonzero electric flux in the deep interior. As a result, we claim that the growth rate of complexity is bounded at late time. The late time complexity growth is
\begin{equation}\label{ca-late}
\lim_{z_{m1}\to z_H}\frac{\mathrm{d}\mathcal{C}_{A}}{\mathrm{d}t}=\frac{\Omega_{2}}{16\pi^2 G_{N}}[-4h(\infty)+(\rho A_{t})|_{\infty}].
\end{equation}
The complexity growth using the CA conjecture for different temperatures in positive $t_{c}$ case is shown in Fig.~\ref{Fig:grca-1}. 
In the normal phase, the complexity growth rate with respect to time increases monotonically and saturates to a temperature-dependent constant. In contrast, there is a time range in which the complexity growth rate of the superconducting phase is decreasing and negative. This feature corresponds to the past joint term $z_{m1}$ that begins to probe the deep interior of the hairy black hole we discussed in Section~\ref{sec:inner}. As shown in Appendix~\ref{CVconj}, in the superconducting phase, the complexity growth rate from CV shares a similar behavior as the normal phase, and thus is not able to characterize the dramatic change of the interior structure of the black hole with scalar hair. We shall leave the relation between the precise feature of the complexity growth rate and the interior dynamics to future work. 

\paragraph{The case of negative critical time}
When the critical time $t_c<0$, the WdW patch of the superconducting phase has two joints $z_{m1}$ and $z_{m2}$ at the boundary time $t=0$ (see the right panel of Fig.~\ref{penrose-ca}). This feature is quite different from all known examples with a spacelike singularity in the literature for which the WdW patch typically has no null joint term at the boundary time $t=0$.

We then find that the terms that contribute to the complexity growth rate by CA conjecture include the bulk term, the joint terms at $z_{m1}$ and $z_{m2}$, and the corresponding counter terms on the null surface before $z_{m2}$ arriving at the spacelike singularity. As time evolves, the future joint $z_{m2}$ will intersect the future singularity. After that, the terms contributing to the complexity growth rate are the bulk term, the GHY boundary term, the joint term at $z_{m1}$, and the corresponding counter terms on the null surface. Thus, we divide the evolution into two periods by the critical time $t=-t_c$ where $-t_{c}$ is the time when the future joint $z_{m2}$ intersects the future singularity. In our present case, the late time complexity growth after the time at which $z_{m2}$ intersects the future singularity is similar to the previous case with positive $t_c$. Nevertheless, the early time growth behavior is different from the former one.
  
Before the time $-t_{c}$, we decompose the bulk region into three portions:
\begin{equation}
\begin{aligned}
S_{\mathrm{bulk}}^{\mathrm{I}}&=\frac{2\Omega_{2}}{16\pi G_{N}} \int_{z_H}^{z_{m2}} \frac{1}{z^{4}} \mathrm{e}^{-\chi(z)/2} I(z)\left(\frac{t}{2}+F(z)-F(0)\right)\mathrm{d}z\,,\\
S_{\mathrm{bulk}}^{\mathrm{II}}&=\frac{4\Omega_{2}}{16\pi G_{N}} \int_{0}^{z_H} \frac{1}{z^{4}} \mathrm{e}^{-\chi(z)/2} I(z)(F(0)-F(z)) \mathrm{d}z\,,\\
S_{\mathrm{bulk}}^{\mathrm{III}}&=\frac{2\Omega_{2}}{16\pi G_{N}} \int_{z_{m1}}^{z_H} \frac{1}{z^{4}} \mathrm{e}^{-\chi(z)/2} I(z)\left(\frac{t}{2}-F(z)+F(0)\right)\mathrm{d}z\,.
\end{aligned}
\end{equation}
Then, we obtain
\begin{equation}
\frac{\mathrm{d}S_{\mathrm{bulk}}}{\mathrm{d}t}=\frac{\Omega_{2}}{16\pi G_{N}} \int_{z_{m1}}^{z_{m2}} \frac{1}{z^{4}}\mathrm{e}^{-\chi(z)/2} I(z)\mathrm{d}z\,.
\end{equation}
In addition, considering other two joint terms and the corresponding counter terms, the total complexity growth rate reads %
\begin{equation}
\begin{split}
\frac{\mathrm{d}\mathcal{C}_{A}}{\mathrm{d}t}=&\frac{\Omega_{2}}{16\pi^2 G_{N}} \int_{z_{m1}}^{z_{m2}} \frac{1}{z^{4}}\mathrm{e}^{-\chi(z)/2} I(z)\mathrm{d}z\\
&+\frac{\Omega_{2}}{16\pi^2 G_{N}} \left[-\frac{\mathrm{d}}{\mathrm{d}z_{m1}}\left(\frac{\ln g_{tt}^2}{z_{m1}^{2}}\right)\frac{\mathrm{d}z_{m1}}{\mathrm{d}t}+4h(z_{m1}) \ln (2z_{m1})\right]\,,\\
&+\frac{\Omega_{2}}{16\pi^2 G_{N}} \left[-\frac{\mathrm{d}}{\mathrm{d}z_{m2}}\left(\frac{\ln g_{tt}^2}{z_{m2}^{2}}\right)\frac{\mathrm{d}z_{m2}}{\mathrm{d}t}-4h(z_{m2}) \ln (2z_{m2})\right].
\end{split}
\end{equation}
It is easy to see that the complexity growth rate equals zero at the starting point $t=0$ because of $z_{m1}=z_{m2}$. 

\begin{figure}[H]
\centering
\includegraphics[width=0.49\textwidth]{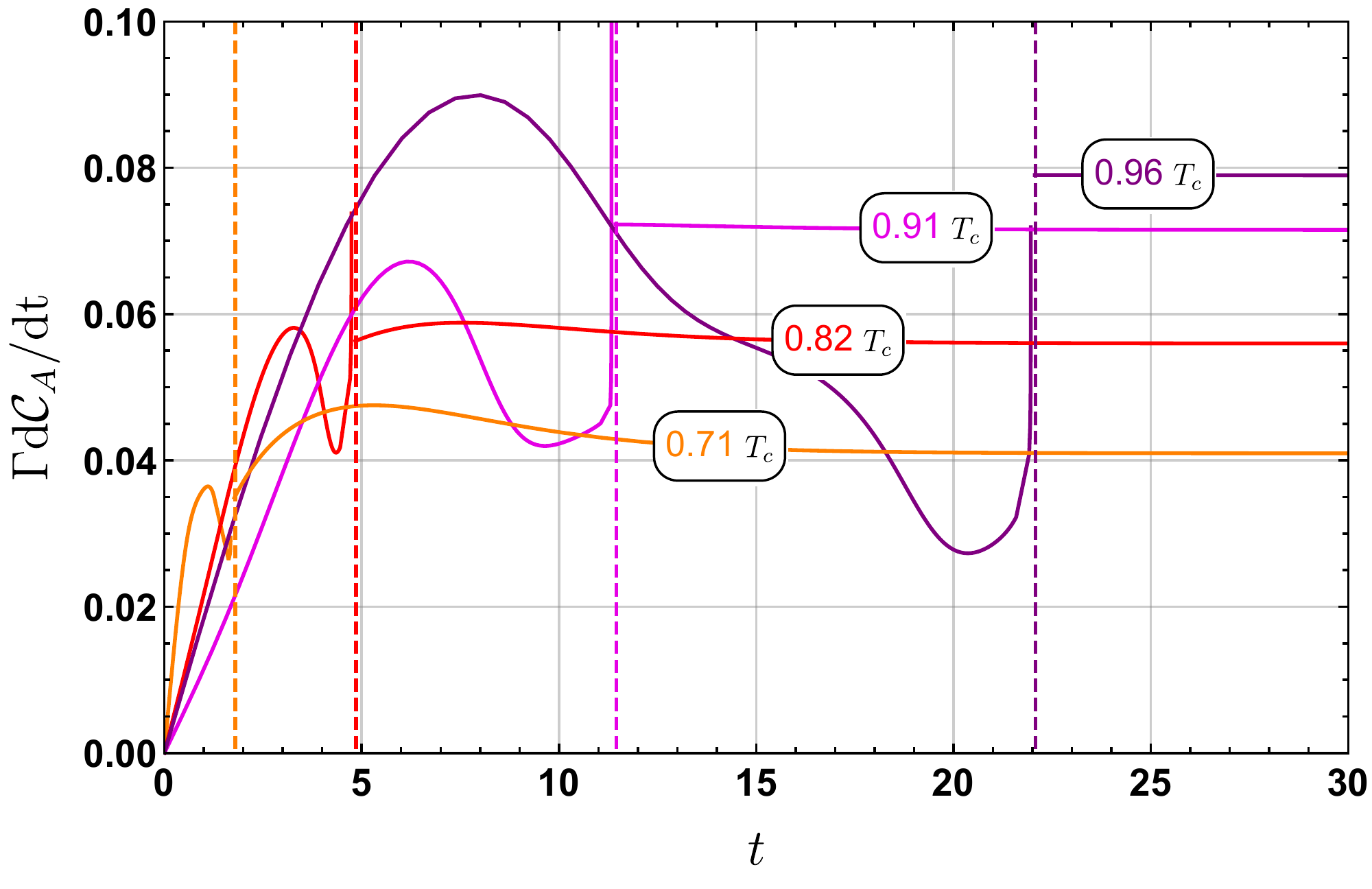}
\includegraphics[width=0.49\textwidth]{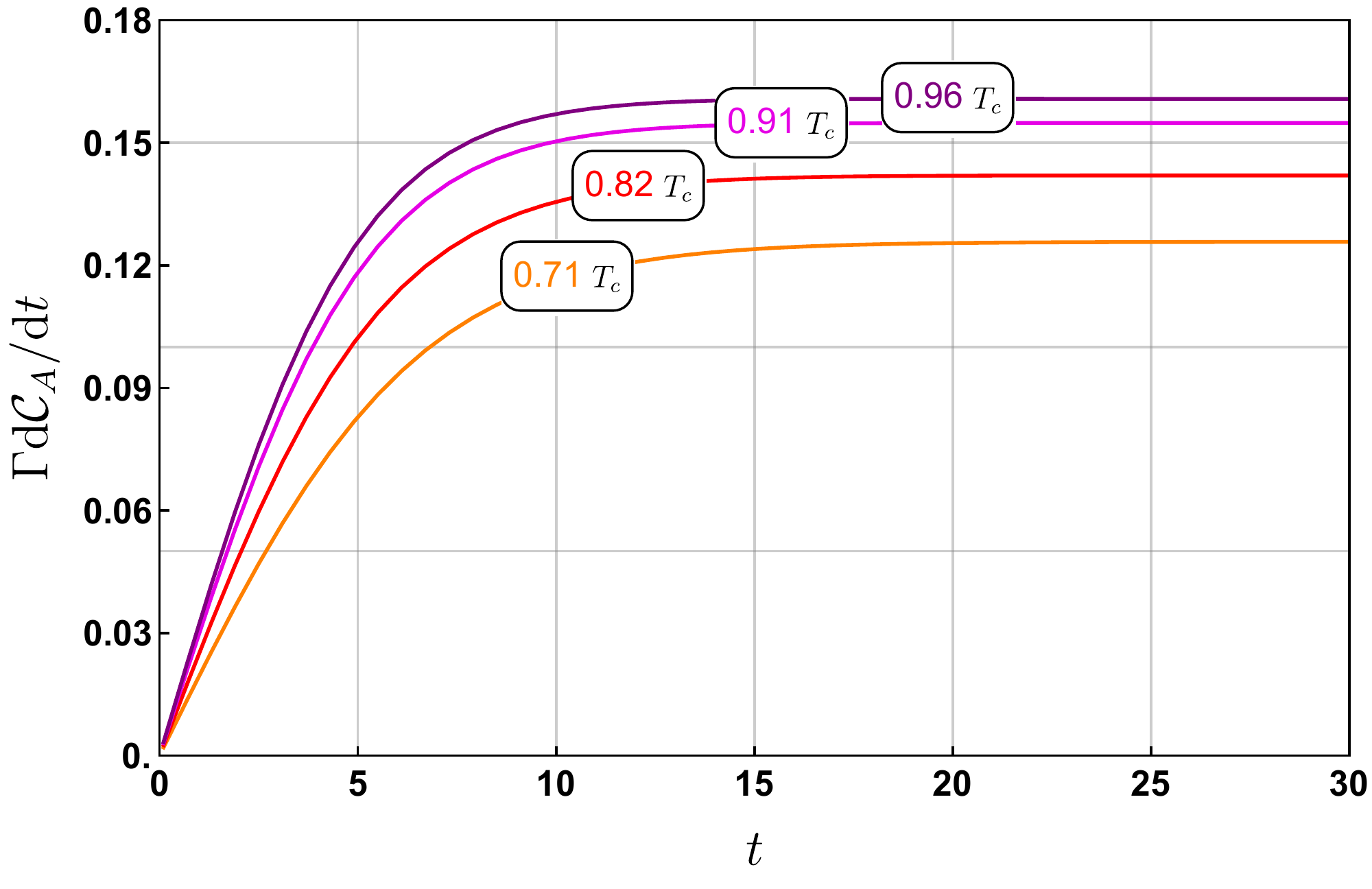}
\caption{The complexity growth rate of the superconducting phase for the negative critical time $t_c<0$ (left panel). The case for the normal phase at the same temperature is shown in the right panel. We choose $\Gamma= 16\pi^2 G_{N}/\Omega_{2}$ and $\mu=1$. }\label{Fig:grca-2}
\end{figure}

Another key situation is when $z_{m2}$ approaches the singularity. The complexity growth rate will diverge for finite $z_{m1}$ and infinite $z_{m2}$.
\begin{equation}
\lim_{z_{m2}\to \infty}\frac{\mathrm{d}\mathcal{C}_{A}}{\mathrm{d}t}=\text{finite terms}+\lim_{z_{m2}\to \infty}\left(-6+\frac{\alpha^2}{2}\right)h(z_{m2})\ln(z_{m2})\to \infty.
\end{equation}
where we have considered the Kasner behavior of our metric. Using the relation between the boundary $t$ and $z_{m2}$~\eqref{bdytime}, it can be seen that this divergence is transient and does not have any influence when doing the average over the thermal time scale as discussed in \cite{Carmi:2017jqz}.
\begin{equation}
  \left[\frac{\mathrm{d} \mathcal{C}_{A}}{\mathrm{d} t}\right]_{\gamma ; \text { avg }}=\frac{1}{\gamma \beta} \int_{t-\gamma \beta / 2}^{t+\gamma \beta / 2} \frac{\mathrm{d} \mathcal{C}_{A}}{\mathrm{d} t^{\prime}} \mathrm{d} t^{\prime},
\end{equation}
where $\gamma$ is some numerical factor of order one. The complexity is still finite even though its derivative is divergent in this transient short time. It should be pointed out that the complexity growth rate will have a violently aperiodic alternation due to the chaotic behavior of $\alpha$ at large $z_{m2}$ limit. Note that this situation does not change the late-time behavior. After this short period in which $z_{m2}$ is large and approaches the singularity, the structure of the WdW patch has a transition, and the behavior of complexity is the same as the first case after $-t_{c}$. The complexity growth rate in function of the time $t$ for different temperatures is shown in Fig.~\ref{Fig:grca-2}. Compared to the normal solution at the same temperature, the complexity growth rate of the superconducting phase shows a rich behavior due to the complicated dynamics inside the hairy black hole we discussed in the last section.

\begin{figure}[H]
\centering
\includegraphics[width=0.49\textwidth]{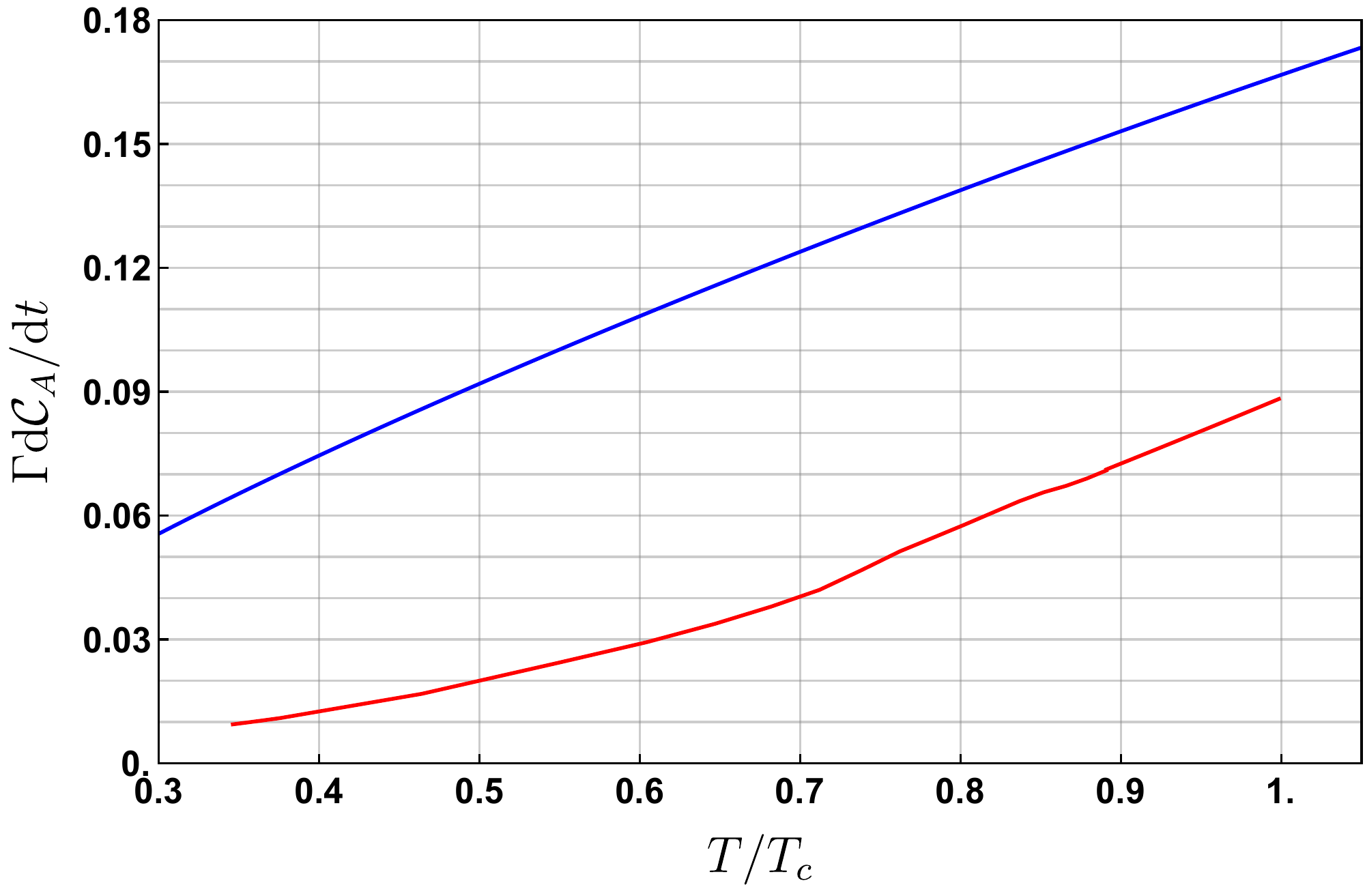}
\includegraphics[width=0.49\textwidth]{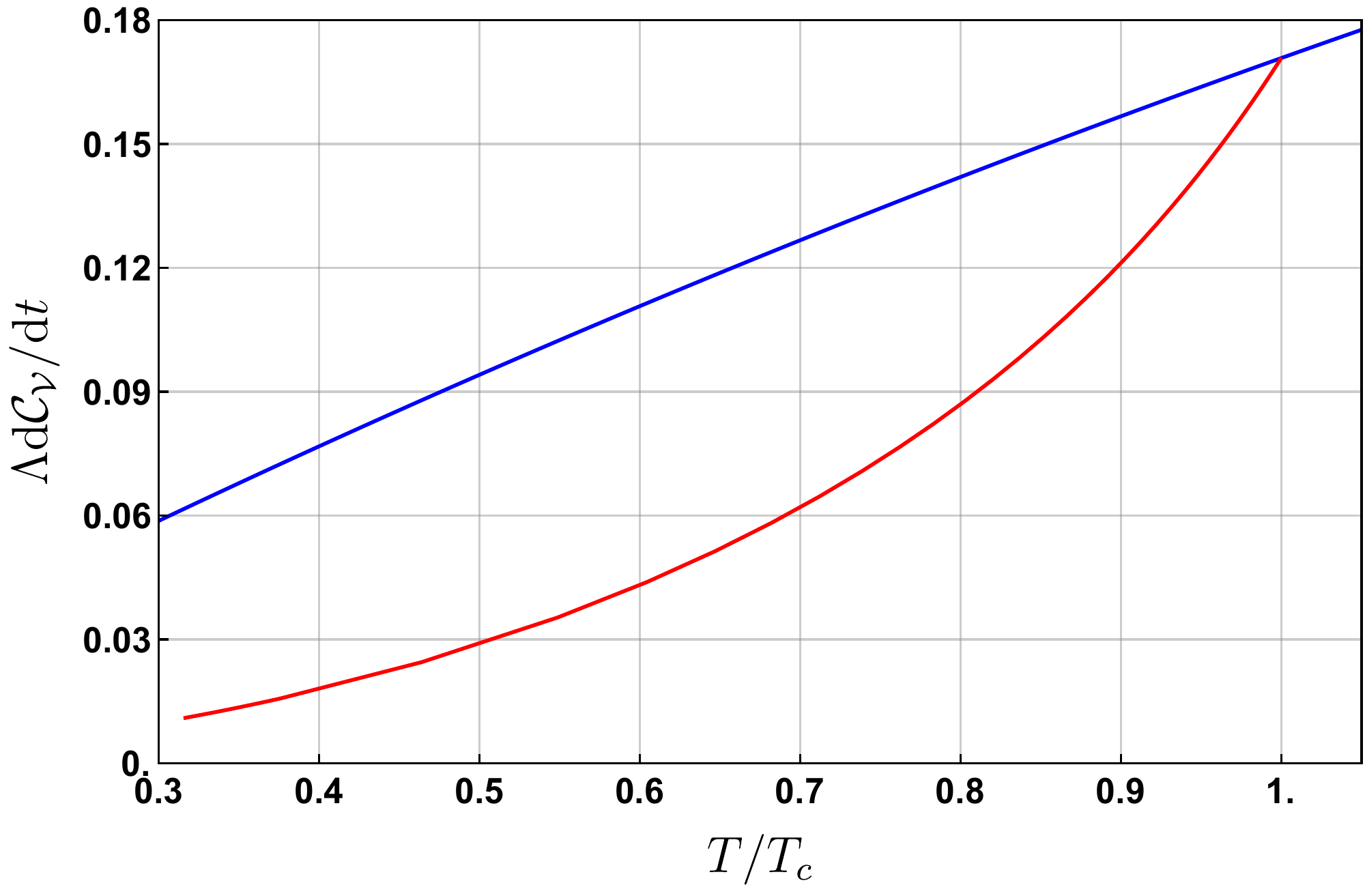}
\caption{The late-time complexity growth rate for the CA duality (left) and the CV duality (right) as a function of temperature. The blue lines correspond to the behavior for the normal phase, while the red lines to the superconducting phase. There is a sharp discontinuity at $T_c$ for the CA case. We choose $\Gamma= 16\pi^2 G_{N}/\Omega_{2}$, $\Lambda=10G_N L/\Omega_{2}$ and $\mu=1$.}\label{Fig:cav-late}
\end{figure}

Note that negative $t_{c}$ corresponds to higher temperatures of the superconducting phase (see Fig.~\ref{ctime-temp}). We then find an interesting feature at the critical temperature $T_c$. Although the phase transition from the normal phase to the superconducting phase is a smooth one, the complexity growth rate at late time has a sharp discontinuity at $T_c$ (see the left panel of Fig.~\ref{Fig:cav-late}). This discontinuity can be understood as follows.

For the superconducting phase, the late time complexity growth is given by
\begin{equation}\label{ca-late}
\frac{\mathrm{d}\mathcal{C}_{A}^{SC}}{\mathrm{d}t}=\frac{\Omega_{2}}{16\pi^2 G_{N}}[-4h(\infty)+(\rho A_{t})|_{\infty}]=\frac{\Omega_{2}}{16\pi^2 G_{N}}\left[\frac{4\pi T}{z_H^2}-(1+\frac{\alpha^2}{4})h(\infty)\right]\,,
\end{equation}
where we have used~\eqref{Th}. For the normal phase, the late time growth is
\begin{equation}\label{ca-late}
\frac{\mathrm{d}\mathcal{C}_{A}^{RN}}{\mathrm{d}t}=\frac{\Omega_{2}}{16\pi^2 G_{N}}\left[\frac{f'(z_I)}{z_I^2}-\frac{f'(z_H)}{z_H^2}\right]=\frac{\Omega_{2}}{16\pi^2 G_{N}}\left[\frac{f'(z_I)}{z_I^2}+\frac{4\pi T}{z_H^2}\right]\,.
\end{equation}
Therefore, the difference in the complexity growth rate between the normal and the superconducting phases at $T_c$ is
\begin{equation}
\Delta\frac{\mathrm{d}\mathcal{C}_{A}}{\mathrm{d}t}=\frac{\Omega_{2}}{16\pi^2 G_{N}} \left[\frac{f'(z_I)}{z_I^2}+(1+\frac{\alpha^2}{4})h(\infty)\right]\Big{|}_{T=T_c}\approx\frac{\Omega_{2}}{16\pi^2 G_{N}} \frac{f'(z_I)}{z_I^2}\Big{|}_{T=T_c}\,,
\end{equation}
where have used the fact that the last term is a small value, see Fig.~\ref{Fig:in-h}. More precisely, we have $z_I/z_H=1.1667$ at $T_{c}\approx0.0208\mu$, therefore
\begin{equation}
 \frac{f^{\prime}(z_I) }{z_I^{2}} =-\frac{3}{z_H z_I^2}\left(\frac{z_I}{z_H}\right)^2\left(1+\frac{z_{H}^{2} \mu^{2}}{4}\right)+\frac{\mu^2 z_{H}}{z_I^2} \left(\frac{z_I}{z_H}\right)^3= 0.07887\mu^2,
\end{equation}
This explains the discontinuity near $T_c$ in Fig.~\ref{Fig:cav-late}.

It is now manifest that the complexity growth rate computed by the CA conjecture is sensitive to the change of the inner structure of a black hole and thus could be a good probe to the black hole interior dynamics. As a comparison, the complexity growth rate by CV is checked in Appandix~\ref{CVconj} and is found to be continuous across the phase transition at $T_c$, see the right panel of Fig.~\ref{Fig:cav-late}.

\section{Conclusion and Discussion}\label{diss-sum}
In the present work, we have investigated the interior dynamics for a class of black holes with charged scalar hair from a holographic superconductor~\eqref{modelst4D} that can be consistently embedded in the M-theory. After proving the no-inner horizon theorem of the hairy black holes~\eqref{bkansatzst} in a generalized St\"{u}ckelberg form~\eqref{modelst}, we obtained both the exterior and interior configuration of the hairy back holes by numerically solving the full equations of motion~\eqref{eom-psi}-\eqref{eom-h}. As the temperature is decreased, the AdS RN solution becomes unstable to forming scalar hair, yielding a second-order superconducting phase transition at the critical temperature $T_c$. Just below $T_c$, the interior dynamics from the top-down theory displays several epochs. No matter how small the scalar is, there is a strong non-linear dynamics near the would-be inner horizon of the RN black hole, characterized by the collapse of the ER bridge and the Josephson oscillations of the scalar condensate. In the far interior, one enters the regime with a never-ending alternation of Kasner epochs toward the spacelike singularity $z\rightarrow \infty$.

We have found self-consistent analytic approximations for the alternation of Kasner epochs guided by numerical exploration. Depending on the value of $\alpha$ defined in~\eqref{kasner}, we are able to predict the new value of $\alpha$ for the next Kasner epoch. The analytic solutions agree with full numerical one quite well, see Fig.~\ref{Fig:tran-fit} for the Kasner transition ($|\alpha|>2$) and Fig.~\ref{Fig:inver-fit} for the Kasner inversion ($|\alpha|<2$). The transformation rule~\eqref{kas-law} of the alternation of Kasner epochs yields that there would be generically a never-ending chaotic sequence of Kasner epochs towards the singularity. As shown in Fig.~\ref{Fig:chaos}, this transformation rule~\eqref{kas-law} corresponds to an underlying pattern of $\alpha$ that is highly sensitive to initial conditions. We have emphasized that the presence of an exponential form of couplings ($\sinh \psi, \cosh\psi$) in the top-down theory~\eqref{modelst4D} plays the key role in triggering such alternation of Kasner epochs. It would be extremely interesting to further understand this behavior with either a more refined numerical approach or more powerful mathematical tools. 

We have then considered the CA complexity as a probe to the internal structure of the hairy black hole. This should be the first work to investigate the complexity growth rate using CA in holographic superconductors. As shown in Fig.~\ref{penrose-ca}, there are two classes of WdW patch depending on the critical time $t_{c}$ at which the null past joint of the WdW patch locates exactly at the past singularity. We have found a sharp discontinuity of the complexity growth rate at $T_c$ where there is a smooth phase transition from the normal phase to the superconducting phase. This discontinuity was shown to be associated with the dramatic change of the inner structure. In contrast, the complexity growth rate by CV was found to be smooth across the phase transition at $T_c$ (see Fig.~\ref{Fig:cav-late}). Therefore, the complexity growth rate by CA conjecture could be a good probe to the black hole interior dynamics. 

We have focused on static charged black holes with planer horizon topology, it would be interesting to generalize our study to other horizon topologies and even to stationary cases.  We shall leave the precise relation between the complexity growth rate and the black hole interior dynamics for future study. While the CA complexity was found to violate the Lloyd's bound in some setup~\cite{HosseiniMansoori:2018gdu,Ageev:2019fxn}, it is desirable to check if the Lloyd's bound will be violated in the top-down model~\eqref{modelst4D} that can be embedded in a UV complete theory. It seems that the chaotic alternation of Kasner epochs we found in the present work is a different new type, thus a deep understanding is required. Moreover, it is interesting to understand the physical consequences of these Kasner epochs in the dual field theory point of view (see, \emph{e.g.}~\cite{Jafferis:2020ora,Leutheusser:2021qhd}). 
 In the present work, we only considered the case in which the scalar hair develops spontaneously, the main features should apply to the hairy black holes with explicit sources. So far, we have been limited to the four-dimensional theory~\eqref{modelst4D}, it is interesting to consider other top-down theories, in particular, the five-dimensional one~\eqref{modelst5D} from IIB supergravity. We hope to report our results in these directions in the near future.

\section*{Acknowledgement}
We thank M. Henneaux for his useful comments and suggestions. This work was partially supported by the National Natural Science Foundation of China Grants No.12122513, No.12075298, No.12005155, No.11991052 and No.12047503, and by the Key Research Program of the Chinese Academy of Sciences (CAS) Grant NO. XDPB15 and the CAS Project for Young Scientists in Basic Research YSBR-006. Yu-Sen An is supported by the Shanghai Post-doctoral Excellence Program No.2021006 and funded by China Postdoctoral Science Foundation No.2022M710801.

\appendix
\section{Configuration inside Hairy Black Holes}\label{alpha-diagram}
\begin{figure}[H]
\centering
\includegraphics[width=0.46\textwidth]{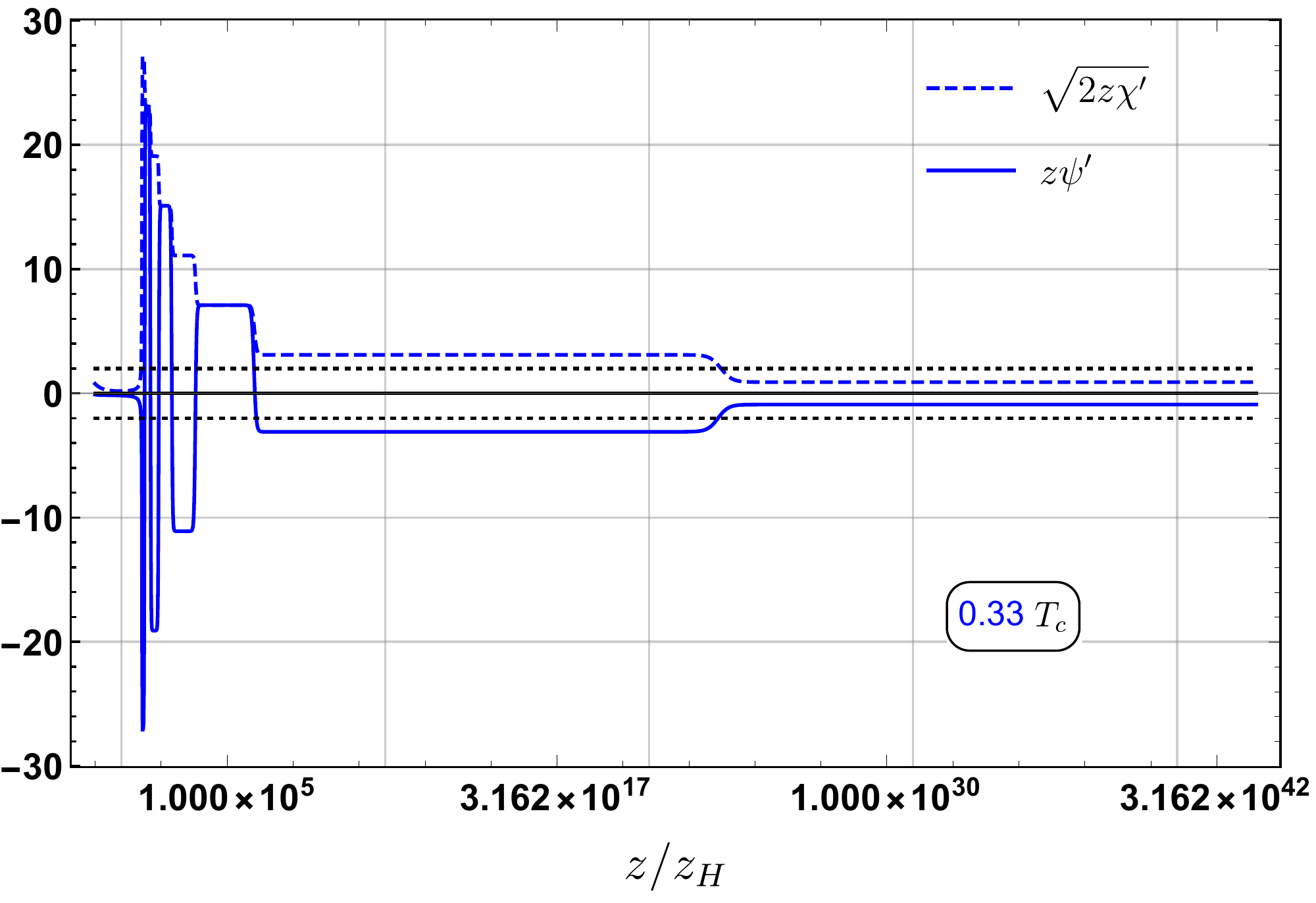}
\includegraphics[width=0.48\textwidth]{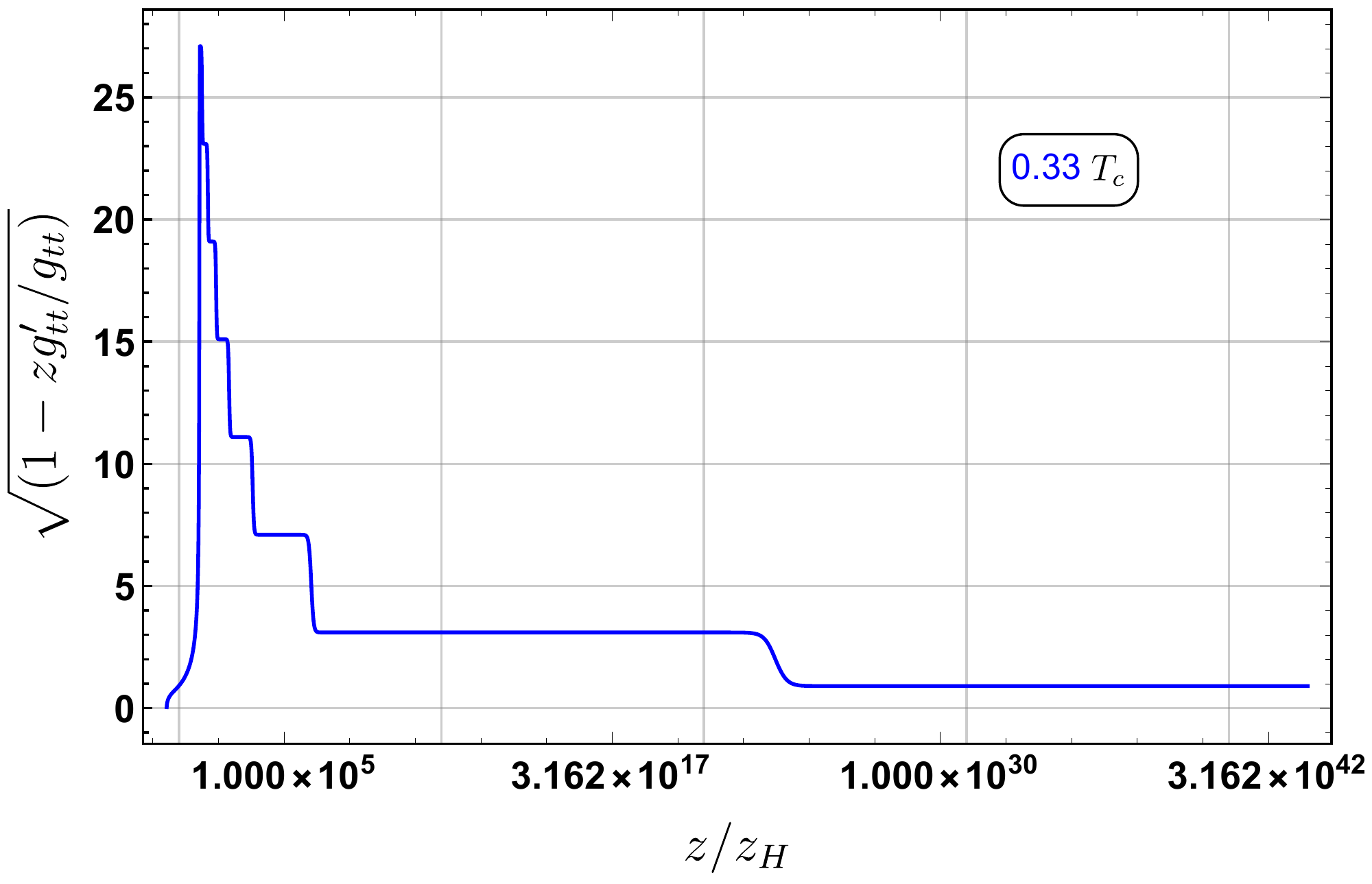}\\
\includegraphics[width=0.46\textwidth]{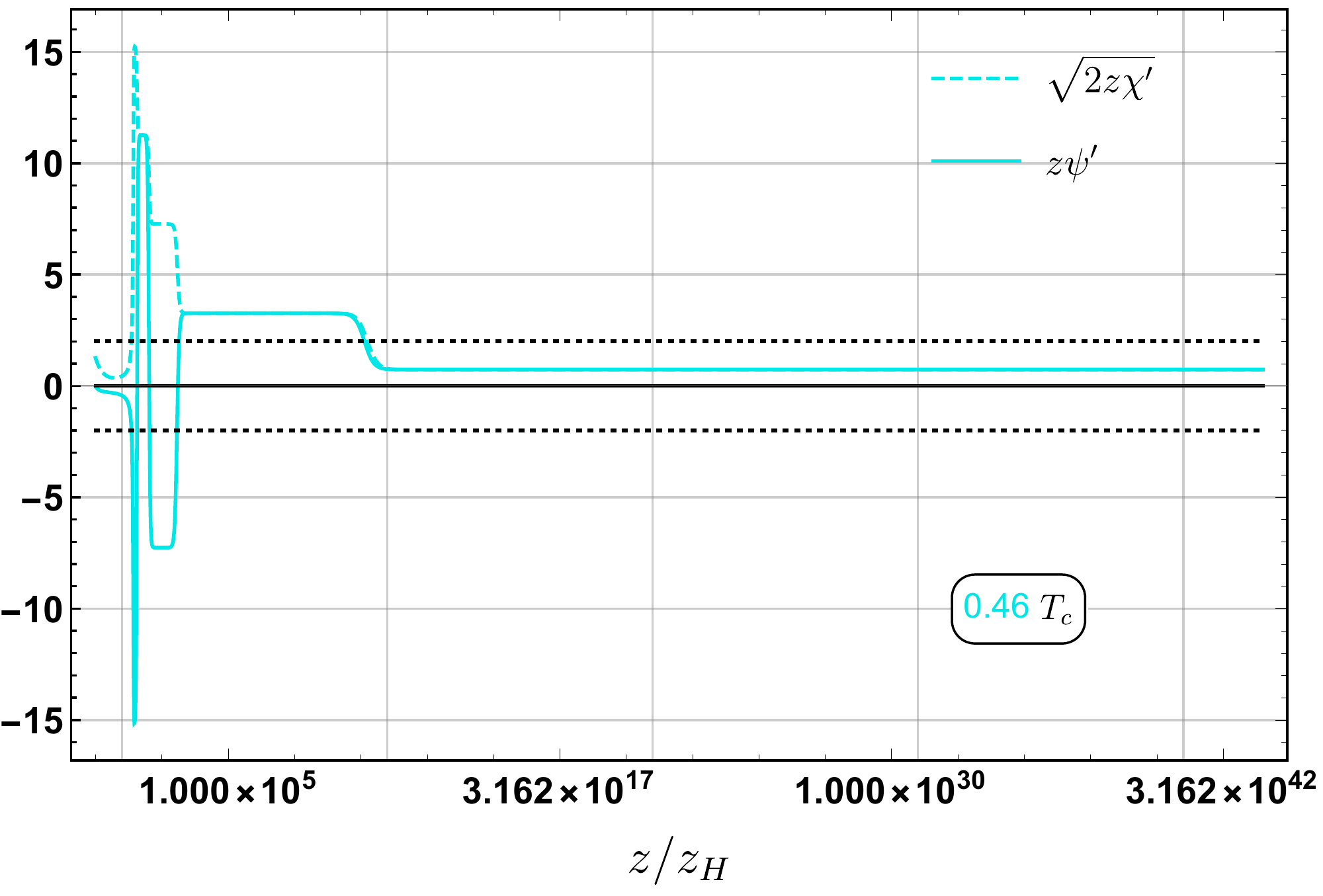}
\includegraphics[width=0.48\textwidth]{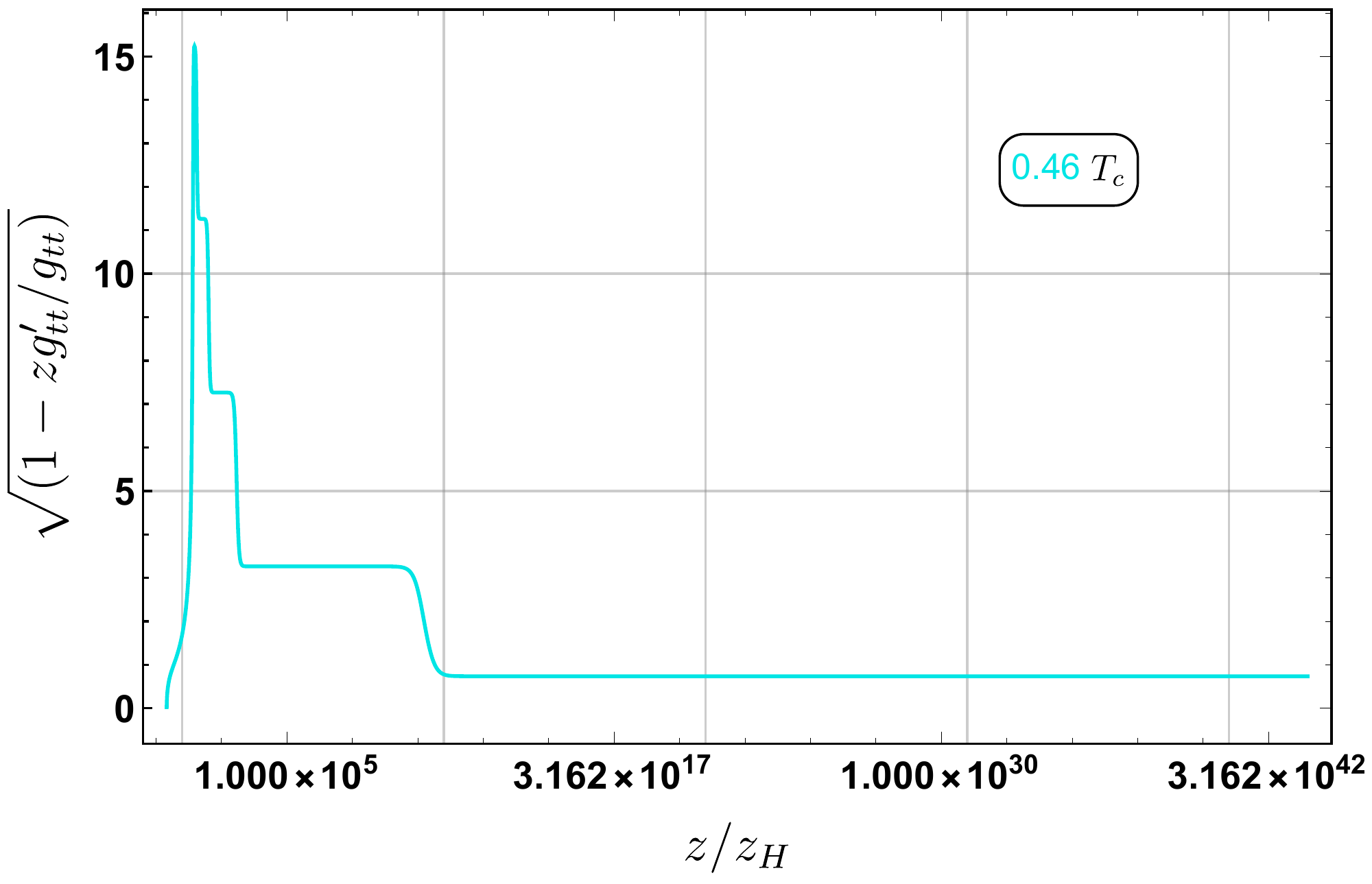}\\
\includegraphics[width=0.46\textwidth]{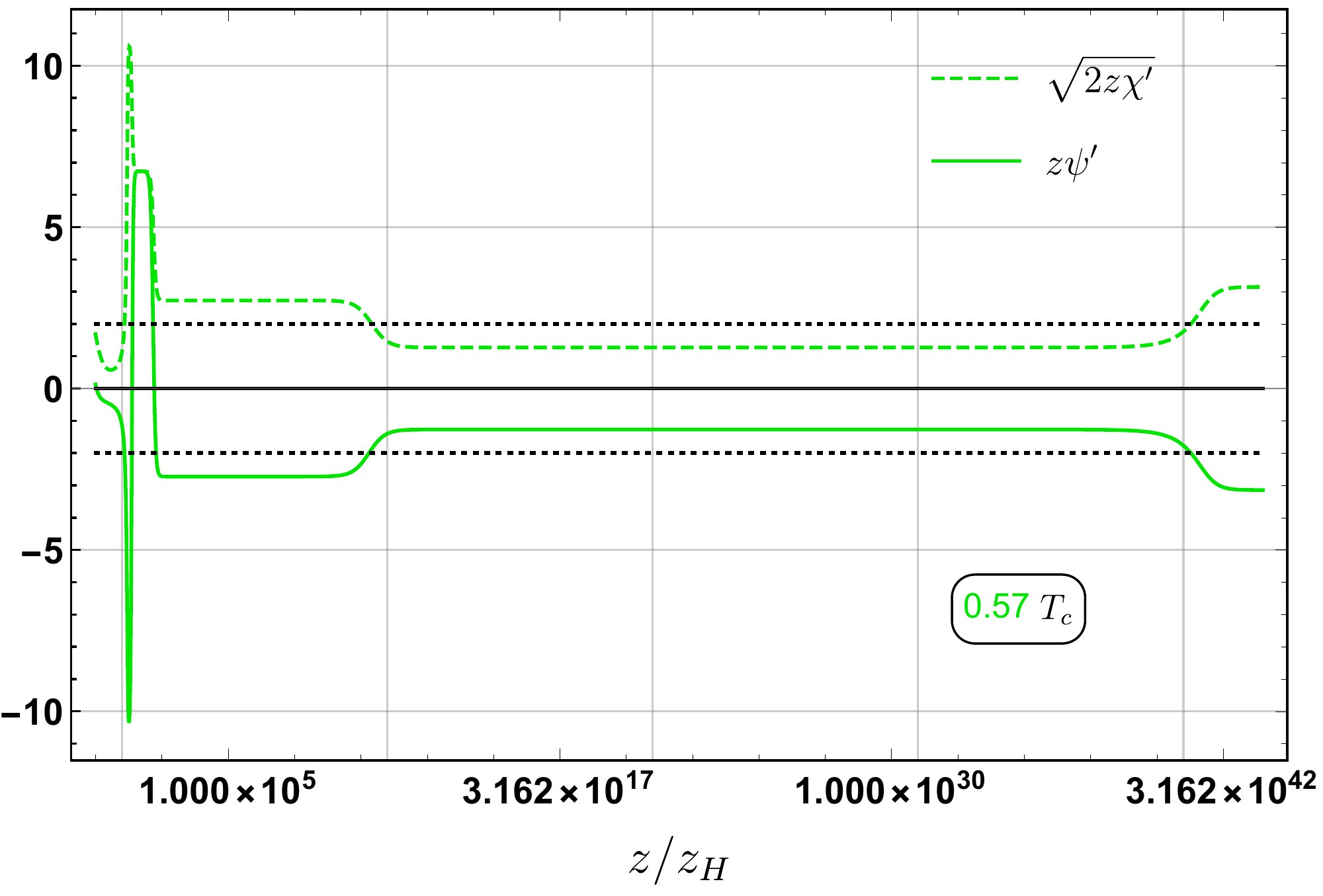}
\includegraphics[width=0.48\textwidth]{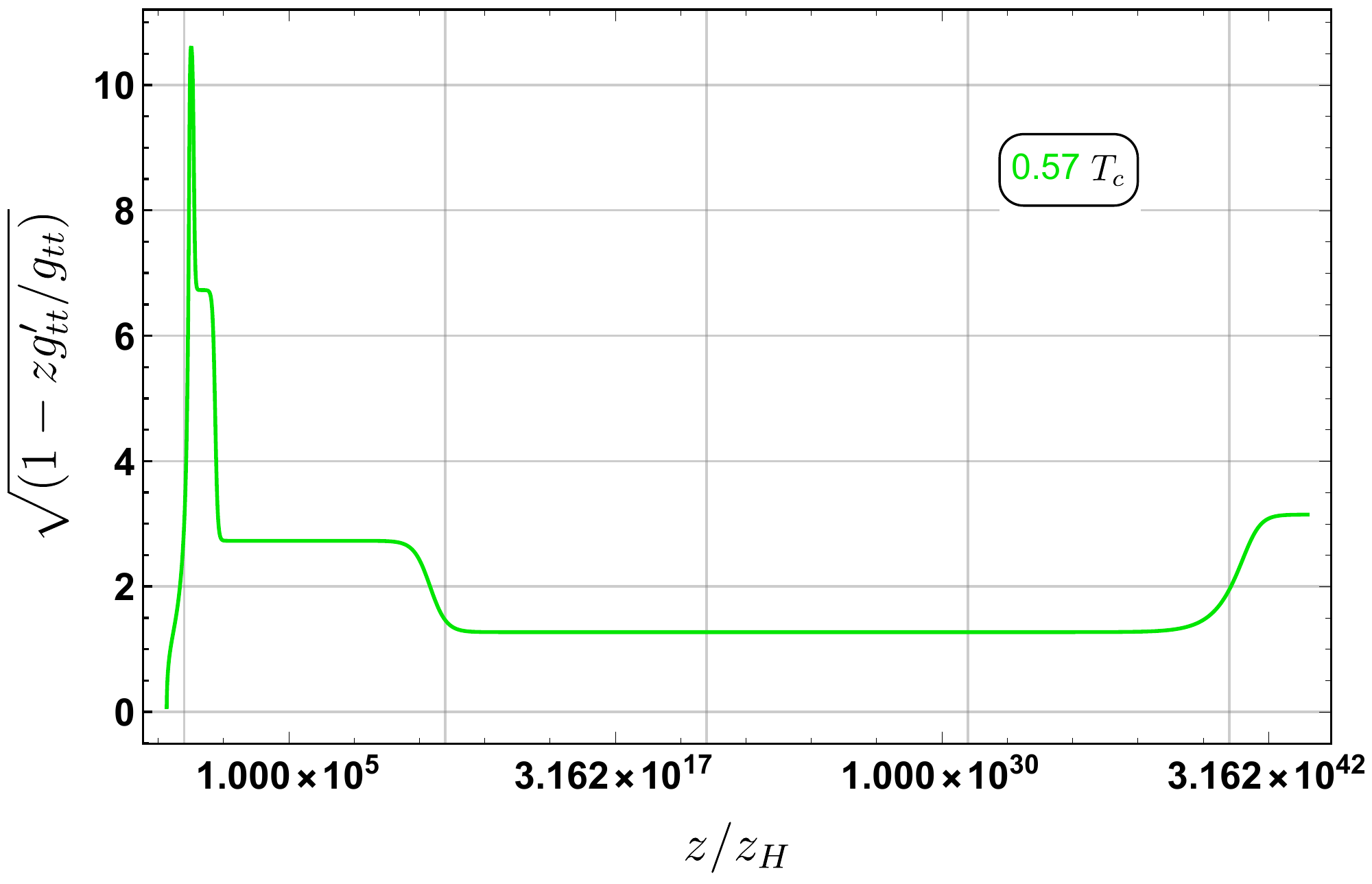}\\
\includegraphics[width=0.46\textwidth]{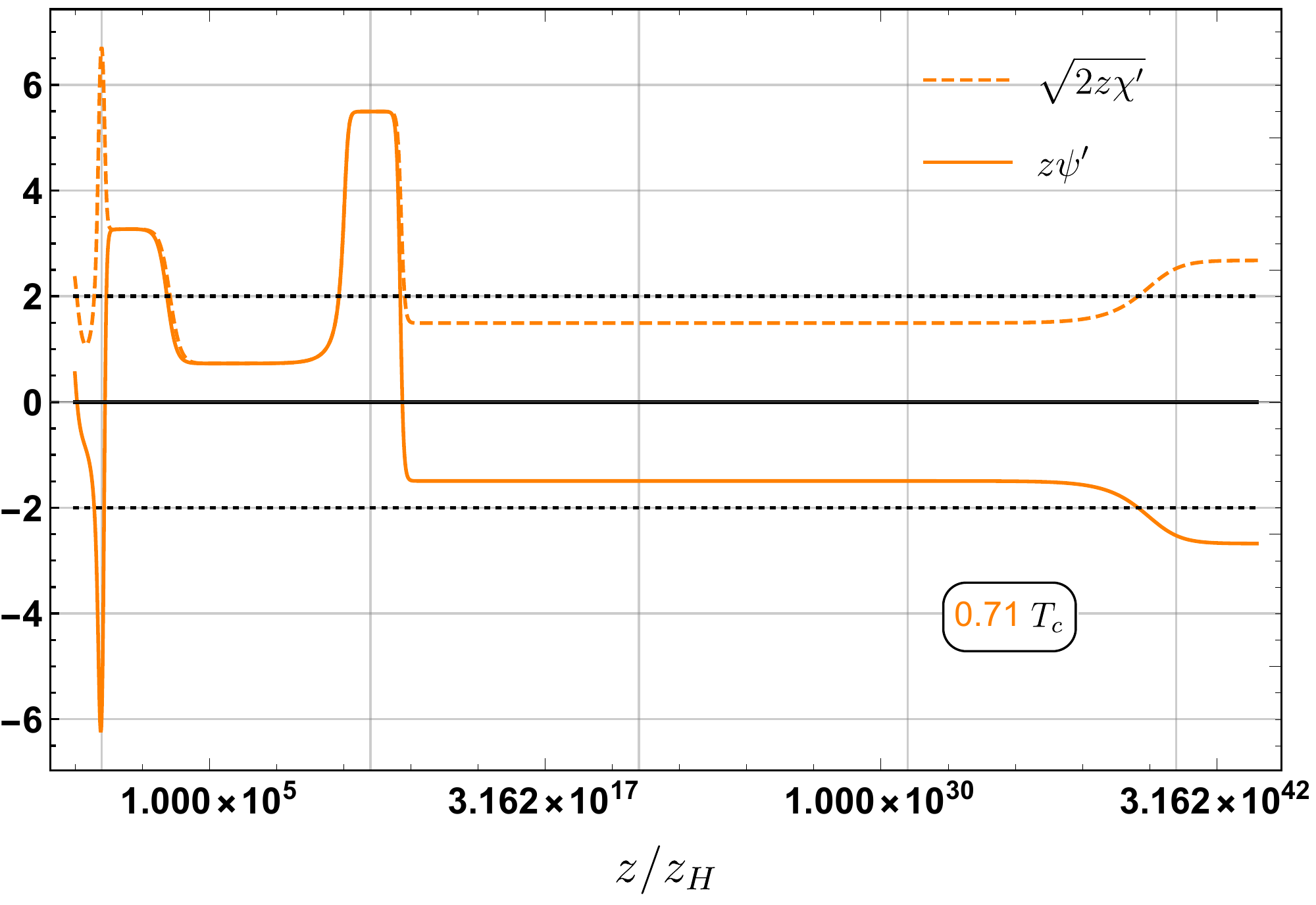}
\includegraphics[width=0.48\textwidth]{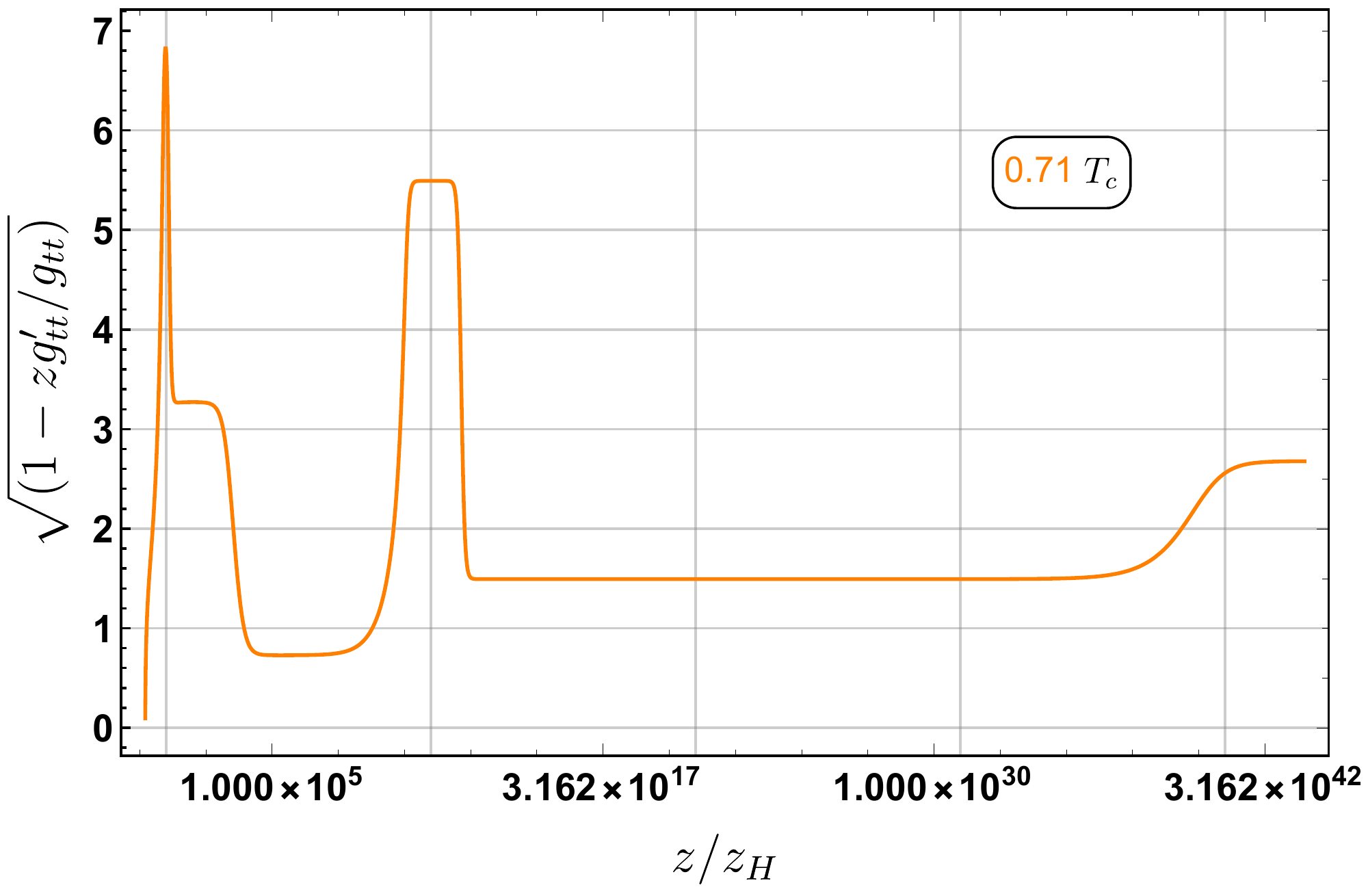}\\
\caption{The configuration of $z\psi'$ and $\sqrt{2z\chi'}$ (left) and $\sqrt{1-zg'_{tt}/g_{tt}}$ (right) inside the superconducting black hole for different temperatures. 
The blue, cyan, green and orange curves correspond to different temperatures at $T/T_{c}=0.33, 0.46, 0.57$ and $0.71$, respectively.
}\label{Fig:alpha-low}
\end{figure}

\begin{figure}[H]
\centering
\includegraphics[width=0.46\textwidth]{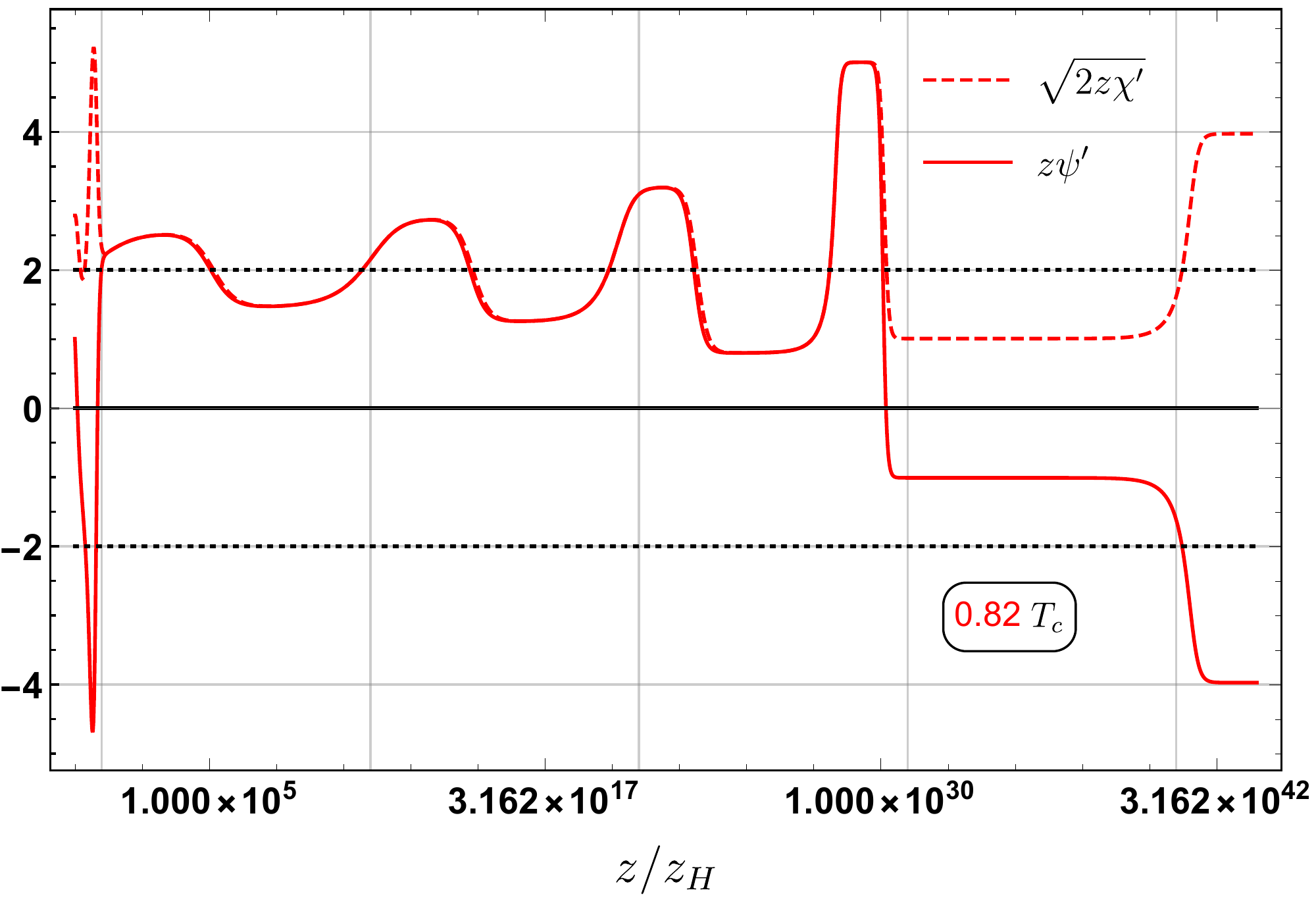}
\includegraphics[width=0.48\textwidth]{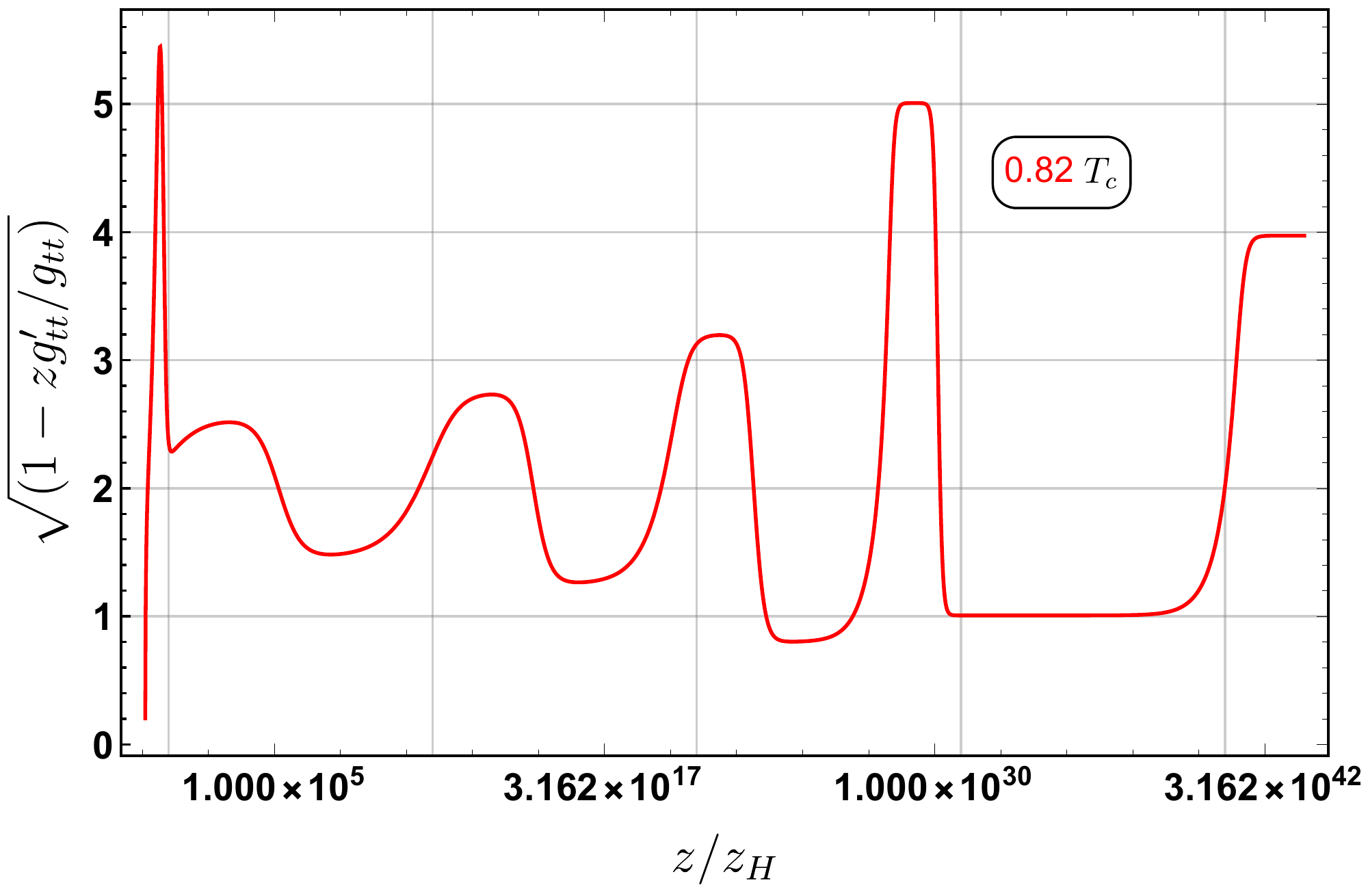}\\
\includegraphics[width=0.46\textwidth]{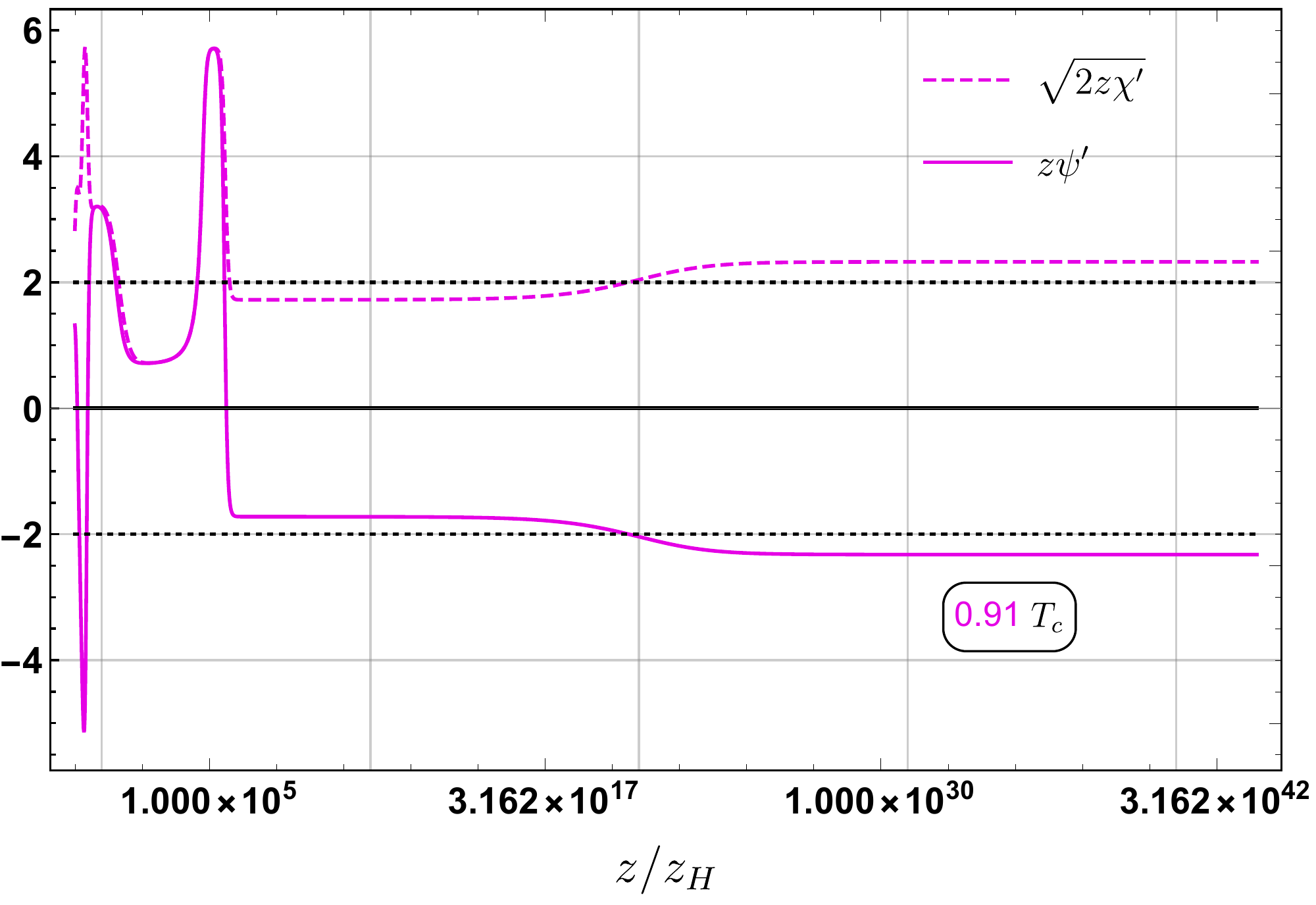}
\includegraphics[width=0.48\textwidth]{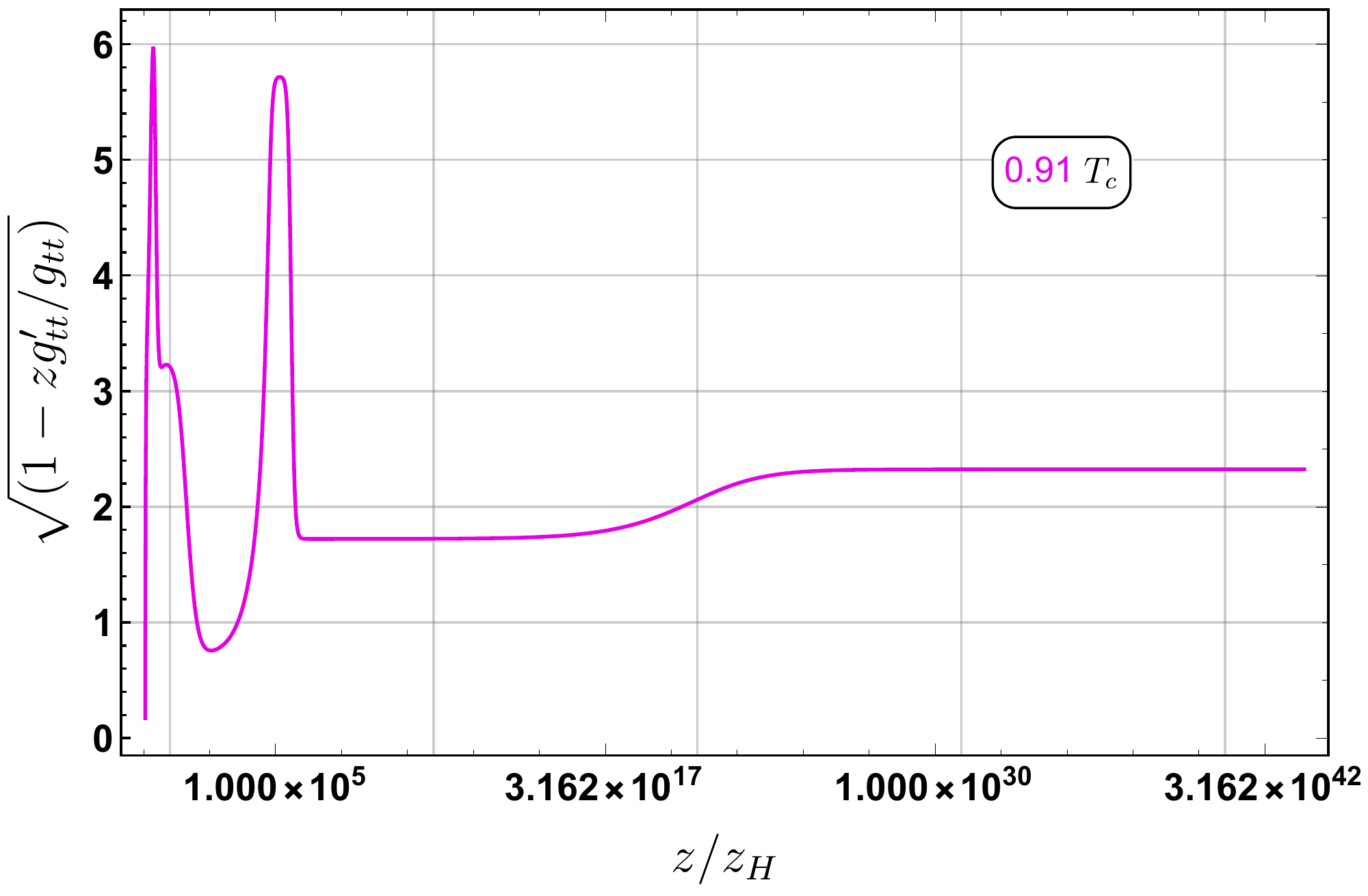}\\
\includegraphics[width=0.46\textwidth]{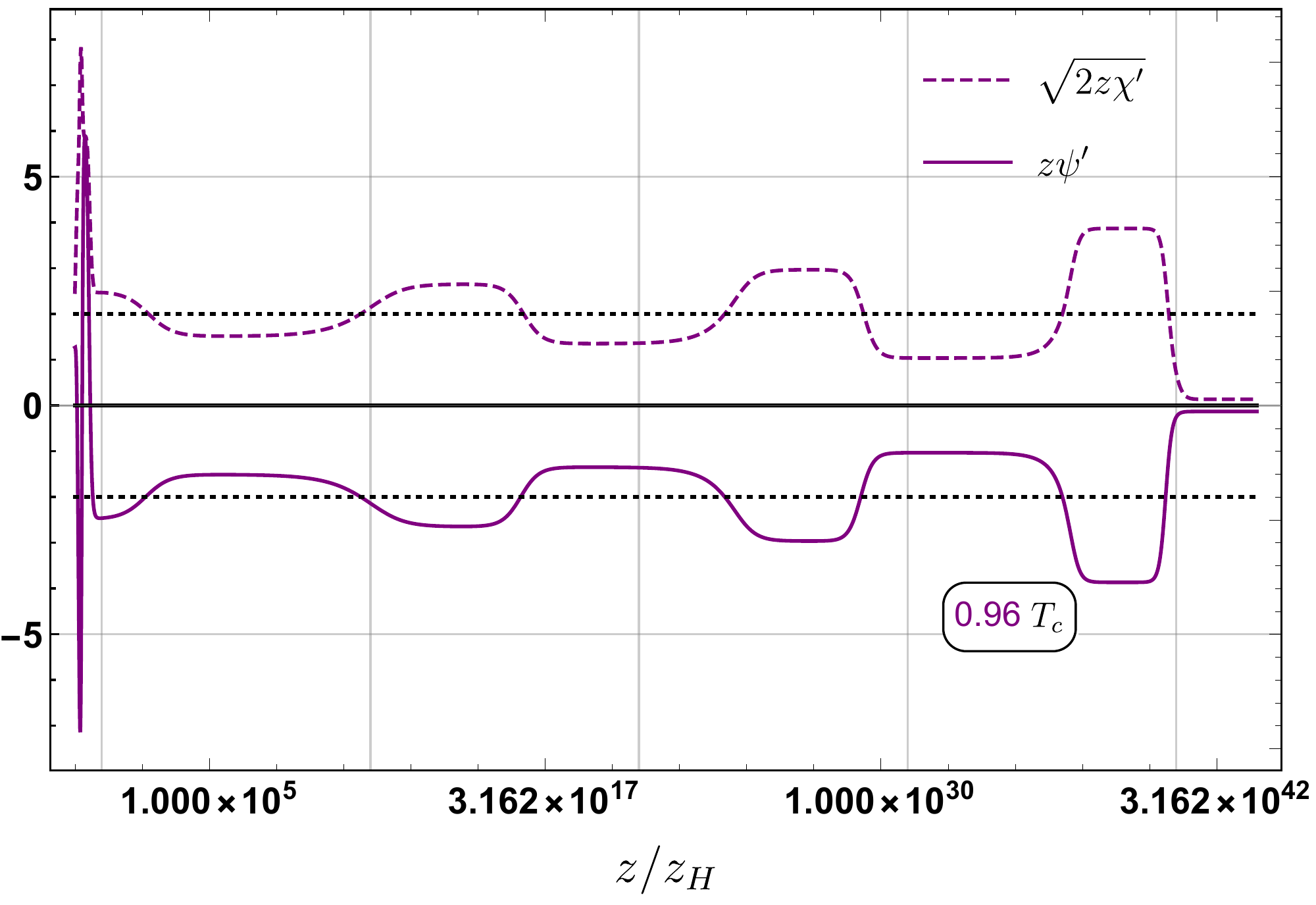}
\includegraphics[width=0.48\textwidth]{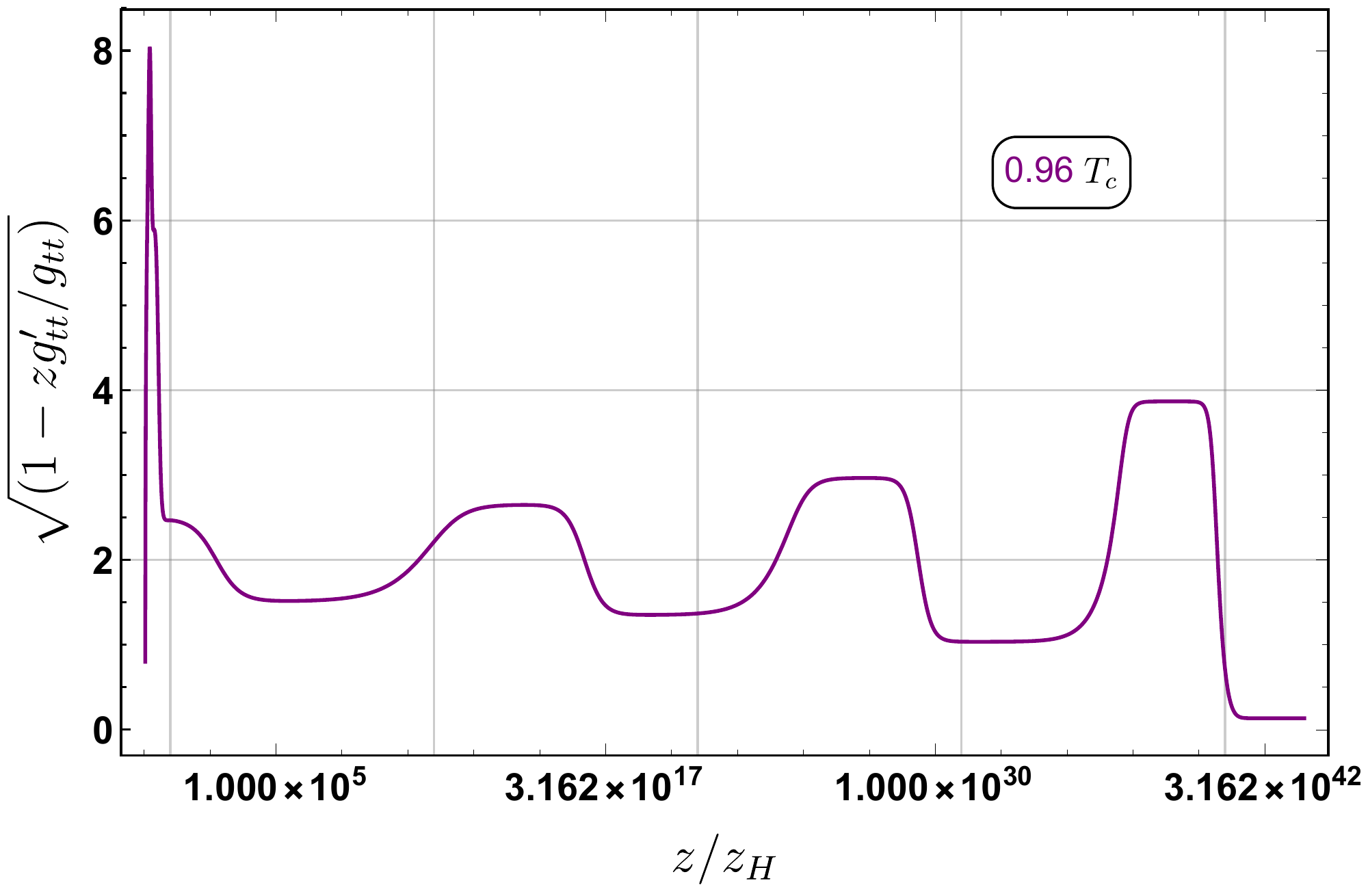}\\
\includegraphics[width=0.46\textwidth]{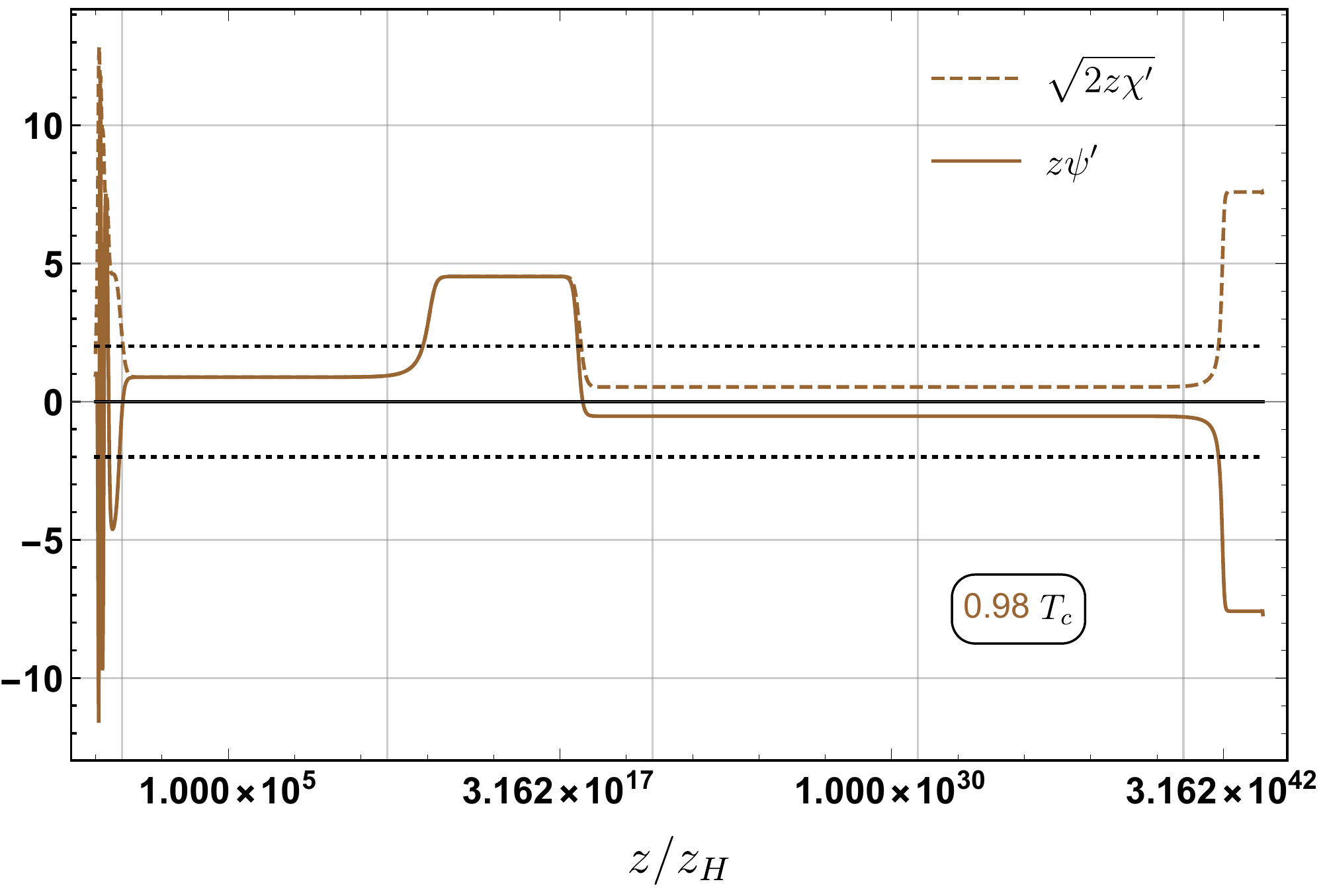}
\includegraphics[width=0.48\textwidth]{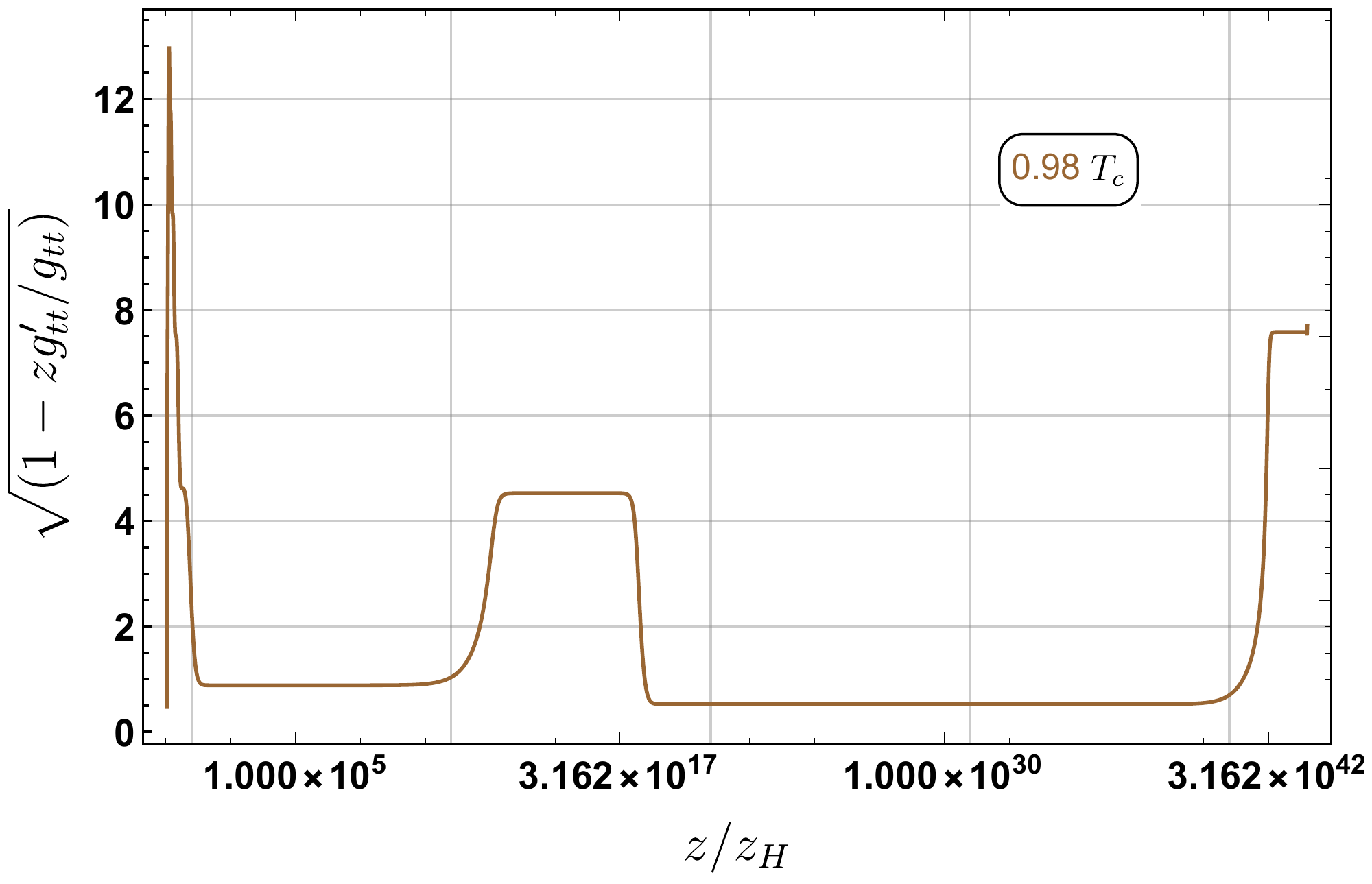}\\
\caption{The configuration of $z\psi'$ and $\sqrt{2z\chi'}$ (left) and $\sqrt{1-zg'_{tt}/g_{tt}}$ (right) inside the superconducting black hole for different temperatures. 
The red, magenta,  purple and brown curves correspond to different temperatures at $T/T_{c}= 0.82, 0.91, 0.96$ and $0.98$, respectively.
}\label{Fig:alpha-high}
\end{figure}

In this section, we present more examples for the interior dynamics inside the hairy black holes by numerically solving the full equations of motion~\eqref{eom-psi}-\eqref{eom-h}.  For a better illustration, we consider $z\psi'$, $\sqrt{2z\chi'}$ and $\sqrt{1-z g'_{tt}/g_{tt}}$ at different temperatures below $T_c$. All of them approach a constant in a Kasner epoch described by~\eqref{kasner}. There are many alternations of plateaus in  Figures.~\ref{Fig:alpha-low} and~\ref{Fig:alpha-high}, characterizing the alternation of Kasner epochs at large $z$.

\renewcommand{\theequation}{A\arabic{equation}}

\section{Relation to Billiard Approach}\label{billard}
\renewcommand{\theequation}{B\arabic{equation}}

In a recent paper~\cite{Henneaux:2022ijt}, the author analyzed the metric behavior near the singularity using the Hamiltonian phase space formalism which offers a clear way to present the transition law of Kanser exponents in different Kasner epochs. It was found that the singularity of a black hole with a massive charged scalar hair or vector hair is of the 4D BKL type transformation law $p_{t} \to \frac{|p_{t}|}{1-2|p_{t}|}$. This approach is also called the ``cosmological billiard approach''~\cite{Damour:2002et,Henneaux:2007ej}. As our Kasner transition result is different from the BKL type, it is necessary to understand where our model is different from the setup of the paper~\cite{Henneaux:2022ijt}.

The author of~\cite{Henneaux:2022ijt} considered the theory with a free charged scalar $\phi$, while our case~\eqref{modelst4D} has non-linear couplings, including $\sinh^{2}\psi A_{\mu}A^{\mu}$ and the potential $\cosh^{2}\frac{\psi}{2}(7-\cosh\psi)$. We now compare our case with the free charged scalar of~\cite{Henneaux:2022ijt}.
\paragraph{For the minimal coupling} The effective scalar potential (rescaled by $\sqrt{g_0}$) for the free charged scalar reads~\cite{Henneaux:2022ijt}
\begin{equation}
V_{\phi}=q^{2}A_{t}^{2}\phi^{\dagger}\phi g^{tt} g_0+m^{2}\phi^{\dagger}\phi g_0\,,
\end{equation}
where $g_0$ is the determinant of the spatial metric. For a Kasner epoch, the geometry in terms of the proper time is given by
\begin{equation}
\begin{split}
\mathrm{d}s^{2}=-\mathrm{d}\tau^{2}+c_{t}\tau^{2p_{t}}\mathrm{d}t^{2}+c_{s}\tau^{2p_{s}}\mathrm{d}\Sigma_{2,k}^{2}, \quad \phi=-\frac{2\alpha}{\frac{\alpha^{2}}{4}+3} \ln \tau\,,\\
p_{t}=\frac{\frac{\alpha^{2}}{4}-1}{\frac{\alpha^{2}}{4}+3},\qquad p_{s}=\frac{2}{\frac{\alpha^{2}}{4}+3}\,,
\end{split}
\end{equation}
where the future singularity is at $\tau=0$ and is approached from negative values of $\tau$.
The determinant of the spatial metric has $g_0\sim \tau^{2(p_t+2 p_s)}= \tau^{2}$. In the asymptotic limit, as $A_{t}$ is approximately a constant, we see that the potential term is of the form
\begin{equation}
    V_{\phi}\sim v_0\, q^2 (\ln\tau)^{2}\tau^{\frac{32}{12+\alpha^{2}}}+m^2\tau^{2}\ln(\tau)^{2}\,,
\end{equation}
with $v_0$ a positive constant. It vanishes for $\tau \to 0$ no matter what $\alpha$ is. As the potential term vanishes asymptotically, it is irrelevant in the asymptotic analysis of the metric behavior and does not affect the dynamics. So, the wall in the ``billiard approach'' method from the scalar potential disappears as close to the singularity.

\paragraph{For the top-down model} The effective potential of the scalar in the top-down model~\eqref{modelst4D} is given by
\begin{equation}\label{potential-vpsi}
\tilde{V}_{\psi}=\frac{1}{2}A_{t}^{2}\sinh^{2}\psi g^{tt} g_0-\cosh^{2}\frac{\psi}{2}(7-\cosh\psi) g_0\,,
\end{equation}
with $g_0$ the determinant of the spatial metric. For a given Kasner epoch, the geometry reads~\eqref{Kasner}
\begin{equation}
\mathrm{d}s^{2}=-\mathrm{d}\tau^{2}+c_{t}\tau^{2p_{t}}\mathrm{d}t^{2}+c_{s}\tau^{2p_{s}}\mathrm{d}\Sigma_{2,0}^{2},\quad \psi\sim-\frac{2\alpha}{\frac{\alpha^{2}}{4}+3} \ln \tau\,,
\end{equation}
for which the determinant of the spatial metric is $g_0 \sim\tau^{2}$. If the potential term vanishes asymptotically, it is irrelevant in the asymptotic analysis of the interior behavior and does not affect the dynamics. For the first term, from~\eqref{potential-vpsi} one can obtain
\begin{equation}\label{kasf-law}
A_{t}^{2}\sinh^{2}\psi  g^{tt} g_0 \sim \tau^{\frac{16(2-|\alpha|)}{12+\alpha^{2}}},
\end{equation}
where we have used that $A_{t}$ is a constant. So in order for this term to vanish as $\tau \to 0$, we must have $|\alpha|<2$. Moreover, the vanishing of second term, $\cosh^{2}\frac{\psi}{2}(7-\cosh\psi)g_0$, as $\tau \to 0$ gives the constraint $|\alpha|<2$ or $|\alpha|>6$. Taking the intersection among them, we find that when $|\alpha|>2$, the effective potential~\eqref{potential-vpsi} will diverge in the $\tau \to 0$ limit. 
Therefore, the effective potential can no longer be irrelevant in the asymptotic limit for any $|\alpha|>2$. This is where the top-down model~\eqref{modelst4D} is different from the model in~\cite{Henneaux:2022ijt}. Therefore, we find the billiard table structure is different from the one in~\cite{Henneaux:2022ijt}, this may be the reason why our new Kasner transition law does not satisfy the rule $p_{t} \to \frac{|p_{t}|}{1-2|p_{t}|}$. We shall leave a thorough study of the transition rules of the top-down model in terms of the billiard approach in future work.

\section{Complexity in CV Conjecture}\label{CVconj} 
\begin{figure}[H]
\centering
\includegraphics[width=0.6\textwidth]{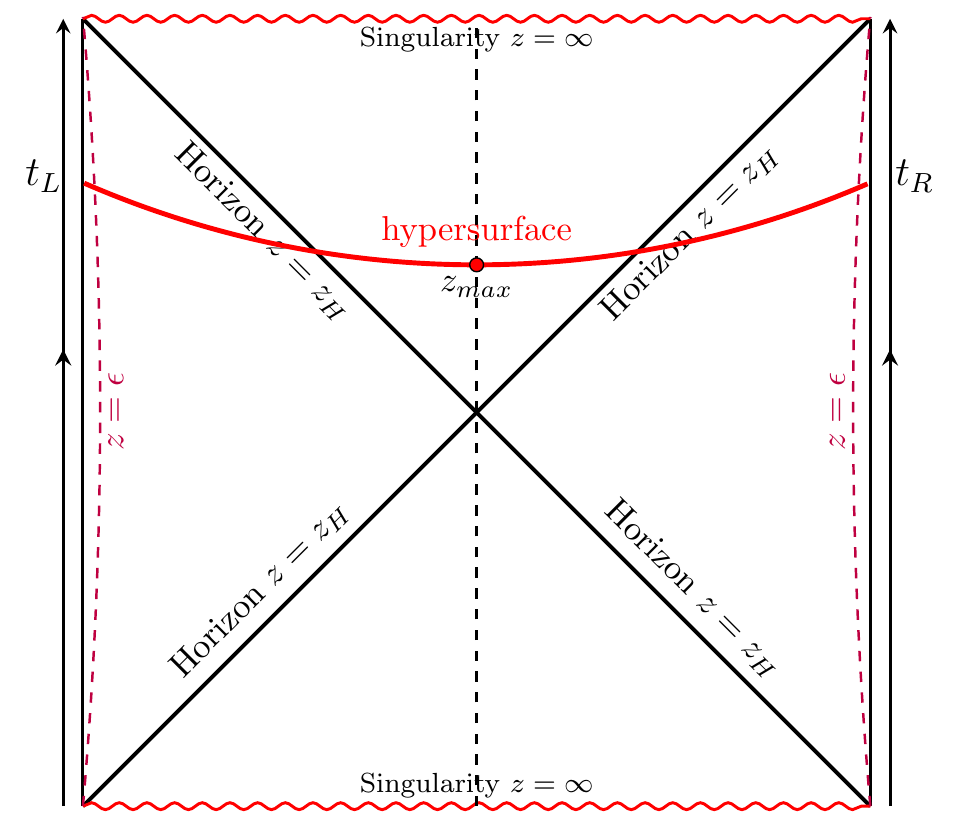}
\caption{A representation of the maximal wormhole connecting the symmetric boundary time $t_L$ and $t_R$. The bridge reaches the maximum distance $z_{max}$ inside the event horizon.}\label{Fig:Cauchy-hypersurface}
\end{figure}

\renewcommand{\theequation}{C\arabic{equation}}
In this appendix, we compute the complexity growth rate using the CV conjecture. The CV duality is given by the following formula:
\begin{equation}
\mathcal{C_{\mathcal{V}}}=\max{\frac{\mathcal{V}}{G_{N}L}}\,,
\end{equation}
where $\mathcal{V}$ is all possible codimension-one surface connecting $t_L$ and $t_R$ at two AdS boundaries, see Fig.~\ref{Fig:Cauchy-hypersurface}. As we will show, the computation of CV complexity is much simpler than the CA one in the main text, as it does not need to know the structure of the interior at large $z$. 
The computation of the extremal volume of the geometry~\eqref{nullcord} manifestly is a variational problem. Due to the planar symmetry, we parameterize the extreme slice by $\{z=z(\lambda), v=v(\lambda)\}$. Then, the maximal volume is obtained by extremizing
\begin{equation}\label{para-volume}
\mathcal{V}=\Omega_{2} \int \mathrm{d}\lambda \frac{1}{z^{3}}\sqrt{-f\mathrm{e}^{-\chi} \dot{v}^{2}-2 \mathrm{e}^{-\chi/2}\dot{v}\dot{z}}=\Omega_{2} \int \mathrm{d}\lambda \mathcal{L}(v,\dot{v})\,,
\end{equation}
where the dots represent the derivative with respect to $\lambda$. As the integrand in the volume does not depend on $v$ explicitly, one can define the conserved quantity $E$ as
\begin{equation}
E=-\frac{\partial\mathcal{L}}{\partial \dot{v}}=\frac{1}{z^{3}} \frac{f\mathrm{e}^{-\chi}\dot{v}+\mathrm{e}^{-\chi/2}\dot{z}}{\sqrt{-f\mathrm{e}^{-\chi}\dot{v}^{2}-2\mathrm{e}^{-\chi/2}\dot{v}\dot{z}}}\,.
\end{equation}
Since the extremal volume~\eqref{para-volume} is parameterization invariant, one can choose special parameter $\lambda$ so that the radial volume element is unit, \emph{i.e.}
\begin{equation}\label{para-equ}
\frac{1}{z^{3}}\sqrt{-f\mathrm{e}^{-\chi}\dot{v}^{2}-2\mathrm{e}^{-\chi/2} \dot{v}\dot{z}}=1\,.
\end{equation}
As a result, the conserved quantity simplifies to
\begin{equation}\label{conser-e}
E=\frac{1}{z^{6}}(f\mathrm{e}^{-\chi}\dot{v}+\mathrm{e}^{-\chi/2}\dot{z})\,.
\end{equation}
Combining~\eqref{para-equ} and~\eqref{conser-e}, one can obtain another equation about $\dot{z}$.
\begin{equation}\label{dotz-equ}
E^{2}z^{6}\mathrm{e}^{\chi}+f=\frac{\dot{z}^{2}}{z^{6}}.
\end{equation}
Substituting~\eqref{para-equ} and~\eqref{dotz-equ} into~\eqref{para-volume}, one finds that the extreme volume reads
\begin{equation}\label{extrm-vol}
\frac{\mathcal{V}}{2\Omega_2}=\int \mathrm{d}\lambda=\int_{z_{\max}}^{\epsilon} \frac{\mathrm{d}z}{\dot{z}}=\int_{z_{\max}}^{\epsilon} \frac{\mathrm{d}z}{z^{6}\mathrm{e}^{\chi/2}\sqrt{E^{2}+f\mathrm{e}^{-\chi}z^{-6}}}\,,
\end{equation}
where $\epsilon$ is the cutoff near the boundary. The maximum value of $z$ in the extreme slice is determined by setting $\dot{z}=0$ in~\eqref{dotz-equ}, which means
\begin{equation}\label{zmax-equ}
   E^2z^6\mathrm{e}^{\chi}+f=0\,.
\end{equation}

Moreover, the turning point $z_{\max}$ is inside the horizon, hence $\dot{z}=0$, $\dot{v}>0$ and $f(z_{\max})<0$. It's obvious that the conserved quantity $E$ is negative at the maximum radius. Considering~\eqref{conser-e}, one can derive from the ingoing coordinate that
\begin{equation}\label{integrate-v}
t_{R}-F(0)+F(z_{\max})=\int_{z_{\max}}^{\epsilon} \mathrm{d}z \frac{\dot{v}}{\dot{z}}=\int_{z_{\max}}^{\epsilon} \mathrm{d}z \left(\frac{E \mathrm{e}^{\chi/2}}{f \sqrt{E^{2}+f\mathrm{e}^{-\chi}z^{-6}}}-\frac{\mathrm{e}^{\chi/2}}{f}\right)\,.
\end{equation}
Then the extreme volume~(\ref{extrm-vol}) can be written in the following form:
\begin{equation}\label{final-vol}
\frac{\mathcal{V}}{2\Omega_2}=\int_{z_{\max}}^{\epsilon}\left(\frac{\mathrm{e}^{\chi/2}\sqrt{E^{2}+f\mathrm{e}^{-\chi}z^{-6}}}{f}-\frac{\mathrm{e}^{\chi/2}E}{f(z)}\right)\mathrm{d}z-E\bigg(t_{R}-F(0)+F(z_{\max})\bigg)\,.
\end{equation}
For simplicity, we consider a particular type of boundary time evolution with $t_{L}=t_{R}=\frac{t}{2}$. With the relation $t_{R}=t/2$ and $\mathcal{C}_{\mathcal{V}}=\mathcal{V}/G_N L$, by taking derivative of equation~\eqref{final-vol} with respect to time $t$ one can obtain
\begin{equation}
\frac{\mathrm{d}\mathcal{C}_{\mathcal{V}}}{\mathrm{d}t}=-\frac{\Omega_2}{G_N L} E\,.
\end{equation}
Here, we have used~\eqref{integrate-v} to simplify the contribution from the derivative acting on $z_{\max}$. Thus, using the conserved quantity $E=-\sqrt{-f(z_{\max})\mathrm{e}^{-\chi(z_{\max})}z_{\max}^{-6}}$, we obtain the growth rate of the complexity
\begin{equation}
\frac{\mathrm{d}\mathcal{C}_{\mathcal{V}}}{\mathrm{d}t}=\frac{\Omega_2}{G_N L} \sqrt{g_{tt}(z_{\max})}z_{\max}^{-2}\,.
\end{equation}
\begin{figure}[H]
\centering
\includegraphics[width=0.49\textwidth]{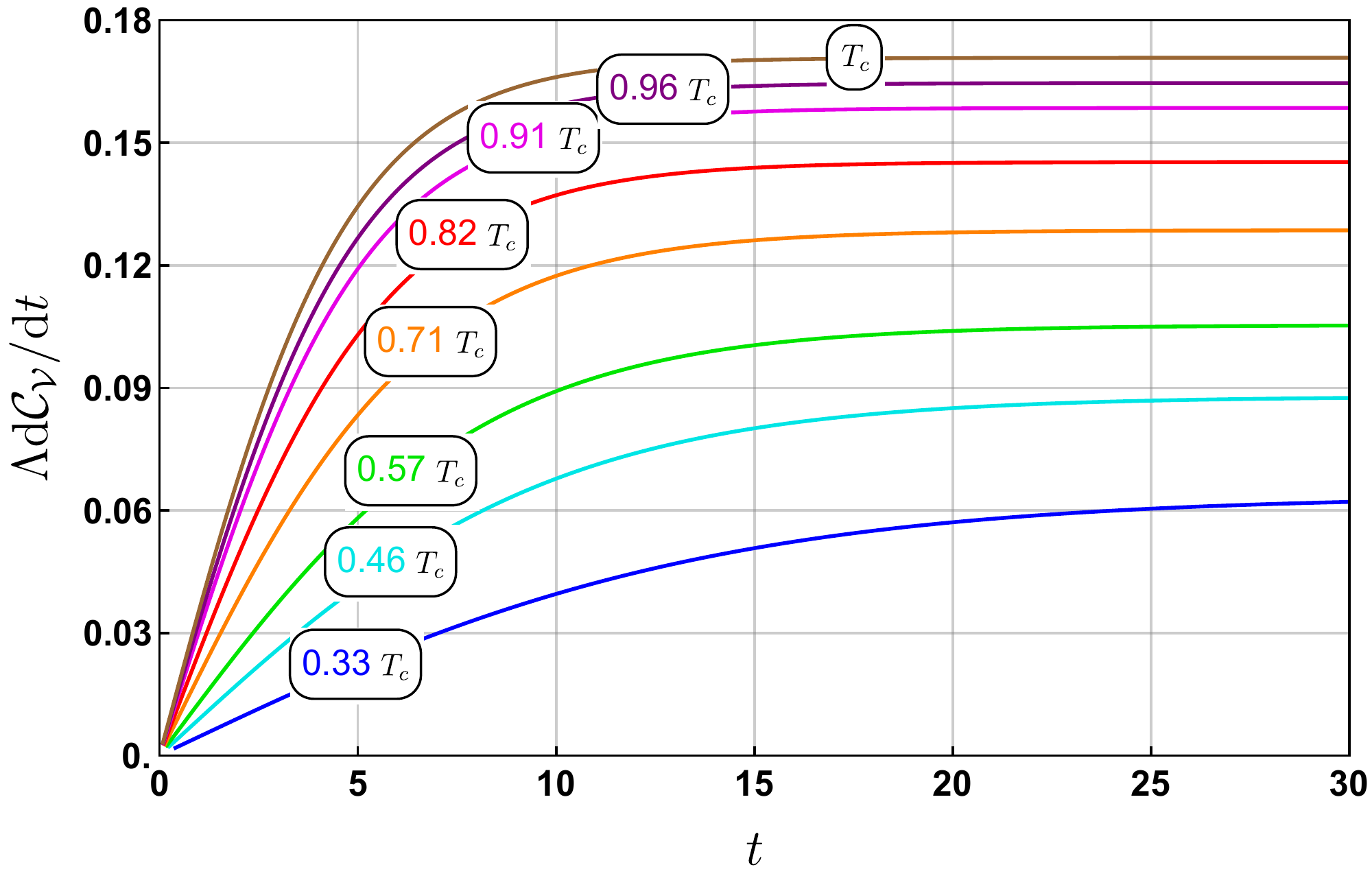}
\includegraphics[width=0.49\textwidth]{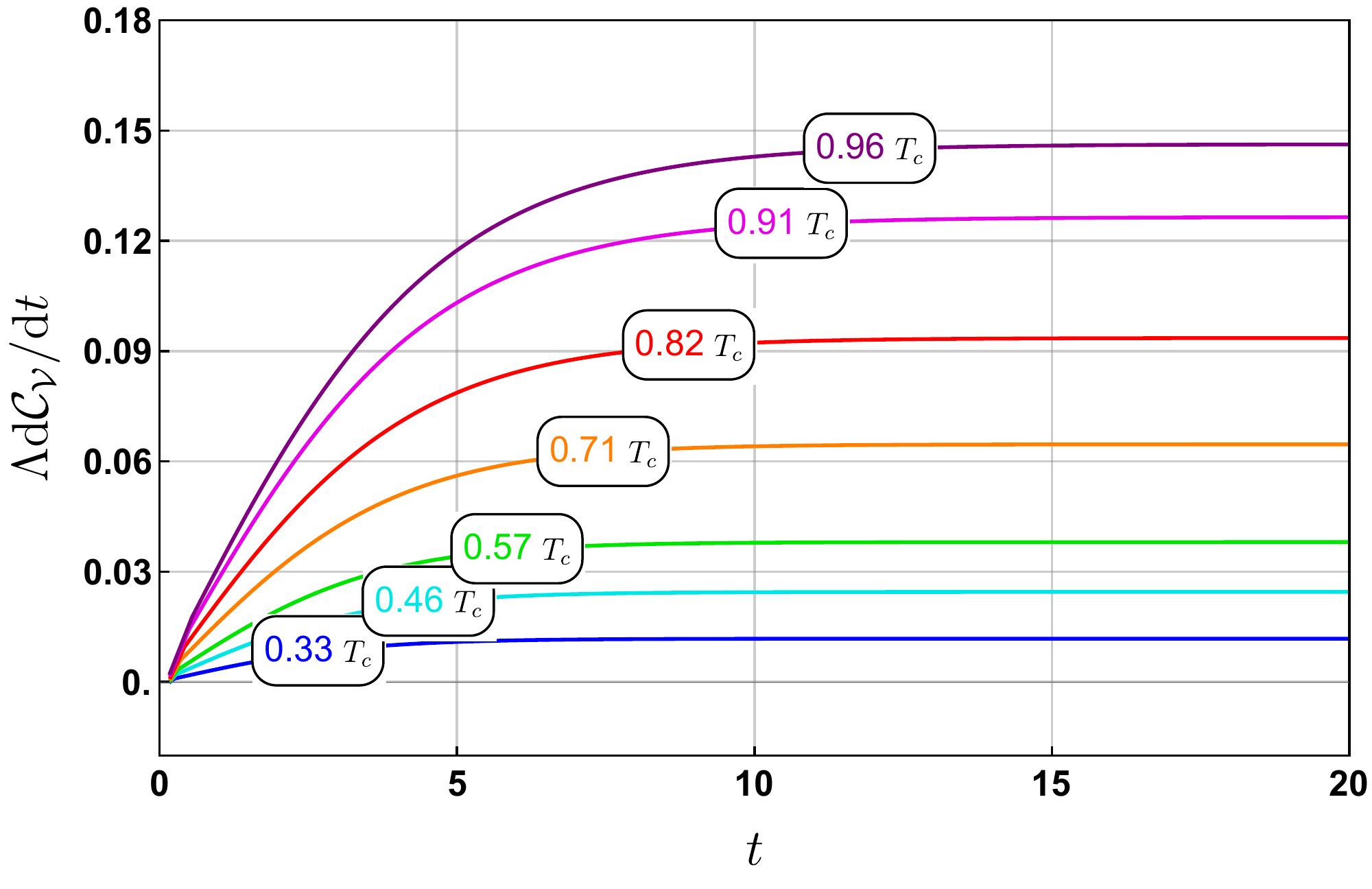}
\caption{The complexity growth rate of CV conjecture for the normal phase (left) and the superconducting phase (right) at different temperatures. The complexcity growth rate of the superconducting phase is smaller than the normal phase below the critical temperature. Here $\Lambda=10G_{N}L/\Omega_{2}$ and $\mu=1$.}
\label{Fig:grcv}
\end{figure}
The complexity growth rate using the CV conjecture for different temperatures is shown in Fig.~\ref{Fig:grcv}. We find that below $T_c$ the complexity growth rate of the superconducting phase is smaller than the normal case. Moreover, at $T=T_{c}$ where the scalar hair is vanishingly small, the complexity growth rate is continuous from the normal phase to the superconducting phase, see the right panel of Fig.~\ref{Fig:cav-late}. This is in contrast to the one by the CA duality in the main text, for which the complexity growth rate is discontinuous at $T_c$ (the left panel of Fig.~\ref{Fig:cav-late}). This is due to the fact that the computation of the CV conjecture is not sensitive to the detailed structure of the singularity, while, for the CA conjecture, one has to know the detailed properties of the singularity as the WdW patch touches the singularity in general.

\bibliographystyle{JHEP}
%\bibliography{ref}

%\providecommand{\href}[2]{#2}\begingroup\raggedright\begin{thebibliography}{1}
\providecommand{\href}[2]{#2}\begingroup\raggedright

\end{document}